\documentclass[11pt]{article}
\usepackage{epsfig,amsmath,graphicx,amssymb,overpic,cite,amsthm}

\def\be{\begin{equation}}
\def\ee{\end{equation}}
\def\bee{\begin{eqnarray}}
\def\ene{\end{eqnarray}}
\def\bes{\begin{subequations}}
\def\ees{\end{subequations}}
\def\no{\nonumber}
\def\Re{{\rm Re}\,}

\def\d{\displaystyle}
\def\l{\left}
\def\r{\right}

\begin{document}

\baselineskip=12pt
\renewcommand {\thefootnote}{\dag}
\renewcommand {\thefootnote}{\ddag}
\renewcommand {\thefootnote}{ }

\pagestyle{plain}

\begin{center}
\baselineskip=16pt \leftline{} \vspace{-.3in} {\Large \bf Initial-boundary value problem for an integrable spin-1 Gross-Pitaevskii system with a $4\times 4$ Lax pair \\ on a finite interval } \\[0.2in]
\end{center}

\begin{center}
 Zhenya Yan\footnote{{\it Email address}: zyyan@mmrc.iss.ac.cn}
 \\[0.08in]
{\it \small Key Laboratory of Mathematics Mechanization, Institute
of Systems Science, AMSS, \\ Chinese Academy of Sciences, Beijing 100190, China\\
School of Mathematical Sciences, University of Chinese Academy of Sciences, Beijing 100049, China}
\end{center}

{\baselineskip=13pt

\vspace{0.1in}
\begin{tabular}{p{15cm}}
 \hline \\
\end{tabular}
\begin{abstract} \small \baselineskip=15pt
     In this paper, we explore the initial-boundary value (IBV) problem for an integrable spin-1 Gross-Pitaevskii system
     with a $4\times 4$ Lax pair on the finite interval $x\in [0, L]$ by extending the Fokas unified transform approach.  The solution of this system can be expressed in terms of the solution of a $4\times 4$ matrix Riemann-Hilbert (RH) problem formulated in the complex $k$-plane. Furthermore, the relevant jump matrices with explicit $(x, t)$-dependence of the matrix RH problem can be explicitly found via three spectral functions $\{s(k),\, S(k),\, S_L(k)\}$ arising from the initial data and the Dirichlet-Neumann boundary conditions at $x=0$ and $x=L$, respectively. The global relation is also found to deduce two distinct but equivalent types of representations (i.e., one via the large $k$ of asymptotics of the eigenfunctions and another one in terms of the Gel'fand-Levitan-Marchenko (GLM) approach) for the Dirichlet and Neumann boundary value problems. In particular, the formulae for IBV problems on the finite interval can reduce to ones on a half-line as the length $L$ of the interval approaches to infinity. Moreover, we also present the linearizable boundary conditions for the GLM representations.


\vspace{0.1in} \noindent  Keywords: Integrable spin-1 Gross-Pitaevskii system; Initial-boundary value problem; Riemann-Hilbert problem; Global relation;  Finite interval; GLM representation; Maps between Dirichlet and Neumann problems

\vspace{0.05in} \noindent Mathematics Subject Classification numbers: 37K15; 35Q15; 34A55
 \vspace{0.05in}
\end{abstract}

\vspace{-0.05in}
\begin{tabular}{p{15cm}}
  \hline \\
\end{tabular}

\vspace{0.15in}

\baselineskip=15pt

\section{Introduction}

\quad At the end of the 1960s, the well-known inverse scattering transform (IST)~\cite{ist} (also called nonlinear Fourier transform) was first presented to solve the {\it initial value problem} for the KdV equation starting from its two linear eigenvalue equations (also called the Lax pair~\cite{lax}). After that, the IST was used to solve many integrable {\it nonlinear} evolution equations (NLEEs) with Lax pairs (see, e.g., Ref.~\cite{soliton,soliton2} and references therein), in which the initial conditions were usually chosen as the constants or plane waves to generate bright or dark solitons and breathers. In the early 1990s, Deift and Zhou~\cite{rh} developed the IST method to present the powerful nonlinear steepest descent method to analytically study the long-time asymptotics of the Cauchy problems of the (1+1)-dimensional integrable NLEEs such as the mKdV equation and nonlinear Schr\"odinger equation. At the end of the 1990s, Fokas~\cite{f1} further extended the idea of the IST to present a unified method studying {\it initial-boundary value problems} for both linear and nonlinear integrable equations with Lax pairs (see Refs.~\cite{f2,f3,f4,f5,f6,f7} for the details). Particularly, the Fokas method is used to study integrable nonlinear PDEs by means of the simultaneous spectral analysis of both parts of the Lax pairs and the global relation among the spectral functions. This differs from the classical IST, in which the spectral analysis of only one part of the Lax pairs was considered. The Fokas method well unites the key ideas of the IST with the Riemann-Hilbert problems~\cite{f4,f6}.

The Fokas method, as a novel and effective method, has been used to study the initial-boundary value (IBV) problems for the linear PDEs and some integrable NLEEs with $2\times 2$ Lax pairs on the half-line and the finite interval such as, the nonlinear Schr\"odinger (NLS) equation~\cite{f1, nls1,nls2,nls3,nls4}, the sine-Gordon equation~\cite{sg,sg2}, the KdV equation~\cite{kdv,kdv2}, the modified KdV equation~\cite{mkdv0, mkdv1,mkdv1b,mkdv2}, the derivative NLS equation~\cite{dnls1}, and etc. (see, e.g., Refs.~\cite{glm1,glm2,glm3,f6,m1,m2,m3} and references therein). Recently, Lenells further developed the Fokas method to analyze the IBV problems for integrable NLEEs
with $3\times 3$ Lax pairs on the half-line~\cite{le12}. After that, the modified approach was applied in the IBV problems of
other integrable NLEEs with $3\times 3$ Lax pairs on the half-line or the finite interval, such as the Degasperis-Procesi equation~\cite{ds}, the Sasa-Satsuma equation~\cite{ss}, the coupled NLS equations~\cite{cnls1,cnls2,cnls3}, and the Ostrovsky-Vakhnenko equation~\cite{os}.

More recently, we~\cite{yangp-h17} successfully extended the ideas of the Fokas method~\cite{f1} for the $2\times 2$ Lax pairs and its extension~\cite{le12} for the $3\times 3$ Lax pairs to study the IBV problems for the integrable spin-1 GP system~\cite{sgp, sgp3,sgp4} with a $4\times 4$ Lax pair on the half-line $0<x<\infty$
\bee
\label{pnls}
\left\{\begin{array}{l}
\d i \frac{\partial q_{1}}{\partial t}+ \frac{\partial^2 q_{1}}{\partial x^2}
-2\alpha\left(|q_1|^{2}+2|q_0|^2\right)q_1-2\alpha\beta q_0^2\bar{q}_{-1}=0, \vspace{0.1in}\\
\d i\frac{\partial q_{0}}{\partial t}+ \frac{\partial^2 q_{0}}{\partial x^2}-2\alpha\left(|q_1|^{2}+|q_0|^2+|q_{-1}|^2\right)q_0-2\alpha\beta q_1q_{-1}\bar{q}_0=0,  \vspace{0.1in}\\
\d i \frac{\partial q_{-1}}{\partial t}+ \frac{\partial^2 q_{-1}}{\partial x^2}-2\alpha\left(2|q_0|^{2}+|q_{-1}|^2\right)q_{-1}-2\alpha\beta q_0^2\bar{q}_1=0, \qquad
 \end{array}\right. \quad \alpha^2=\beta^2=1,
 \ene
with the IBV conditions
\bee\label{ibv0}
 \left\{\!\!\!\begin{array}{lll}
 {\rm Initial\,\, conditions:} & q_j(x, t=0)=q_{0j}(x),& \! j=1,0,-1,\quad 0<x<\infty, \vspace{0.1in} \\
 {\rm Dirichlet \,\, boundary \,\, conditions:} & q_j(x=0, t)=u_{0j}(t), &\! j=1,0,-1,\quad 0<t<\infty, \vspace{0.1in} \\
 {\rm Neumann \,\, boundary \,\, conditions:} & q_{jx}(x=0, t)=u_{1j}(t),&\! j=1,0,-1,\quad 0<t<\infty,
 \end{array}\right.
  \ene
 where the overbar stands for the complex conjugate. The spin-1 GP system (\ref{pnls}) can describe soliton dynamics of an $F=1$ spinor Bose-Einstein condensates~\cite{sgp}. The four types of parameters: $(\alpha, \beta)=\{(1,1), (1, -1), (-1, 1), (-1, -1)\}$ in the spin-1 GP system (\ref{pnls})
 correspond to the four roles of the self-cross-phase modulation (nonlinearity) and spin-exchange modulation,  respectively, that is, (attractive, attractive), (attractive, repulsive), (repulsive, attractive), and (repulsive, repulsive).

 In this paper, we extend the idea in Ref.~\cite{yangp-h17} from the half-line to the finite interval $0<x<L<\infty$.
The aim of this paper is to develop a methodology for analyzing the integrable spin-1 GP system (\ref{pnls}) with  the following IBV problem
\bee\label{ibv}
 \left\{\begin{array}{lll}
 \!\!\!{\rm Initial\,\, data:} &\!\! q_j(x, t=0)=q_{0j}(x),&   0<x<L,  \vspace{0.1in} \\
 \!\!\!{\rm Dirichlet \,\, boundary \,\, data:} & \!\! q_j(x=0, t)=u_{0j}(t), & q_j(x=L, t)=v_{0j}(t), \,\, 0<t<T, \vspace{0.1in} \\
 \!\!\!{\rm Neumann \,\, boundary \,\, data:} &  q_{jx}(x=0, t)=u_{1j}(t),& q_{jx}(x=L, t)=v_{1j}(t), \,\, 0<t<T,
 \end{array}\right. 
  \ene
$j=1,0,-1$, on the finite interval
\bee
\Omega=\left\{(x,t)\,|\, x\in [0, L],\, t\in [0, T]\right\}
\ene with $L>0$ and $T>0$ being the fixed finite length and time, respectively, the initial data $q_{0j}(x),\,j=1,0,-1$  and boundary data $\{u_{0j}(t), \, u_{1j}(t),\, j=1,0, -1\}$ are sufficiently smooth and compatible at points $(x,t)=(0, 0)$.
where the initial data $q_{0j}(x),\,j=1,0,-1$  and boundary data $\{u_{sj}(t), \, v_{sj}(t),\, s=0,1; \, j=1,0,-1\}$ are sufficiently smooth and compatible at points $(x,t)=(0, 0),\, (L, 0)$, respectively.

The main steps for analyzing the IBV problem for the integrable spin-1 GP system (\ref{pnls}) with Eq.~(\ref{ibv}) are listed as follows:

{\it Step 1.} Suppose that a sufficiently smooth solution $\{q_j(x,t),\, j=1,0,-1\},\, 0<x<L,\, 0<t<T$ of the integrable spin-1 GP system (\ref{pnls}) exists, we implement the direct spectral analysis of the corresponding Lax pair in order to explore these points:

\begin{itemize}

\item {} Introduce appropriate solutions of the exact one-form of the modified Lax pair (\ref{mulax}), which are bounded and analytic for the isospectral parameter $k$ in domains to form a partition of the Riemann sphere.

\item{} Introduce the following matrix-valued  spectral functions

\begin{itemize}

\item [(a)] $s(k)$ is determined using the initial data $q_j(x, 0)=q_{0j}(x),\, j=1,0,-1,\, 0<x<L$;

\item[(b)] $S(k)$ is generated using the boundary data at $x=0$,  $q_j(x=0, t)=u_{0j}(t),\, q_{jx}(x=0, t)=u_{1j}(t),\,  j=1,0,-1,\, 0<t<T$;

\item[(c)] $S_L(k)$ is determined using the boundary data  at $x=L$, $q_j(x=L, t)=v_{0j}(t),\, q_{jx}(x=L, t)=v_{1j}(t),\,  j=1,0,-1,\, 0<t<T$;
 \end{itemize}

\item {} Show that these above-mentioned spectral functions satisfy a global relation, which implies that
 the initial-boundary value conditions can not be chosen arbitrary.
 \end{itemize}

{\it Step 2}. Use the spectral functions $\{s(k),\, S(k),\, S_L(k)\}$ to determine a regular Reimann-Hilbert problem, whose solution can generate a solution of the spin-1 GP system (\ref{pnls}).

{\it Step 3}. The Gel'fand-Levitan-Marchenko (GLM) representations can also given for the IBV problem of system (\ref{pnls}).

The rest of this paper is organized as follows. In Sec. 2, we introduce the $4\times 4$ Lax pair of Eq.~(\ref{pnls}) and explore
its spectral analysis such as the eigenfunctions, the jump matrices, and the global relation. Sec. 3 exhibits the corresponding
$4\times 4$ matrix RH problem in terms of the jump matrices found in Sec. 2. The global relation is found to generate the map between the Dirichlet and Neumann boundary values in Sec. 4. Particularly, the relevant formulae for boundary value problems on the finite interval can reduce to ones on the half-line as the length of the interval approaches to infinity. In Sec. 5, we give the GLM representations of the eigenfunctions in terms of the global relation. Moreover, we also show that the GLM representations are equivalent to  ones in Sec. 4 and present the linearizable boundary conditions for the GLM representations. Finally, we give the conclusions and discussions.

\section{A $4 \times 4$ Lax pair and its spectral analysis}

\subsection*{\it 2.1.\, The closed one-form}

\quad The integrable spin-1 GP system  (\ref{pnls}) can be regarded as a compatibility condition of the following $4 \times 4$ Lax pair formulation~\cite{sgp,yangp-h17}
\bee \label{lax}
\left\{\begin{array}{l}
                \psi_x+ik\sigma_4\psi=U(x,t)\psi,    \vspace{0.1in}  \\
                \psi_t+2ik^2\sigma_4\psi=V(x,t,k)\psi,
                 \end{array}\right.
    \ene
where $\psi\equiv \psi(x,t,k)$ is a $4\times 1$ column vector-valued or 4$\times$4 matrix-valued eigenfunction, $k\in \mathbb{C}$ is an iso-spectral parameter, $\sigma_4={\rm diag}(1,1,-1,-1)$, and the $4 \times 4$ matrix-valued functions $U(x,t)$ and $V(x,t,k)$ are given as
\bee\label{uv}
U(x,t)=\left(\begin{array}{cccc}
            0 & 0 & q_1(x,t) & q_0(x,t) \vspace{0.05in}\\
            0&  0 & \beta q_0(x,t) & q_{-1}(x,t) \vspace{0.05in}\\
            \alpha \bar{q}_1(x,t) & \alpha\beta \bar{q}_0(x,t) & 0 & 0 \vspace{0.05in}\\
            \alpha \bar{q}_0(x,t) & \alpha \bar{q}_{-1}(x,t) & 0 & 0
            \end{array}\right), \quad \bar{U}^T(x,t)=\alpha U(x,t),
            \ene
and
\bee\begin{array}{l}
 V(x,t,k)=2kU+i\sigma_4(U_{x}-U^2) \vspace{0.1in}\\
    \quad =\left(\begin{array}{cccc}
                  -i\alpha(|q_1|^2+|q_0|^2) & -i\alpha(\beta q_1\bar{q}_0+q_0\bar{q}_{-1}) & iq_{1x}+2kq_1 & iq_{0x}+2kq_0 \vspace{0.1in}\\
      -i\alpha(\beta q_0\bar{q}_1+q_{-1}\bar{q}_0) & -i\alpha(|q_{-1}|^2+|q_0|^2) & \beta (iq_{0x}+2kq_0)  & iq_{-1x}+2kq_{-1} \vspace{0.1in}\\
      \alpha(-i\bar{q}_{1x}+2k\bar{q}_1) & \alpha\beta(-i\bar{q}_{0x}+2k\bar{q}_0) & i\alpha(|q_1|^2+|q_0|^2) & i\alpha(\beta q_{-1}\bar{q}_0+q_0\bar{q}_{1})  \vspace{0.1in}\\
           \alpha (-i\bar{q}_{0x}+2k\bar{q}_0) & \alpha(-i\bar{q}_{-1x}+2k\bar{q}_{-1}) & i\alpha(\beta q_0\bar{q}_{-1}+q_{1}\bar{q}_0) & i\alpha(|q_{-1}|^2+|q_0|^2)
            \end{array}\right).
           \end{array}
            \ene
where the potential function $\{q_j(x,t),\, j=1,0,-1\}$ satisfies the spin-1 GP equations (\ref{pnls}).

 Introduce a new eigenfunction $\mu(x,t,k)$ defined by
 \bee\label{mud}
 \mu(x,t,k)=\psi(x,t,k)e^{i(kx+2k^2t)\sigma_4},
 \ene
such that the Lax pair (\ref{lax}) becomes an equivalent form
\bee\label{mulax}
    \left\{     \begin{array}{l}
                 \mu_x+ik[\sigma_4,\mu]= U(x,t)\mu,    \vspace{0.1in}            \\
                 \mu_t+2ik^2[\sigma_4,\mu]=V(x,t,k)\mu,
                 \end{array}\right.
    \ene
where $[\sigma_4, \mu]\equiv\sigma_4\mu-\mu\sigma_4$.

Let $\hat{\sigma}_4$ denote the commutator with respect to $\sigma_4$ and the operator acting on a $4\times 4$ matrix $A$ by
$\hat{\sigma}_4A=[\sigma_4, A]=\sigma_4A-A\sigma_4$ such that $e^{\hat{\sigma}_4}A=e^{\sigma_4}Ae^{-\sigma_4}$, then the Lax pair (\ref{mulax}) can be written as a full derivative form
\bee
d\left[e^{i(kx+2k^2t)\hat{\sigma}_4}\mu(x,t,k)\right]=W(x,t,k),
\label{dform}
\ene
where $W(x,t,k)$ is the exact one-form defined by
\bee \label{w}
 W(x,t,k)=e^{i(kx+2k^2t)\hat{\sigma}_4}\l[U(x,t)\mu(x,t,k)dx+V(x,t,k)\mu(x,t,k)dt\r].
 \ene

Notice that Eq.~(\ref{dform}) can be used to obtain an expression for $\mu(x,t,k)$ via the fundamental theorem of calculus.

\subsection*{\it 2.2. \, The eigenfunctions $\{\mu_j(x,t,k)\}_1^4$ }

\quad For any point $(x,t)\in \Omega=\{(x,t)| x\in [0, L],\, t\in [0, T]\}$,  we assume that $\{\gamma_j\}_1^4$ denote the four contours connecting
fours vertexes
\bee \no
(x_1, t_1)=(0, T),\quad  (x_2, t_2)=(0, 0),\quad  (x_3, t_3)=(L, 0),\quad  (x_4, t_4)=(L, T),
\ene
 of the rectangle $\Omega$ to the point $(x,t)$ (see Fig.~\ref{ga}). Thus for any point $(\xi, \tau) \in \gamma_j,\, j=1,2,3,4$, we have the relations on the contours:
\bee\label{gammad}
 \begin{array}{rll}
 \gamma_1: & x-\xi \geq 0, & t-\tau \leq 0, \vspace{0.1in}\\
 \gamma_2: & x-\xi \geq 0, & t-\tau \geq 0, \vspace{0.1in}\\
 \gamma_3=-\gamma_1: & x-\xi \leq 0, & t-\tau \geq 0, \vspace{0.1in}\\
 \gamma_4=-\gamma_2: & x-\xi \leq 0, & t-\tau \leq 0,
 \end{array}
\ene
where the negative sign in $\gamma_3=-\gamma_1$ and $\gamma_4=-\gamma_2$ denotes the opposite directions.

\begin{figure}
\begin{center}
{\scalebox{0.7}[0.7]{\includegraphics{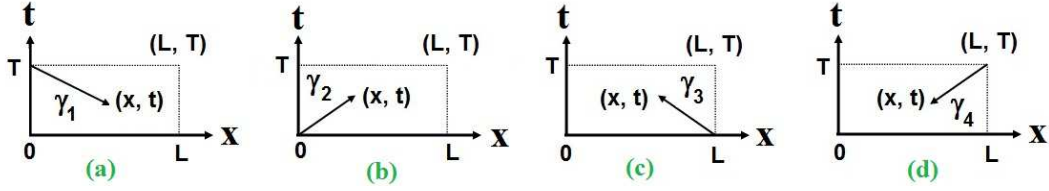}}}
\end{center}
\vspace{-0.15in}\caption{The contours $\gamma_j \, (j=1,2,3,4)$ exhibited in the finite region $\Omega=\{(x,t)| x\in [0, L],\, t\in [0, T]\}$. }
\label{ga}
\end{figure}

We assume that system (\ref{pnls}) possesses a smooth complex-valued solution $\{q_j(x,t),\, j=1,0,-1\}$ in the domain $\Omega$ (if $T=\infty$ then we assume that the solution $\{q_j(x,t),\, j=1,0,-1\}$ is a sufficient decay as $t\to \infty$). It follows from the Lax pair (\ref{mulax}) that we can define its four eigenfunctions (sectionally holomorphic functions) $\{\mu_j(x,t,k)\}_1^4$ on the four contours $\{\gamma_j\}_1^4$
\bee \label{mus}
\begin{array}{rl}
\mu_j(x,t,k)=&\!\!\! \d\mathbb{I}+\int_{\gamma_j}e^{-i(kx+2k^2t)\hat{\sigma}_4}W_j(\xi,\tau,k)\vspace{0.1in}\\
=&\!\!\!\d\mathbb{I}+\int_{(x_j, t_j)}^{(x,t)}e^{-i(kx+2k^2t)\hat{\sigma}_4}W_j(\xi,\tau,k),\quad j=1,2,3,4.
\end{array}
\ene
in terms of the Volterra integral equations, where $\mathbb{I}={\rm diag}(1,1,1,1)$, the integral is over a piecewise smooth curve from $(x_j, t_j)$ to $(x,t)$,  $W_j(x,t,k)$'s are defined by Eq.~(\ref{w}) with $\mu(x,t,k)$ replaced by $\mu_j(x,t,k)$'s. Since the one-forms $W_j(x,t,k)$'s are closed, thus $\mu_j(x,t,k)$'s are independent of the path of integration. The integral Eq.~(\ref{mus}) reduces to
\bee\label{musg}
\begin{array}{rl}
\mu_j(x,t,k)=&\d \mathbb{I}+\int_{x_j}^x e^{-ik(x-\xi)\hat{\sigma}_4}(U\mu_j)(\xi,t,k)d\xi \vspace{0.1in}\\
   & \d + e^{-ik(x-x_j)\hat{\sigma}_4}\int_{t_j}^te^{-2ik^2(t-\tau)\hat{\sigma}_4} (V\mu_j)(x_j,\tau,k)d\tau, \quad j=1,2,3,4.
   \end{array}
\ene
if the paths of integration are chosen to be parallel to the $x$ and $t$ axes.

Eq.~(\ref{musg}) implies that the four columns of the matrix $\mu_j(x,t,k)$ contain, respectively, these exponentials
\bes\label{muc} \bee
&\quad [\mu_j(x,t,k)]_s : \,\,  e^{2i[k(x-\xi)+2k^2(t-\tau)]},\quad e^{2i[k(x-\xi)+2k^2(t-\tau)]}, \quad s=1,2; j=1,2,3,4, \vspace{0.1in}\\
&\qquad [\mu_j(x,t,k)]_s : \,\, e^{-2i[k(x-\xi)+2k^2(t-\tau)]},\quad e^{-2i[k(x-\xi)+2k^2(t-\tau)]}, \quad s=3,4; j=1,2,3,4,
\ene\ees

To analyze the bounded regions of the eigenfunctions $\{\mu_j(x,t,k)\}_1^4$, we need to use the curve
\bee\no
 K=\{k\in \mathbb{C} |  (\Re f(k))(\Re g(k))=0,\, f(k)=ik,\, g(k)=ik^2\},
  \ene
 to separate the complex $k$-plane into four domains (see Fig.~\ref{kplane}):
\bee\left\{
\begin{array}{l}
D_1=\{k\in\mathbb{C} \,|\, \Re f(k)<0 \,\, {\rm and} \,\,  \Re g(k)<0\}, \vspace{0.1in} \\
D_2=\{k\in\mathbb{C} \,|\, \Re f(k)<0 \,\, {\rm and} \,\,  \Re g(k)>0\}, \vspace{0.1in}\\
D_3=\{k\in\mathbb{C} \,|\,\Re f(k)>0 \,\, {\rm and} \,\, \Re g(k)<0\}, \vspace{0.1in}\\
D_4=\{k\in\mathbb{C} \,|\, \Re f(k)>0\,\, {\rm and} \,\,  \Re g(k)>0\},
\end{array}\right.
\label{d}
\ene
which imply that $D_1$ and $D_3$ ($D_2$ and $D_4$) are symmetric about the origin of the complex $k$-plane.

Therefore, it follows from Eqs.~(\ref{gammad}), (\ref{muc}) and (\ref{d}) that the regions, where the distinct columns of eigenfunctions $\{\mu_j(x,t,k)\}_1^4$ are bounded and analytic in the complex $k$-plane, are given below:
\bee \label{muregion}
\left\{\begin{array}{l}
 \mu_1(x,t,k): (f_- \cap g_+,\, f_- \cap g_+,\, f_+ \cap g_-,\, f_+ \cap g_-)=: (D_2, D_2, D_3, D_3), \vspace{0.1in} \\
 \mu_2(x,t,k): (f_- \cap g_-,\, f_- \cap g_-,\, f_+ \cap g_+,\, f_+ \cap g_+)=: (D_1, D_1, D_4, D_4), \vspace{0.1in}\\
 \mu_3(x,t,k): (f_+ \cap g_-,\, f_+ \cap g_-,\, f_- \cap g_+,\, f_- \cap g_+)=: (D_3, D_3, D_2, D_2), \vspace{0.1in}\\
 \mu_4(x,t,k): (f_+ \cap g_+,\, f_+ \cap g_+,\, f_- \cap g_-,\, f_- \cap g_-)=: (D_4, D_4, D_1, D_1),
\end{array}\right.
\ene
where $f_+=: \Re f(k)>0,\, f_-=:\Re f(k)<0,\, g_+=: \Re g(k)>0$, and $g_-=:\Re g(k)<0$.

\begin{figure}[!t]
\begin{center}
{\scalebox{0.2}[0.2]{\includegraphics{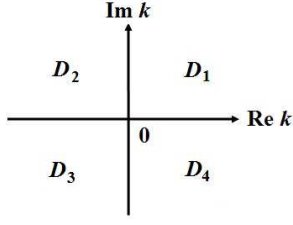}}}
\end{center}
\vspace{-0.25in}\caption{The regions $D_n,\, (n=1,2,3,4)$ separating the complex $k$-plane. }
\label{kplane}
\end{figure}

\subsection*{\it 2.3. \, The new matrix-valued functions $M_n(x,t,k)$'s}

\quad We introduce the matrix-valued solutions $M_n(x,t,k), \, n=1,2,3,4$ of Eq.~(\ref{mulax}) in the form
\bee \label{mn}
(M_n)_{lj}(x,t,k)=\delta_{lj}+\int_{(\gamma^n)_{lj}}\left[e^{-i(kx+2k^2t)\hat{\sigma}_4}W_n(\xi,\tau,k)\right]_{lj}, \quad  l,j=1,2,3,4, \quad k\in D_n,
\ene
 via the Volterra integral equations, where $\delta_{ij}=1$ for $l=j$ and $\delta_{ij}=0$ for $l\not =j$,  $W_n(x,t,k)$ is defined by
 \bee \label{wn}
 W_n(x,t,k)=e^{i(kx+2k^2t)\hat{\sigma}_4}\l[U(x,t)M_n(x,t,k)dx+V(x,t,k)M_n(x,t,k)dt\r].
 \ene
  and the contours $(\gamma^n)_{lj}$'s are defined as
\bee\label{gamma}
(\gamma^n)_{lj}=\left\{ \begin{array}{l}
\gamma_1,\,\,\, {\rm if} \,\,  \Re f_l(k)< \Re f_j(k) \,\,  {\rm and} \,\, \Re g_l(k)\geq \Re g_j(k), \vspace{0.1in} \\
\gamma_2,\,\,\, {\rm if}  \,\, \Re f_l(k)< \Re f_j(k) \,\, {\rm and} \,\, \Re g_l(k) < \Re g_j(k), \vspace{0.1in}\\
\gamma_ 3,\,\,\, {\rm if} \,\,  \Re f_l(k) \geq \Re f_j(k) \,\, {\rm and}\,\,  \Re g_l(k) \leq \Re g_j(k), \vspace{0.1in}\\
\gamma_4, \,\,\, {\rm if} \,\,  \Re f_l(k)\geq \Re f_j(k)  \,\, {\rm and} \,\, \Re g_l(k)\geq \Re g_j(k),
\end{array}\right. \quad k\in D_n,\quad l,j=1,2,3,4,
\ene
where $f_{1,2}(k)=-f_{3,4}(k)=-ik$ and $g_{1,2}(k)=-g_{3,4}(k)=-2ik^2$.

{\bf Remark.} To distinguish $(\gamma^n)_{lj}$'s to be the contour $\gamma_3$ or $\gamma_4$ for the special cases, $\Re f_l(k)=\Re f_j(k)$ and $\Re g_l(k)=\Re g_j(k)$, we choose them as $\gamma_3$ (or $\gamma_4)$ in these cases if we can determine that $\gamma_3$ (or $\gamma_4)$ must appear in other positions of the matrices $\gamma^{n}$.

The definition (\ref{gamma}) of $(\gamma^n)_{lj}$  can generate the  explicit expressions of $\gamma^n\, (n=1,2,3,4)$  as
\bee\begin{array}{rr}
\gamma^1=\left(
\begin{array}{cccc}
 \gamma_4 & \gamma_4 & \gamma_4 & \gamma_4 \\
 \gamma_4 & \gamma_4 & \gamma_4 & \gamma_4 \\
 \gamma_2 & \gamma_2 & \gamma_4 & \gamma_4 \\
 \gamma_2 & \gamma_2 & \gamma_4 & \gamma_4
 \end{array}
\right), &
\gamma^2=\left(
\begin{array}{cccc}
 \gamma_3 & \gamma_3 & \gamma_3 & \gamma_3 \\
 \gamma_3 & \gamma_3 & \gamma_3 & \gamma_3 \\
 \gamma_1 & \gamma_1 & \gamma_3 & \gamma_3 \\
 \gamma_1 & \gamma_1 & \gamma_3 & \gamma_3
 \end{array}
\right), \vspace{0.1in} \\
\gamma^3=(\gamma^2)^T=\left(
\begin{array}{cccc}
 \gamma_3 & \gamma_3 & \gamma_1 & \gamma_1 \\
 \gamma_3 & \gamma_3 & \gamma_1 & \gamma_1 \\
 \gamma_3 & \gamma_3 & \gamma_3 & \gamma_3 \\
 \gamma_3 & \gamma_3 & \gamma_3 & \gamma_3
 \end{array}
\right), &
\gamma^4=(\gamma^1)^T=\left(
\begin{array}{cccc}
 \gamma_4 & \gamma_4 & \gamma_2 & \gamma_2 \\
 \gamma_4 & \gamma_4 & \gamma_2 & \gamma_2 \\
 \gamma_4 & \gamma_4 & \gamma_4 & \gamma_4 \\
 \gamma_4 & \gamma_4 & \gamma_4 & \gamma_4
 \end{array}
\right),
\end{array}\ene

\vspace{0.1in}
\noindent {\bf Proposition 2.1.} {\it For the matrix-valued functions $M_n(x,t,k),\, n=1,2,3,4$ defined by Eq.~(\ref{mn}) for $k\in \bar{D}_n$ and $(x,t)\in \Omega$ and any fixed point $(x_0,t_0)$, $M_n(x_0,t_0,k)$'s are the bounded and analytic function of $k\in D_n$ away from a possible discrete set of singularity $\{k_j\}$ at which the Fredholm determinants vanish. Furthermore, $M_n(x,t,k)$'s also have the bounded and continuous extensions to $\bar{D}_n$ and
\bee\label{ml}
 M_n(x,t,k)=\mathbb{I}+O\left(\frac{1}{k}\right),\,\,k\in D_n,\,\,  n=1,2,3,4,\,\ k\to\infty,
\ene}

\noindent {\bf Proof.} Following the proof for the $3\times 3$ Lax pair in Ref.~\cite{le12}, we can also show
the bounedness and analyticity of the $4\times 4$ matrix $M_n(x,t,k)$. The substitution of the eigenfunction
\bee\no
\mu(x,t,k)=M_n(x,t,k)=M_n^{(0)}(x,t)+\sum_{j=1}^{\infty}\frac{M_n^{(j)}(x,t)}{k^j},\quad k \to \infty,\,\,  n=1,2,3,4,
\ene
into the $x$-part of the Lax pair (\ref{mulax}) can obtain Eq.~(\ref{ml}). $\square$ \\

Notice that the above-defined functions $M_n(x,t,k)$'s can be used to formulate a $4\times 4$ matrix Riemann-Hilbert problem.

\subsection*{\it 2.4. \,  The minors of eigenfunctions and Lax pair}

\quad The cofactor matrix $X^A$ (or the transpose of the adjugate) of a $4\times 4$ matrix $X$ is given by
\bee
{\rm adj}(X)^T=X^A=\left(\begin{array}{rrrr}
 m_{11}(X) & -m_{12}(X) &  m_{13}(X) &  -m_{14}(X) \vspace{0.05in}\\
 -m_{21}(X) & m_{22}(X) &  -m_{23}(X) &  m_{24}(X) \vspace{0.05in}\\
 m_{31}(X) & -m_{32}(X) &  m_{33}(X) &  -m_{34}(X) \vspace{0.05in}\\
 -m_{41}(X) & m_{42}(X) &  -m_{43}(X) &  m_{44}(X)
\end{array}\right),
\ene
where $m_{ij}(X)$ denotes the $(ij)$th minor of $X$,  and $(X^A)^TX ={\rm adj}(X) X=\det X$.

It follows from the Lax pair (\ref{mulax}) that the cofactor matrices $\{\mu_j^A(x,t,k)\}_1^4$ of the matrices $\{\mu_j(x,t,k)\}_1^4$ satisfy the modified Lax equation
\bee\label{mualax}
    \left\{   \begin{array}{l}
                 \mu_{j,x}^A(x,t,k)-ik\hat{\sigma}_4\mu_j^A(x,t,k)= -U^T(x,t)\mu_j^A(x,t,k),    \vspace{0.1in}            \\
                 \mu_{j,t}^A(x,t,k)-2ik^2\hat{\sigma}_4\mu_j^A(x,t,k)=-V^T(x,t,k)\mu_j^A(x,t,k),
                 \end{array} \right.
                 \ene
whose solutions can also be expressed as
\bee\begin{array}{rl}
\mu_j^A(x,t,k)=&\!\!\!\mathbb{I}-\d\int_{\gamma_j}e^{i[k(x-\xi)+2k^2(t-\tau)]\hat{\sigma}_4}\left[U^T(\xi, \tau)d\xi+V^T(\xi, \tau, k)d\tau\right]\mu_j^A(\xi, \tau, k) \vspace{0.1in}\\
=&\!\!\! \d \mathbb{I}-\int_{x_j}^x e^{ik(x-\xi)\hat{\sigma}_4}(U^T\mu_j^A)(\xi,t,k)d\xi  \vspace{0.1in}\\ &\quad \d -e^{ik(x-x_j)\hat{\sigma}_4}\int_{t_j}^te^{2ik^2(t-\tau)\hat{\sigma}_4} (V^T\mu_j^A)(x_j,\tau,k)d\tau,\quad j=1,2,3,4
\end{array}
\ene
using the Volterra integral equations, where $U^T(x,t,k)$ and $V^T(x,t,k)$ denote the  transposes of $U(x,t,k)$ and $V(x,t,k)$ given by Eq.~(\ref{uv}), respectively.

Therefore, the regions of boundedness of $\mu_j^A(x,t,k),\, j=1,2,3,4$ are given by
\bee \no \left\{\begin{array}{l}
 \mu_1^A(x,t,k) {\rm \,\, is \,\, bounded\,\, for\,\,} k\in (D_3, D_3, D_2, D_2), \vspace{0.1in}\\
 \mu_2^A(x,t,k) {\rm \,\, is \,\, bounded\,\, for\,\,} k\in (D_4, D_4, D_1, D_1), \vspace{0.1in}\\
 \mu_3^A(x,t,k) {\rm \,\, is \,\, bounded\,\, for\,\,} k\in (D_2, D_2, D_3, D_3), \vspace{0.1in}\\
 \mu_4^A(x,t,k) {\rm \,\, is \,\, bounded\,\, for\,\,} k\in (D_1, D_1, D_4, D_4),
\end{array}\right.
\ene
which are symmetric ones of $\mu_j$ about the $\Re k$-axis (cf. Eq.~(\ref{muregion})).

\subsection*{\it 2.5.\, Symmetries of eigenfunctions}

\quad Let
\bee
\check{U}(x,t, k)=-ik\sigma_4+U(x,t,k),\quad \check{V}(x,t, k)=-2ik^2\sigma_4+V(x,t,k).
\ene
Since
\bee
P_{\pm}\overline{\check{U}(x,t, \bar{k})}P_{\pm}=-\check{U}(x,t,k)^T, \quad P_{\pm}\overline{\check{V}(x,t, \bar{k})}P_{\pm}=-\check{V}(x,t,k)^T,
\ene
where
\bee
P_{\pm}=\left(\begin{array}{cccc} \pm\alpha & 0 & 0 & 0 \\
                            0 & \pm\alpha & 0 & 0 \\
                            0 & 0 & \mp 1 & 0 \\
                            0 & 0  & 0 & \mp 1
                            \end{array}\right), \quad P_{\pm}^2=\mathbb{I},\quad \alpha^2=1
\ene

According to Eq.~(\ref{mualax}), we have the following proposition:

\vspace{0.1in}
\noindent {\bf Proposition 2.2.} {\it The eigenfunction $\psi(x,t,k)$ of the Lax pair (\ref{lax}) and $\mu_j(x,t,k)$ of the Lax pair (\ref{mulax}) both possess the same symmetric relation
\bee\label{symmetry}
\begin{array}{l}
\psi^{-1}(x,t,k)=P_{\pm}\overline{\psi(x,t,\bar{k})}^TP_{\pm},\vspace{0.1in}\\
 \mu_j^{-1}(x,t,k)=P_{\pm}\overline{\mu_j(x,t,\bar{k})}^TP_{\pm},\quad j=1,2,3,4,
\end{array}
\ene }

Moreover, in the domains where $\mu_j$ is bounded, we have
\bee\no
 \mu_j(x,t,k)=\mathbb{I}+O\left(\frac{1}{k}\right),\quad k\to \infty, \quad j=1,2,3, 4
\ene
and
\bee\no
 {\rm det} [\mu_j(x,t,k)]=1, \quad j=1,2,3, 4
\ene
since the traces of the matrices  $U(x,t)$ and $V(x,t,k)$ are zero.

\subsection*{\it 2.6. \, The spectral functions}

\begin{figure}[!t]
\begin{center}
{\scalebox{0.3}[0.3]{\includegraphics{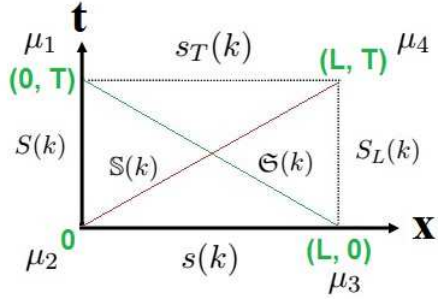}}}
\end{center}
\vspace{-0.2in}\caption{The relations among $\mu_j(x,t,k),\, j=1,2,3, 4$. }
\label{mu}
\end{figure}

\quad Since $\mu_j(x,t,k)$ are linearly dependent, thus we can have six relations $\{S(k), s(k), \mathbb{S}(k), S_L(k), s_T(k), \mathfrak{S}(k)\}$ between any two eigenfunctions $\mu_j$ (see Fig.~\ref{mu}), however, we know that these six relations are not dependent such that we only introduce three of those, that is, the three $4\times 4$ matrix-valued functions
$S(k),\, s(k)$ and $\mathbb{S}(k)$ between $\mu_j(x,t,k),\,j=1,3,4$ and $\mu_2(x,t,k)$ as
\bee\label{mu1234}\begin{array}{l}
\mu_1(x,t,k)=\mu_2(x,t,k)e^{-i(kx+2k^2t)\hat{\sigma}_4}S(k), \vspace{0.1in}\\
\mu_3(x,t,k)=\mu_2(x,t,k)e^{-i(kx+2k^2t)\hat{\sigma}_4}s(k), \vspace{0.1in}\\
\mu_4(x,t,k)=\mu_2(x,t,k)e^{-i(kx+2k^2t)\hat{\sigma}_4}\mathbb{S}(k),
\end{array}
\ene

Evaluating  system (\ref{mu1234}) at $(x,t)=(0,0)$ and the three equations in system (\ref{mu1234}) at $(x, t)=(0, T), (L, 0), (L, T)$, respectively, we find
\bee\label{sss}
\begin{array}{l}
 S(k)=\mu_1(0,0,k)=e^{2ik^2T\hat{\sigma}_4}\mu_2^{-1}(0,T,k),\vspace{0.1in}\\
 s(k)=\mu_3(0,0,k)=e^{ikL\hat{\sigma}_4}\mu_2^{-1}(L,0,k),\vspace{0.1in}\\
\mathbb{S}(k)=\mu_4(L,0,k)=e^{2ik^2T\hat{\sigma}_4}\mu_3^{-1}(L,T,k),
\end{array}
\ene

It follows from Eqs.~(\ref{mu1234}) and (\ref{sss}) that we can find the relations among $\{S(k), s(k), \mathbb{S}(k), S_L(k), s_T(k), \mathfrak{S}(k)\}$:
\begin{itemize}
\item [(I)] the relation between $\mu_3(x,t,k)$ and $\mu_1(x,t,k)$
\bee\begin{array}{rl}
\mu_3(x,t,k)=&\!\!\! \mu_1(x,t,k)e^{-i(kx+2k^2t)\hat{\sigma}_4}\mathfrak{S}(k),
\end{array}
\ene
with
\bee\label{sf}
\mathfrak{S}(k)=S^{-1}(k)s(k),
\ene
which generates the relation of the three edges of the triangle consisting of $(x_j, t_j),\, j=1,2,3$.

\item[(II)] the relation between $\mu_4(x,t,k)$ and $\mu_3(x,t,k)$
\bee\begin{array}{rl}
\mu_4(x,t,k)=&\!\!\! \mu_3(x,t,k)e^{-i[k(x-L)+2k^2(t-T)]\hat{\sigma}_4}\mu_3^{-1}(L, T, k) \vspace{0.1in}\\
=&\!\!\! \mu_3(x,t,k)e^{-i(kx+2k^2t)\hat{\sigma}_4}[s^{-1}(k)\mathbb{S}(k)] \vspace{0.1in}\\
=&\!\!\! \mu_3(x,t,k)e^{-i[k(x-L)+2k^2t]\hat{\sigma}_4}S_L(k),
\end{array}
\ene
with
\bee \label{sl}
\label{sr} S_L(k)=\mu_4(L, 0,k)=e^{2ik^2T\hat{\sigma}_4}\mu_3^{-1}(L, T, k)=e^{-ikL\hat{\sigma}_4}[s^{-1}(k)\mathbb{S}(k)],
\ene
that is,
\bee
 \mathbb{S}(k)=s(k)e^{ikL\hat{\sigma}_4}S_L(k),
\ene
which generates the relation of the three edges of the triangle consisting of $(x_j, t_j),\, j=2,3,4$.

\item[(III)] the relation between $\mu_1(x,t,k)$ and $\mu_4(x,t,k)$
\bee\begin{array}{rl}
\mu_4(x,t,k)=&\!\!\! \mu_1(x,t,k)e^{-i(kx+2k^2t)\hat{\sigma}_4}s_T(k),
\end{array}
\ene
with
\bee\label{st}
s_T(k)=S^{-1}(k)\mathbb{S}(k)=\mathfrak{S}(k)e^{ikL\hat{\sigma}_4}S_L(k)=S^{-1}(k)s(k)e^{ikL\hat{\sigma}_4}S_L(k),
\ene
\end{itemize}
which generates the relations of the three edges of the triangle consisting of $(x_j, t_j),\, j=1,2,4$, of the three edges of the triangle consisting of $(x_j, t_j),\, j=1,3,4$, and of the four edges of the rectangle consisting of $(x_j, t_j),\, j=1,2,3,4$, respectively.

According to the definition (\ref{musg}) of $\mu_j$ and Eqs.~(\ref{sss}) and (\ref{sl}), we have
\bee\label{musg1}
\begin{array}{rl}
S(k)=&\!\!\!\d \mathbb{I}-\int_0^T e^{2ik^2\tau\hat{\sigma}_4}(V\mu_1)(0, \tau,k)d\xi=\left[\mathbb{I}+\int_0^T e^{2ik^2\tau\hat{\sigma}_4}(V\mu_2)(0,\tau,k)d\tau\right]^{-1},  \vspace{0.1in}\\
s(k)=&\!\!\! \d \mathbb{I}-\int_0^L e^{ik\xi\hat{\sigma}_4}(U\mu_3)(\xi,0,k)d\xi = \left[\mathbb{I}+\int_0^L e^{ik\xi\hat{\sigma}_4}(U\mu_2)(\xi, 0,k)d\xi\right]^{-1} ,  \vspace{0.1in}\\
S_L(k)=&\!\!\! \d \mathbb{I}-\int_0^T e^{2ik^2\tau\hat{\sigma}_4}(V\mu_4)(L,\tau,k)d\tau =\left[\mathbb{I}+\int_0^T e^{2ik^2\tau\hat{\sigma}_4}(V\mu_3)(L,\tau,k)d\tau\right]^{-1}, \vspace{0.1in}\\
 \end{array}
\ene
\bee\label{musg12}
\begin{array}{rl}
\mathbb{S}(k)=&\!\!\!\d \mathbb{I}-\int_0^L e^{ik\xi\hat{\sigma}_4}(U\mu_4)(\xi,0,k)d\xi
      -e^{ikL\hat{\sigma}_4}\int_0^T e^{2ik^2\tau\hat{\sigma}_4}(V\mu_4)(L,\tau,k)d\tau\vspace{0.1in}\\
=&\!\!\!\d \left[\mathbb{I}+e^{2ik^2T\hat{\sigma}_4}\int_0^L e^{ik\xi\hat{\sigma}_4}(U\mu_2)(\xi,T,k)d\xi+\int_0^T e^{2ik^2\tau\hat{\sigma}_4}(V\mu_2)(0,\tau,k)d\tau\right]^{-1}\vspace{0.1in}\\
=&\!\!\!\d \l[\mathbb{I}-\int_0^L e^{ik\xi\hat{\sigma}_4}(U\mu_3)(\xi,0,k)d\xi\r]
  e^{ikL\hat{\sigma}_4}\l[\mathbb{I}-\int_0^T e^{2ik^2\tau\hat{\sigma}_4}(V\mu_4)(L,\tau,k)d\tau\r]\vspace{0.1in}\\
  =&\!\!\!\d \left[\mathbb{I}+\int_0^L e^{ik\xi\hat{\sigma}_4}(U\mu_2)(\xi, 0,k)d\xi\right]^{-1}
  e^{ikL\hat{\sigma}_4}\left[\mathbb{I}+\int_0^T e^{2ik^2\tau\hat{\sigma}_4}(V\mu_3)(L,\tau,k)d\tau\right]^{-1},\vspace{0.1in} \\
 \end{array}
\ene
\bee\label{musg13}
\begin{array}{rl}
\mathfrak{S}(k)=&\!\!\!\d \l[\mathbb{I}-\int_0^T e^{2ik^2\tau\hat{\sigma}_4}(V\mu_1)(0, \tau,k)d\xi\r]^{-1}
          \l[ \mathbb{I}-\int_0^L e^{ik\xi\hat{\sigma}_4}(U\mu_3)(\xi,0,k)d\xi \r]\vspace{0.1in}\\
          =&\!\!\!\d \left[\mathbb{I}+\int_0^T e^{2ik^2\tau\hat{\sigma}_4}(V\mu_2)(0,\tau,k)d\tau\right]\left[\mathbb{I}+\int_0^L e^{ik\xi\hat{\sigma}_4}(U\mu_2)(\xi, 0,k)d\xi\right]^{-1}, \vspace{0.1in}\\
 \end{array}
\ene
\bee\label{musg14}
\begin{array}{rl}
s_T(k)=&\!\!\!\d \l[\mathbb{I}-\int_0^T e^{2ik^2\tau\hat{\sigma}_4}(V\mu_1)(0, \tau,k)d\xi\r]^{-1}
 \l[ \mathbb{I}-\int_0^L e^{ik\xi\hat{\sigma}_4}(U\mu_3)(\xi,0,k)d\xi \r] \vspace{0.1in}\\
 &\!\!\!\d \times e^{ikL\hat{\sigma}_4}\l[\mathbb{I}-\int_0^T e^{2ik^2\tau\hat{\sigma}_4}(V\mu_4)(L,\tau,k)d\tau\r],
  \end{array}
\ene
where $\mu_{j_2}(0,t,k),\ j_2=1,2,\, \mu_{j_3}(L, t, k),\, j_3=3,4,\, \mu_{j_1}(x,0,k),\, j_1=2,3,4,\, 0<x<L,\, 0<t<T$ satisfy the integral equations
\bee \begin{array}{l}
\mu_j(0,t,k)=\d \mathbb{I}+\int_T^te^{-2ik^2(t-\tau)\hat{\sigma}_4} (V\mu_j)(0,\tau,k)d\tau,\quad j=1,2,\vspace{0.1in}\\
\mu_j(L,t,k)=\d \mathbb{I}+\int_0^te^{-2ik^2(t-\tau)\hat{\sigma}_4} (V\mu_j)(L,\tau,k)d\tau, \quad j=3,4,\vspace{0.1in}\\
\mu_j(x,0,k)=\d \mathbb{I}+\int_0^x e^{ik\xi\hat{\sigma}_4}(U\mu_j)(\xi,0,k)d\xi,\quad j=2,3,\vspace{0.1in}\\
\mu_4(x,0,k)=\d \mathbb{I}+\int_L^x e^{ik\xi\hat{\sigma}_4}(U\mu_4)(\xi,0,k)d\xi
 -e^{-ik(x-L)\hat{\sigma}_4}\int_0^Te^{2ik^2\tau\hat{\sigma}_4} (V\mu_4)(L,\tau,k)d\tau.
\end{array}
\ene

It follows from Eqs.~(\ref{sss}) and (\ref{musg1}) that $s(k),\, S(k)$ and $S_L(k)$ are determined by $U(x,0,k),\,
 V(0,t,k)$, and  $V(L,t,k)$, that is, $s(k)$ is decided by the initial data $q_{j}(x, t=0)$, and $S(k)$ and $S_L(k)$ are
 found by the Dirichlet-Neumann boundary data $q_{j}(x, t)$ and $q_{jx}(x, t),\, j=1,0,-1$ at $x=0$ and $x=L$, respectively. In fact, $\mu_3(x,0,k)$ and $\{\mu_{1}(0,t,k),\,\mu_{4}(L,t,k)\} $ satisfy the $x$-part and $t$-part of the Lax pair (\ref{mulax}) at $t=0$ and $x=0,L$, respectively, that is,
\bee
x-{\rm part}:\, \left\{\begin{array}{l}
\mu_{x}(x,0,k)+ik\hat{\sigma}_4\mu(x,0,k)=U(x, t=0)\mu(x,0,k),\quad 0<x<L, \qquad \qquad\qquad \vspace{0.1in}\\
\d \mu(L,0,k)=\mathbb{I},
\end{array}\right.
\ene
\bee
t-{\rm part}:\, \left\{\begin{array}{l}
\mu_{t}(x_j,t,k)+2ik^2\hat{\sigma}\mu(x_j,t,k)=V(x=0,t,k)\mu(x-j,t,k), \,\, 0<t<T,\, j=1,4,\vspace{0.1in}\\
\mu(x_j,0,k)=\mathbb{I}, \quad \mu(x_j,T,k)=\mathbb{I},\quad x_1=0,\, x_4=L,
\end{array}\right.
\ene

It follows from the properties of $\mu_j$ and $\mu_j^A$ that the functions  $\{S(k),\, s(k),\, \mathbb{S}(k),\, S_L(k),
\mathfrak{S}(k),\, s_T(k)\}$ and $\{S^A(k),\, s^A(k),\, \mathbb{S}^A(k),\, S_L^A(k),\,
\mathfrak{S}^A(k),\, s_T^A(k)\}$ have the following boundedness:
 \bee\no
\left\{\begin{array}{l}
 S(k){\rm \,\, is \,\, bounded\,\, for\,\,} k\in  (D_2\cup D_4, D_2\cup D_4, D_1\cup D_3, D_1\cup D_3), \vspace{0.1in} \\
 s(k){\rm \,\, is \,\, bounded\,\, for\,\,} k\in  (D_3\cup D_4, D_3\cup D_4, D_1\cup D_2, D_1\cup D_2), \vspace{0.1in}\\
 \mathbb{S}(k){\rm \,\, is \,\, bounded\,\, for\,\,} k\in  (D_4, D_4, D_1, D_1), \vspace{0.1in}\\
 S_L(k){\rm \,\, is \,\, bounded\,\, for\,\,} k\in  (D_2\cup D_4, D_2\cup D_4, D_1\cup D_3, D_1\cup D_3), \vspace{0.1in}\\
 \mathfrak{S}(k){\rm \,\, is \,\, bounded\,\, for\,\,} k\in  (D_4, D_4, D_1, D_1), \vspace{0.1in}\\
 s_T(k){\rm \,\, is \,\, bounded\,\, for\,\,} k\in  (D_4, D_4, D_1, D_1), \vspace{0.1in}\\
\end{array} \right.
\ene
and 
 \bee\no
\left\{\begin{array}{l}
 S^A(k){\rm \,\, is \,\, bounded\,\, for\,\,} k\in  (D_1\cup D_3, D_1\cup D_3, D_2\cup D_4, D_21\cup D_4), \vspace{0.1in} \\
 s^A(k){\rm \,\, is \,\, bounded\,\, for\,\,} k\in  (D_1\cup D_2, D_1\cup D_2, D_3\cup D_4, D_3\cup D_4), \vspace{0.1in}\\
 \mathbb{S}^A(k){\rm \,\, is \,\, bounded\,\, for\,\,} k\in  (D_2, D_2, D_3, D_3), \vspace{0.1in}\\
 S_L^A(k){\rm \,\, is \,\, bounded\,\, for\,\,} k\in  (D_2\cup D_4, D_2\cup D_4, D_1\cup D_3, D_1\cup D_3), \vspace{0.1in}\\
 \mathfrak{S}^A(k){\rm \,\, is \,\, bounded\,\, for\,\,} k\in  (D_2, D_2, D_3, D_3), \vspace{0.1in}\\
 s_T^A(k){\rm \,\, is \,\, bounded\,\, for\,\,} k\in  (D_2, D_2, D_3, D_3),
\end{array} \right.
\ene

\vspace{0.1in}
\noindent {\bf Proposition 2.3.} {\it The new matrix-valued functions $S_n(k)=(S_n^{ij}(k))_{4\times 4},\, n=1,2,3,4$ introduced by
\bee\label{mns}
 M_n(x,t,k)=\mu_2(x,t,k)e^{-i(kx+2k^2t)\hat{\sigma}_4}S_n(k), \quad k\in D_n,
 \ene
can be found uing the entries of the matrix-valued spectral functions $s(k)=(s_{ij}(k))_{4\times 4},\, S(k)=(S_{ij}(k))_{4\times 4}$, and $\mathbb{S}(k)=(\mathbb{S}_{ij}(k))_{4\times 4}=s(k)e^{ikL\hat{\sigma}_4}S_L(k)$  given by Eq.~(\ref{sss}) as follows:
\bee\label{sn}
\begin{array}{ll}
S_1(k)=\left(\begin{array}{cccc}
 \dfrac{m_{22}(\mathbb{S}(k))}{n_{33,44}(\mathbb{S}(k))} &  \dfrac{m_{21}(\mathbb{S}(k))}{n_{33,44}(\mathbb{S}(k))} &  \mathbb{S}_{13}(k) & \mathbb{S}_{14}(k) \vspace{0.1in}\\
 \dfrac{m_{12}(\mathbb{S}(k))}{n_{33,44}(\mathbb{S}(k))} &  \dfrac{m_{11}(\mathbb{S}(k))}{n_{33,44}(\mathbb{S}(k))} &  \mathbb{S}_{23}(k) & \mathbb{S}_{24}(k) \vspace{0.1in}\\
 0 &  0 &  \mathbb{S}_{33}(k) & \mathbb{S}_{34}(k) \vspace{0.1in}\\
 0 &  0 &  \mathbb{S}_{43}(k) & \mathbb{S}_{44}(k)
 \end{array}
\right), \vspace{0.15in}\\
S_2(k)=\left(\begin{array}{cccc}
 S_2^{11}(k) &  S_2^{12}(k) & s_{13}(k) & s_{14}(k) \vspace{0.1in}\\
 S_2^{21}(k) &  S_2^{22}(k) &  s_{23}(k) & s_{24}(k) \vspace{0.1in}\\
  S_2^{31}(k) &  S_2^{32}(k)  &  s_{33}(k) & s_{34}(k) \vspace{0.1in}\\
  S_2^{41}(k) &  S_2^{42}(k) &  s_{43}(k) & s_{44}(k)
 \end{array}
\right), \vspace{0.15in}\\
S_3(k)=\left(\begin{array}{cccc}
 s_{11}(k) &  s_{12}(k) & S_3^{13}(k) & S_3^{14}(k) \vspace{0.1in}\\
 s_{21}(k) &  s_{22}(k) &  S_3^{23}(k) & S_3^{24}(k) \vspace{0.1in}\\
 s_{31}(k) &  s_{32}(k)  & S_3^{33}(k) & S_3^{34}(k) \vspace{0.1in}\\
 s_{41}(k) &  s_{42}(k) & S_3^{43}(k) & S_3^{44}(k)
 \end{array}
\right),\vspace{0.15in}\\
S_4(k)=\left(\begin{array}{cccc}
 \mathbb{S}_{11}(k) & \mathbb{S}_{12}(k) & 0  & 0 \vspace{0.1in}\\
 \mathbb{S}_{21}(k) & \mathbb{S}_{22}(k) & 0 &  0 \vspace{0.1in}\\
  \mathbb{S}_{31}(k) & \mathbb{S}_{32}(k) & \dfrac{m_{44}(\mathbb{S}(k))}{n_{11,22}(\mathbb{S}(k))} &  \dfrac{m_{43}(\mathbb{S}(k))}{n_{11,22}(\mathbb{S}(k))} \vspace{0.1in}\\
 \mathbb{S}_{41}(k) & \mathbb{S}_{42}(k) & \dfrac{m_{34}(\mathbb{S}(k))}{n_{11,22}(\mathbb{S}(k))} &  \dfrac{m_{33}(\mathbb{S}(k))}{n_{11,22}(\mathbb{S}(k))}    \end{array}
\right),
\end{array}
\ene
where $n_{i_1j_1,i_2j_2}(X)$ denotes the determinant of the sub-matrix generated by choosing the cross elements of $i_{1,2}$th rows and $j_{1,2}$th columns of a $4\times 4$ matrix $X$, that is,
\bee
 n_{i_1j_1,i_2j_2}(X)=\left|\begin{array}{cc}
  X_{i_1j_1} & X_{i_1j_2}\vspace{0.1in}\\
  X_{i_2j_1} & X_{i_2j_2}\end{array}\right|.
\ene
and
\bee
\left\{\begin{array}{l}
S_2^{1j}(k)=\dfrac{n_{1j,2(3-j)}(S)m_{2(3-j)}(s)+n_{1j,3(3-j)}(S)m_{3(3-j)}(s)+n_{1j,4(3-j)}(S)m_{4(3-j)}(s)}{\mathcal{N}([S]_1[S]_2[s]_3[s]_4)}, \vspace{0.1in}\\
S_2^{2j}(k)=\dfrac{n_{2j,1(3-j)}(S)m_{1(3-j)}(s)+n_{2j,3(3-j)}(S)m_{3(3-j)}(s)+n_{2j,4(3-j)}(S)m_{4(3-j)}(s)}{\mathcal{N}([S]_1[S]_2[s]_3[s]_4)}, \vspace{0.1in}\\
S_2^{3j}(k)=\dfrac{n_{3j,1(3-j)}(S)m_{1(3-j)}(s)+n_{3j,2(3-j)}(S)m_{2(3-j)}(s)+n_{3j,4(3-j)}(S)m_{4(3-j)}(s)}{\mathcal{N}([S]_1[S]_2[s]_3[s]_4)}, \vspace{0.1in}\\
S_2^{4j}(k)=\dfrac{n_{4j,1(3-j)}(S)m_{1(3-j)}(s)+n_{4j,2(3-j)}(S)m_{2(3-j)}(s)+n_{4j,3(3-j)}(S)m_{3(3-j)}(s)}{\mathcal{N}([S]_1[S]_2[s]_3[s]_4)},
\end{array}\right.\quad j=1,2,
\ene
\bee\left\{\begin{array}{l}
S_3^{1j}(k)=\dfrac{n_{1j,2(7-j)}(S)m_{2(7-j)}(s)+n_{1j,3(7-j)}(S)m_{3(7-j)}(s)+n_{1j,4(7-j)}(S)m_{4(7-j)}(s)}{\mathcal{N}([s]_1[s]_2[S]_3[S]_4)},\vspace{0.1in}\\
S_3^{2j}(k)=\dfrac{n_{2j,1(7-j)}(S)m_{1(7-j)}(s)+n_{2j,3(7-j)}(S)m_{3(7-j)}(s)+n_{2j,4(7-j)}(S)m_{4(7-j)}(s)}{\mathcal{N}([s]_1[s]_2[S]_3[S]_4)}, \vspace{0.1in}\\
S_3^{3j}(k)=\dfrac{n_{3j,1(7-j)}(S)m_{1(7-j)}(s)+n_{3j,2(7-j)}(S)m_{2(7-j)}(s)+n_{3j,4(7-j)}(S)m_{4(7-j)}(s)}{\mathcal{N}([s]_1[s]_2[S]_3[S]_4)}, \vspace{0.1in}\\
S_3^{4j}(k)=\dfrac{n_{4j,1(7-j)}(S)m_{1(7-j)}(s)+n_{4j,2(7-j)}(S)m_{2(7-j)}(s)+n_{4j,3(7-j)}(S)m_{3(7-j)}(s)}{\mathcal{N}([s]_1[s]_2[S]_3[S]_4)},
\end{array}\right.\quad j=3,4,
\ene
where $\mathcal{N}([S]_1[S]_2[s]_3[s]_4)={\rm det}([S]_1, [S]_2, [s]_3, [s]_4)$ denotes the determinant of the matrix generated by choosing the first and second columns of $S(k)$ and the third and fourth columns of $s(k)$, and $\mathcal{N}([s]_1[s]_2[S]_3[S]_4)={\rm det}([s]_1, [s]_2, [S]_3, [S]_4)$, that is,
\bee
 \mathcal{N}([S]_1[S]_2[s]_3[s]_4)={\rm det}([S]_1, [S]_2, [s]_3, [s]_4)=\left|\begin{array}{cccc}
  S_{11}(k) & S_{12}(k) & s_{13}(k) & s_{14}(k)\vspace{0.1in}\\
  S_{21}(k) & S_{22}(k) & s_{23}(k) & s_{24}(k) \vspace{0.1in}\\
  S_{31}(k) & S_{32}(k) & s_{33}(k) & s_{34}(k) \vspace{0.1in}\\
  S_{41}(k) & S_{42}(k) & s_{43}(k) & s_{44}(k)
  \end{array}\right|.
\ene
\bee
 \mathcal{N}([s]_1[s]_2[S]_3[S]_4)={\rm det}([s]_1, [s]_2, [S]_3, [S]_4)=\left|\begin{array}{cccc}
  s_{11}(k) & s_{12}(k) & S_{13}(k) & S_{14}(k)\vspace{0.1in}\\
  s_{21}(k) & s_{22}(k) & S_{23}(k) & S_{24}(k) \vspace{0.1in}\\
  s_{31}(k) & s_{32}(k) & S_{33}(k) & S_{34}(k) \vspace{0.1in}\\
  s_{41}(k) & s_{42}(k) & S_{43}(k) & S_{44}(k)
  \end{array}\right|.
\ene
}

\noindent {\bf Proof.}\, Introduce the $4\times 4$ matrix-valued functions $R_n(k), S_n(k), T_n(k)$, and $P_n(k)$ by the eigenfunctions $M_n(x,t,k)$ and $\mu_j(x,t,k),\,j=1,2,3,4$
\bee\label{rstp}
\left\{\begin{array}{l}
M_n(x,t,k)=\mu_1(x,t,k)e^{-i(kx+2k^2t)\hat{\sigma}_4}R_n(k), \vspace{0.1in}\\
M_n(x,t,k)=\mu_2(x,t,k)e^{-i(kx+2k^2t)\hat{\sigma}_4}S_n(k), \vspace{0.1in}\\
M_n(x,t,k)=\mu_3(x,t,k)e^{-i(kx+2k^2t)\hat{\sigma}_4}T_n(k), \vspace{0.1in}\\
M_n(x,t,k)=\mu_4(x,t,k)e^{-i(kx+2k^2t)\hat{\sigma}_4}P_n(k),
\end{array}\right.
\ene

It follows from Eq.~(\ref{rstp}) that we have the relations
\bee
\left\{\begin{array}{l}
R_n(k)=e^{2ik^2T\hat{\sigma}_4}M_n(0,T,k),\vspace{0.1in}\\
S_n(k)=M_n(0,0,k),\vspace{0.1in}\\
T_n(k)=e^{ikL\hat{\sigma}_4}M_n(L,0,k),\vspace{0.1in} \\
P_n(k)=e^{i(kL+2k^2T)\hat{\sigma}_4}M_n(L,T,k),
\end{array}\right.
\ene
and
\bee \label{sneq}
\left\{\begin{array}{l}
S(k)=\mu_1(0,0,k)=S_n(k)R_n^{-1}(k), \vspace{0.1in}\\
s(k)=\mu_3(0,0,k)=S_n(k)T_n^{-1}(k), \vspace{0.1in}\\
\mathbb{S}(k)=\mu_4(0,0,k)=S_n(k)P_n^{-1}(k),
\end{array}\right.
\ene
which can in general deduce the functions $\{R_n, S_n, T_n, P_n\}$ for the given functions $\{s(k), S(k), \mathbb{S}(k)\}$.

Moreover, we can also determine some entries of $\{R_n, S_n, T_n, P_n\}$ by using Eqs.~(\ref{mn}) and (\ref{rstp})
\bee\label{rstp2}
\left\{
\begin{array}{l}
(R_n(k))_{ij}=0, \,\,\,\,\, {\rm if} \,\,\, (\gamma^n)_{ij}=\gamma_1, \vspace{0.1in} \\
(S_n(k))_{ij}=0,  \,\,\,\,\, {\rm if} \,\,\,  (\gamma^n)_{ij}=\gamma_2, \vspace{0.1in}\\
(T_n(k))_{ij}=\delta_{ij}, \,\,\, {\rm if} \,\,\,  (\gamma^n)_{ij}=\gamma_3, \vspace{0.1in}\\
(P_n(k))_{ij}=\delta_{ij},  \,\,\, {\rm if} \,\,\,  (\gamma^n)_{ij}=\gamma_4,
\end{array}
\right.
 \ene
Thus it follows from systems (\ref{sneq}) and (\ref{rstp2}) that we can deduce
Eq.~(\ref{sn}) by the direct calculation. $\square$

\subsection*{\it 2.7.\, The residue conditions}

\quad Since $\mu_2(x,t,k)$ is an entire function, it follows from Eq.~(\ref{mns}) that $M_n(x,t,k)$'s only have singularities at the points where
the $S_n$'s have singularities. We know from the expressions of $S_n$ given by  Eq.~(\ref{sn}) that the possible singularities of $M_n$ in the complex $k$-plane are as follows:

\begin{itemize}

\item {} $[M_1(x,t,k)]_j,\, j=1,2$ could possess poles in $D_1$ at the zeros of $n_{33,44}(\mathbb{S})(k)$;
\item {} $[M_2(x,t,k)]_j,\,j=1,2$ could admit poles in $D_2$ at the zeros of $\mathcal{N}([S]_1[S]_2[s]_3[s]_4)(k)$;
\item {} $[M_3(x,t,k)]_j,\, j=3,4$ could have poles in $D_3$ at the zeros of $\mathcal{N}([s]_1s]_2[S]_3[S]_4)(k)$;
\item {} $[M_4(x,t,k)]_j,\, j=3,4$ could be of poles in $D_4$ at the zeros of $n_{11,22}(\mathbb{S})(k)$.
\end{itemize}

We introduce the above-mentioned possible zeros by $\{k_j\}_1^N$ and suppose that they satisfy the following assumption.

\vspace{0.1in}
\noindent {\bf Assumption 2.4.} {\it Suppose  that
\begin{itemize}

\item {} $n_{33,44}(\mathbb{S})(k)$ admits $n_1$ possible simple zeros $\{k_j\}_1^{n_1}$ in $D_1$;

\item {} $\mathcal{N}([S]_1[S]_2[s]_3[s]_4)(k)$ has $n_2-n_1$ possible simple zeros $\{k_j\}_{n_1+1}^{n_2}$ in $D_2$;

\item {} $\mathcal{N}([s]_1[s]_2[S]_3[S]_4)(k)$ is of $n_3-n_2$ possible simple zeros $\{k_j\}_{n_2+1}^{n_3}$ in $D_3$;

\item {} $n_{11,22}(\mathbb{S})(k)$ has $N-n_3$ possible simple zeros $\{k_j\}_{n_3+1}^N$ in $D_4$,
\end{itemize}
and that none of these zeros coincide. Moreover, none of these functions are assumed to have zeros on the boundaries
on the $D_n$'s ($n=1,2,3,4)$.}

\vspace{0.1in}
\noindent {\bf Lemma 2.5.} {\it For a $4\times 4$ matrix $X=(X_{ij})_{4\times 4}$, $e^{\theta\hat{\sigma}_4}X$ is given by
\bee\no
  e^{\theta\hat{\sigma}_4}X=e^{\theta\sigma_4}Xe^{-\theta\sigma_4}=\left(\begin{array}{llll}
   X_{11} & X_{12} & X_{13}e^{2\theta} & X_{14}e^{2\theta} \vspace{0.05in}\\
   X_{21} & X_{22} & X_{23}e^{2\theta} & X_{24}e^{2\theta} \vspace{0.05in}\\
   X_{31}e^{-2\theta} & X_{32}e^{-2\theta} & X_{33} &X_{34} \vspace{0.05in}\\
   X_{41}e^{-2\theta} & X_{42}e^{-2\theta} & X_{43} &X_{44}
   \end{array} \right),
\ene
}

We can deduce the residue conditions at these zeros in the following expressions:

\vspace{0.1in}
\noindent {\bf Proposition 2.6. } {\it Let $\{M_n(x,t,k)\}_1^4$ be the eigenfunctions given by Eq.~(\ref{mn}) and suppose that the set $\{k_j\}_1^N$ of singularities are as the above-mentioned Assumption 2.4. Then we have the following residue conditions:
\bee
 \label{rm1a} \begin{array}{rl}
\d{\rm Res}_{k=k_j}[M_1(x,t,k)]_l=&\!\! \dfrac{m_{2(3-l)}(\mathbb{S})(k_j)\mathbb{S}_{24}(k_j)-m_{1(3-l)}(\mathbb{S})(k_j)\mathbb{S}_{14}(k_j)}
            {\dot{n}_{33,44}(\mathbb{S})(k_j)n_{13,24}(\mathbb{S})(k_j)e^{2\theta(k_j)}}[M_1(x,t,k_j)]_3 \vspace{0.1in}\\
        &    +\dfrac{m_{1(3-l)}(\mathbb{S})(k_j)\mathbb{S}_{13}(k_j)-m_{2(3-l)}(\mathbb{S})(k_j)\mathbb{S}_{23}(k_j)}
          {\dot{n}_{33,44}(\mathbb{S})(k_j)n_{13,24}(\mathbb{S})(k_j)e^{2\theta(k_j)}}[M_1(x,t,k_j)]_4,  \vspace{0.1in}\\
        & \quad {\rm for}\quad 1\leq j\leq n_1,\quad l=1,2,\quad k\in D_1,
        \end{array}
\ene

\bee
 \label{rm2a} \begin{array}{rl}
 {\rm Res}_{k=k_j}[M_2(x,t,k)]_l=&\!\! \dfrac{S_{2}^{1l}(k_j)s_{24}(k_j)-S_2^{2l}(k_j)s_{14}(k_j)}
          {\dot{\mathcal{N}}([S]_1[S]_2[s]_3[s]_4)(k_j)n_{13,24}(s)(k_j)e^{2\theta(k_j)}}[M_2(x,t,k_j)]_3 \vspace{0.1in}\\
       &   +\dfrac{S_2^{2l}(k_j)s_{13}(k_j)-S_2^{1l}(k_j)s_{23}(k_j)}
          {\dot{\mathcal{N}}([S]_1[S]_2[s]_3[s]_4)(k_j)n_{13,24}(s)(k_j)e^{2\theta(k_j)}}[M_2(x,t,k_j)]_4, \vspace{0.1in}\\
        & \quad {\rm for}\quad n_1+1\leq j\leq n_2,\quad l=1,2, \quad k\in D_2, \vspace{0.1in}\\
   \end{array}
\ene
\bee
 \label{rm3a} \begin{array}{rl}
 {\rm Res}_{k=k_j}[M_3(x,t,k)]_l=&\!\! \dfrac{S_{3}^{1l}(k_j)s_{22}(k_j)-S_3^{2l}(k_j)s_{12}(k_j)}
          {\dot{\mathcal{N}}([s]_1[s]_2[S]_3[S]_4)(k_j)n_{11,22}(s)(k_j)}e^{2\theta(k_j)}[M_3(x,t,k_j)]_1 \vspace{0.1in}\\
      &    +\dfrac{S_3^{2l}(k_j)s_{11}(k_j)-S_3^{1l}(k_j)s_{21}(k_j)}
          {\dot{\mathcal{N}}([s]_1[s]_2[S]_3[S]_4)(k_j)n_{11,22}(s)(k_j)}e^{2\theta(k_j)}[M_3(x,t,k_j)]_2, \vspace{0.1in}\\
      & \quad {\rm for}\quad n_2+1\leq j\leq n_3,\quad l=3,4,\quad k\in D_3, \vspace{0.1in}\\
   \end{array}
\ene
\bee
\label{rm4a} \begin{array}{rl}
 {\rm Res}_{k=k_j}[M_4(x,t,k)]_l=&\!\! \dfrac{m_{4(7-l)}(\mathbb{S})(k_j)\mathbb{S}_{42}(k_j)-m_{3(7-l)}(\mathbb{S})(k_j)\mathbb{S}_{32}(k_j)}
          {\dot{n}_{11,22}(\mathbb{S})(k_j)n_{31,42}(\mathbb{S})(k_j)}e^{2\theta(k_j)}[M_4(x,t,k_j)]_1 \vspace{0.1in}\\
        &  +\dfrac{m_{3(7-l)}(\mathbb{S})(k_j)\mathbb{S}_{31}(k_j)-m_{4(7-l)}(\mathbb{S})(k_j)\mathbb{S}_{41}(k_j)}
          {\dot{n}_{11,22}(\mathbb{S})(k_j)n_{31,42}(\mathbb{S})(k_j)}e^{2\theta(k_j)}[M_4(x,t,k_j)]_2, \vspace{0.2in}\\
    & \quad {\rm for}\quad n_3+1\leq j\leq N,\quad l=3,4,\quad k\in D_4,
   \end{array}
\ene
where the overdot denotes the derivative with resect to the parameter $k$ and $\theta(k)=-i(kx+2k^2t)$.} \\

\noindent {\bf Proof.}\, According to Lemma 2.5, it follows from Eqs.~(\ref{mns}) and (\ref{sn}) that the four columns of $M_1(x,t,k)$ are obtained by the matrices $\mu_2$ and $S_1(k)$
\bes \label{m1}\bee
\label{m1a}  &[M_1]_1=[\mu_2]_1 \dfrac{m_{22}(\mathbb{S})}{n_{33,44}(\mathbb{S})}+[\mu_2]_2\dfrac{m_{12}(\mathbb{S})}{n_{33,44}(\mathbb{S})}, \vspace{0.1in}\\
\label{m1b} &[M_1]_2=[\mu_2]_1 \dfrac{m_{21}(\mathbb{S})}{n_{33,44}(\mathbb{S})}+[\mu_2]_2\dfrac{m_{11}(\mathbb{S})}{n_{33,44}(\mathbb{S})}, \vspace{0.1in}\\
\label{m1c} &[M_1]_3=[\mu_2]_1 \mathbb{S}_{13}e^{2\theta} +[\mu_2]_2\mathbb{S}_{23}e^{2\theta}+[\mu_2]_3\mathbb{S}_{33}+[\mu_2]_4\mathbb{S}_{43}, \vspace{0.1in}\\
\label{m1d} &[M_1]_4=[\mu_2]_1 \mathbb{S}_{14}e^{2\theta} +[\mu_2]_2\mathbb{S}_{24}e^{2\theta}+[\mu_2]_3\mathbb{S}_{34}+[\mu_2]_4\mathbb{S}_{44},
  \ene\ees
the four columns of $M_2(x,t,k)$ are found by  the matrices $\mu_2$ and $S_2(k)$
\bes \label{m2}\bee
\label{m2a} &[M_2]_1=[\mu_2]_1 S_2^{11} +[\mu_2]_2S_2^{21}+[\mu_2]_3e^{-2\theta}S_2^{31}+[\mu_2]_4e^{-2\theta}S_2^{41}, \vspace{0.1in}\\
\label{m2b} &[M_2]_2=[\mu_2]_1 S_2^{12} +[\mu_2]_2S_2^{22}+[\mu_2]_3e^{-2\theta}S_2^{32}+[\mu_2]_4e^{-2\theta}S_2^{42}, \vspace{0.1in}\\
\label{m2c} &[M_2]_3=[\mu_2]_1e^{2\theta} s_{13} +[\mu_2]_2e^{2\theta}s_{23}+[\mu_2]_3s_{33}+[\mu_2]_4s_{43}, \vspace{0.1in}\\
\label{m2d} &[M_2]_4=[\mu_2]_1e^{2\theta} s_{14} +[\mu_2]_2e^{2\theta}s_{24}+[\mu_2]_3s_{34}+[\mu_2]_4s_{44},
  \ene\ees
the four columns of $M_3(x,t,k)$ are given by  the matrices $\mu_2$ and $S_3(k)$
\bes \label{m3}\bee
\label{m3a} &[M_3]_1=[\mu_2]_1 s_{11} +[\mu_2]_2s_{21}+[\mu_2]_3e^{-2\theta}s_{31}+[\mu_2]_4e^{-2\theta}s_{41}, \vspace{0.1in}\\
\label{m3b} &[M_3]_2=[\mu_2]_1 s_{12} +[\mu_2]_2s_{22}+[\mu_2]_3e^{-2\theta}s_{32}+[\mu_2]_4e^{-2\theta}s_{42},\vspace{0.1in} \\
\label{m3c} &[M_3]_3=[\mu_2]_1e^{2\theta} S_3^{13} +[\mu_2]_2e^{2\theta}S_3^{23}+[\mu_2]_3S_3^{33}+[\mu_2]_4S_3^{43}, \vspace{0.1in}\\
\label{m3d} &[M_3]_4=[\mu_2]_1e^{2\theta} S_3^{14} +[\mu_2]_2e^{2\theta}S_3^{24}+[\mu_2]_3S_3^{34}+[\mu_2]_4S_3^{44},
  \ene\ees
and the four columns of $M_4(x,t,k)$ are given by  the matrices $\mu_2$ and $S_4(k)$
\bes \label{m4}\bee
\label{m4a} &[M_4]_1=[\mu_2]_1 \mathbb{S}_{11} +[\mu_2]_2\mathbb{S}_{21}+[\mu_2]_3\mathbb{S}_{31}e^{-2\theta}+[\mu_2]_4\mathbb{S}_{41}e^{-2\theta}, \vspace{0.1in}\\
\label{m4b} &[M_4]_2=[\mu_2]_1 \mathbb{S}_{12} +[\mu_2]_2\mathbb{S}_{22}+[\mu_2]_3\mathbb{S}_{32}e^{-2\theta}+[\mu_2]_4\mathbb{S}_{42}e^{-2\theta}, \vspace{0.1in}\\
\label{m4c} &[M_4]_3=[\mu_2]_3 \dfrac{m_{44}(\mathbb{S})}{n_{11,22}(\mathbb{S})}+[\mu_2]_4\dfrac{m_{34}(\mathbb{S})}{n_{11,22}(\mathbb{S})}, \vspace{0.1in}\\
\label{m4d} &[M_4]_4=[\mu_2]_3 \dfrac{m_{43}(\mathbb{S})}{n_{11,22}(\mathbb{S})}+[\mu_2]_4\dfrac{m_{33}(\mathbb{S})}{n_{11,22}(\mathbb{S})},
  \ene\ees

For the case that $k_j\in D_1$ is a simple zero of $n_{33,44}(\mathbb{S})(k)$, it follows from Eqs.~(\ref{m1c}) and (\ref{m1d}) that we have $[\mu_2]_1$ and $[\mu_2]_2$
\bee\begin{array}{l}
\d[\mu_2]_1=\frac{[M_1]_3\mathbb{S}_{24}-[M_1]_4\mathbb{S}_{23}
  +[\mu_2]_3(\mathbb{S}_{23}\mathbb{S}_{34}-\mathbb{S}_{33}\mathbb{S}_{24})
  +[\mu_2]_4(\mathbb{S}_{23}\mathbb{S}_{44}-\mathbb{S}_{43}\mathbb{S}_{24})
   }{\mathbb{S}_{13}\mathbb{S}_{24}-\mathbb{S}_{14}\mathbb{S}_{23}}e^{-2\theta},\vspace{0.1in}\\
 \d [\mu_2]_2=\frac{[M_1]_3\mathbb{S}_{14}-[M_1]_4\mathbb{S}_{13}
  +[\mu_2]_3(\mathbb{S}_{14}\mathbb{S}_{33}-\mathbb{S}_{13}\mathbb{S}_{34})
  +[\mu_2]_4(\mathbb{S}_{14}\mathbb{S}_{43}-\mathbb{S}_{13}\mathbb{S}_{44})
   }{\mathbb{S}_{13}\mathbb{S}_{24}-\mathbb{S}_{14}\mathbb{S}_{23}}e^{-2\theta},
\end{array}\ene

and then substitute them into Eqs.~(\ref{m1a}) and (\ref{m1b})
yields
\bes\bee \nonumber
& [M_1]_1=\dfrac{m_{22}(\mathbb{S})\mathbb{S}_{24}-m_{12}(\mathbb{S})\mathbb{S}_{14}}
          {n_{33,44}(\mathbb{S})n_{13,24}(\mathbb{S})}e^{-2\theta}[M_1]_3
          +\dfrac{m_{12}(\mathbb{S})\mathbb{S}_{13}-m_{22}(\mathbb{S})\mathbb{S}_{23}}
          {n_{33,44}(\mathbb{S})n_{13,24}(\mathbb{S})}e^{-2\theta}[M_1]_4 \vspace{0.1in}\\
  &        +\dfrac{m_{42}(\mathbb{S})}{n_{13,24}(\mathbb{S})}e^{-2\theta}[\mu_2]_3
          +\dfrac{m_{32}(\mathbb{S})}{n_{13,24}(\mathbb{S})}e^{-2\theta}[\mu_2]_4,\vspace{0.1in}\\
&\nonumber
[M_1]_2=\dfrac{m_{21}(\mathbb{S})\mathbb{S}_{24}-m_{11}(\mathbb{S})\mathbb{S}_{14}}
          {n_{33,44}(\mathbb{S})n_{13,24}(\mathbb{S})}e^{-2\theta}[M_1]_3
          +\dfrac{m_{11}(\mathbb{S})\mathbb{S}_{13}-m_{21}(\mathbb{S})\mathbb{S}_{23}}
          {n_{33,44}(\mathbb{S})n_{13,24}(\mathbb{S})}e^{-2\theta}[M_1]_4 \vspace{0.1in}\\
&  +\dfrac{m_{41}(\mathbb{S})}{n_{13,24}(\mathbb{S})}e^{-2\theta}[\mu_2]_3
          +\dfrac{m_{31}(\mathbb{S})}{n_{13,24}(\mathbb{S})}e^{-2\theta}[\mu_2]_4,
\ene\ees
whose residues at $k_j$ yield Eq.~(\ref{rm1a}) for $k_j\in D_1$, respectively.

Similarly, we solve Eqs.~(\ref{m2c}) and (\ref{m2d}) for $[\mu_2]_1$ and $[\mu_2]_2$,
  and then substitute them into Eqs.~(\ref{m2a}) and (\ref{m2b}) to yield
\bes\bee \nonumber
& [M_2]_1=\dfrac{S_{2}^{11}s_{24}-S_2^{21}s_{14}}
          {\mathcal{N}([S]_1[S]_2[s]_3[s]_4)n_{13,24}(s)}e^{-2\theta}[M_2]_3
          +\dfrac{S_2^{21}s_{13}-S_2^{11}s_{23}}
          {\mathcal{N}([S]_1[S]_2[s]_3[s]_4)n_{13,24}(s)}e^{-2\theta}[M_2]_4 \vspace{0.1in}\\
  &        +\dfrac{m_{42}(s)}{n_{13,24}(s)}e^{-2\theta}[\mu_2]_3
          +\dfrac{m_{32}(s)}{n_{13,24}(s)}e^{-2\theta}[\mu_2]_4, \vspace{0.1in}\\
& \nonumber
[M_2]_2=\dfrac{S_2^{12}s_{24}-S_2^{22}s_{14}}
          {\mathcal{N}([S]_1[S]_2[s]_3[s]_4)n_{13,24}(s)}e^{-2\theta}[M_2]_3
          +\dfrac{S_2^{22}s_{13}-S_2^{12}s_{23}}
         {\mathcal{N}([S]_1[S]_2[s]_3[s]_4)n_{13,24}(s)}e^{-2\theta}[M_2]_4 \vspace{0.1in}\\
         & +\dfrac{m_{41}(s)}{n_{13,24}(s)}e^{-2\theta}[\mu_2]_3
          +\dfrac{m_{31}(s)}{n_{13,24}(s)}e^{-2\theta}[\mu_2]_4,
\ene\ees
whose residues at $k_j$ yield Eq.~(\ref{rm2a}) for $k_j\in D_2$, respectively.

Similarly, we can verify Eq.~(\ref{rm3a}) for $k_j\in D_3$ and Eq.~(\ref{rm4a}) for $k_j\in D_4$ by studying Eq.~(\ref{m3a})-(\ref{m4d}).  $\square$

\subsection*{\it 2.8.\, The global relation}

\quad The definitions of the above-mentioned spectral functions $S(k), s(k), S_L(k)$, and $\mathbb{S}(k)$ imply that they are dependent. It follows from Eqs.~(\ref{mu1234}) and (\ref{sr}) that
\bee \label{srr} \begin{array}{rl}
\mu_4(x,t,k)=&\mu_2(x,t,k)e^{-i(kx+2k^2t)\hat{\sigma}_4}\mathbb{S}(k) \vspace{0.1in} \\
         =&\mu_2(x,t,k)e^{-i(kx+2k^2t)\hat{\sigma}_4}[s(k)e^{ikL\hat{\sigma}_4}S_L(k)] \vspace{0.1in}\\
         =&\mu_1(x,t,k)e^{-i(kx+2k^2t)\hat{\sigma}_4}[S^{-1}(k)s(k)e^{ikL\hat{\sigma}_4}S_L(k)],
        \end{array}
\ene
which leads to the  global relation
\bee\label{srr2}
c(T,k)=\mu_4(0, T, k)=e^{-2ik^2T\hat{\sigma}_4}[S^{-1}(k)s(k)e^{ikL\hat{\sigma}_4}S_L(k)],
\ene
by evaluating Eq.~(\ref{srr}) at the point $(x,t)=(0, T)$ and using $\mu_1(0, T, k)=\mathbb{I}$.

\subsection*{\it 2.9.\, The jump matrices with explicit $(x, t)$-dependence }

\quad It follows from Eq.~(\ref{mns}) that the spectral functions $S_n(k),\, n=1,2,3, 4$ is given by
\bee
 S_n(k)=M_n(x=0,t=0,k), \quad k\in D_n,\quad n=1,2,3,4.
\ene
Let $M(x,t,k)$ stand for the sectionally analytic function on the Riemann $k$-spere, which is equivalent to $M_n(x,t,k)$ for $k\in D_n$. Then $M(x,t,k)$ solves the jump equations
\bee\label{jumpc}
 M_n(x,t,k)=M_m(x,t,k)J_{mn}(x,t,k),\quad k\in \bar{D}_n\cap \bar{D}_m, \quad n,m=1,2,3,4,\quad n\not= m,
\ene
with the jump matrices $J_{mn}(x,t,k)$ defined by
\bee\label{jump}
J_{mn}(x,t,k)=e^{-i(kx+2k^2t)\hat{\sigma}_4}[(S_m^{-1}(k)S_n(k)].
\ene

\section{The $4\times 4$ matrix Riemann-Hilbert problem}

\quad By using the district contours $\gamma_j\, (j=1,2,3,4)$, the integral solutions of the revised Lax pair (\ref{mulax}), and $S_n(k),\, n=1,2,3,4$ due to $\{S(k), s(k), \mathbb{S}(k), S_L(k)\}$, we have defined the sectionally analytic function $M_n(x,t,k)\, (n=1,2,3,4)$, which solves a
$4\times 4$ matrix Riemann-Hilbert (RH) problem. This matrix RH problem can be formulated on basis of the initial and boundary data of the functions $q_j(x,t),\ j=1,0,-1$. Thus the solution of Eq.~(\ref{pnls}) for all values of $x,t$ can be refound by solving the RH problem.

\vspace{0.1in}
\noindent {\bf Theorem 3.1.}\, {\it Suppose that $\{q_j(x,t),\,\, j=1,0,-1\}$ is a solution of system (\ref{pnls}) in the interval region  $\Omega=\{(x,t)| x\in [0, L],\, t\in [0, T]\}$. Then it can be refound from the initial data defined by
\bee\no
 q_j(x, t=0)=q_{0j}(x),&  j=1,0,-1,
\ene
and Dirichlet and Neumann boundary values defined by
 \bee\no
 \begin{array}{lll}
  {\it Dirichlet \,\, boundary \,\, data:} & q_j(x=0, t)=u_{0j}(t), & q_j(x=L, t)=v_{0j}(t), \,\,\, j=1,0,-1, \vspace{0.1in} \\
 {\it Neumann \,\, boundary \,\, data:} &  q_{jx}(x=0, t)=u_{1j}(t),& q_{jx}(x=L, t)=v_{1j}(t), \,\,\, j=1,0,-1,
 \end{array} \ene
We use the initial and boundary data to construct the jump matrices $J_{mn}(x, t, k),\, n, m = 1,..., 4$, by Eq.~(\ref{jump}) as well as the spectral functions $S(k), \, s(k)$ and  $(\mathbb{S}_{ij}(k))_{4\times 4}=s(k)e^{ikL\hat{\sigma}_4}S_L(k)$  by Eq. (\ref{sss}). Assume that the possible zeros $\{k_j\}^N_1$ of the functions $n_{33,44}(\mathbb{S})(k),\,  \mathcal{N}(S_1S_2s_3s_4)(k),\, \mathcal{N}(s_1s_2S_3S_4)(k)$ and $n_{11,22}(\mathbb{S})(k)$ are as in Assumption 2.4. Then the solution $(q_1(x,t),\, q_0(x,t), q_{-1}(x,t))$ of system (\ref{pnls}) is found using $M(x,t,k)$ in the form
\bee\label{solu}
\left\{\begin{array}{l}
q_1(x,t)=\d 2i\lim_{k\to \infty}(kM(x,t,k))_{13}, \vspace{0.1in}\\
 q_0(x,t)=\d 2i\lim_{k\to \infty}(kM(x,t,k))_{14}=2i\beta\lim_{k\to \infty}(kM(x,t,k))_{23},\vspace{0.1in}\\
q_{-1}(x,t)=\d 2i\lim_{k\to \infty}(kM(x,t,k))_{24},
\end{array}\right.
\ene
where $M(x,t,k)$ satisfies the following $4\times 4$ matrix Riemann-Hilbert problem:

\begin{itemize}
\item {} $M(x,t,k)$ is sectionally meromorphic on the Riemann $k$-sphere with jumps
 across the contours $\bar{D}_n\cup \bar{D}_m$, $(n, m = 1,..., 4)$ (see Fig.~\ref{kplane}).

\item {} Across the contours $\bar{D}_n\cup \bar{D}_m\, (n, m = 1,..., 4)$, $M(x, t, k)$ satisfies the
 jump condition (\ref{jumpc}).

\item {} The residue conditions of $M(x,t,k)$ are satisfied in Proposition 2.6.

\item {} $M(x, t, k) = \d\mathbb{I}+O(k^{-1})$ as $k\to\infty$.

\end{itemize}
}

\noindent {\bf Proof.}\, System~(\ref{solu}) can be deduced from the large $k$ asymptotics of the eigenfunctions.
We can follow the similar one in Refs.~\cite{f3, nls3}  to complete the rest proof of the Theorem. $\square$

\section{Nonlinearizable boundary conditions}

\quad The key difficulty of initial-boundary value problems is to find the boundary values for a well-posed problem.
 All boundary conditions are required for the definition of $S(k)$ and $S_L(k)$, and hence for the formulating the
 Riemann-Hilbert problem. Our main conclusion exhibits the unknown boundary condition on basis of the prescribed boundary condition and the initial condition
 in terms of the solution of a system of nonlinear integral equations.

\subsection*{\it 4.1.\, The revisited global relation}

\quad By evaluating Eqs.~(\ref{srr}) and (\ref{srr2}) at the point $(x,t)=(0, t)$, we have
\bee
 c(t,k)=\mu_2(0,t,k)e^{-2ik^2t\hat{\sigma}_4}[s(k)e^{ikL\hat{\sigma}_4}S_L(k)],
 \ene
which and Eq.~(\ref{sl}) lead to
\bee\label{gr}
\begin{array}{rl}
  c(t,k)=&\mu_2(0,t,k)e^{-2ik^2t\hat{\sigma}_4}[s(k)e^{ikL\hat{\sigma}_4}e^{2ik^2t\hat{\sigma}_4}\mu_3^{-1}(L,t,k)],\vspace{0.1in}
  \\
 =&\mu_2(0,t,k)[e^{-2ik^2t\hat{\sigma}_4}s(k)][e^{ikL\hat{\sigma}_4}\mu_3^{-1}(L,t,k)],
\end{array}
\ene
Let $\check{s}(t,k)=(\check{s}_{ij}(t,k))_{4\times4}=e^{-2ik^2t\hat{\sigma}_4}s(k)$ and $\mu_3(L,t,k)=(\phi_{ij}(t,k))_{4\times 4}$, then Eq.~(\ref{gr}) can be expanded as
\bes\bee
& \label{c12o}
\begin{array}{rl}
[c(t,k)]_l=&\bar{\phi}_{l1}(t,\bar{k})\sum_{j=1}^4[\mu_2(0,t,k)]_j\check{s}_{j1}(t,k)
+\bar{\phi}_{l2}(t,\bar{k})\sum_{j=1}^4[\mu_2(0,t,k)]_j\check{s}_{j2}(t,k) \vspace{0.1in} \\
 & -\alpha e^{-2ikL}\bar{\phi}_{l3}(t,\bar{k})\sum_{j=1}^4[\mu_2(0,t,k)]_j\check{s}_{j3}(t,k) \vspace{0.1in}\\
 &   -\alpha e^{-2ikL}\bar{\phi}_{l4}(t,\bar{k})\sum_{j=1}^4[\mu_2(0,t,k)]_j\check{s}_{j4}(t,k),\quad l=1,2, \vspace{0.1in} \\
 \end{array} \\
& \label{c34o}
\begin{array}{rl}
  [c(t,k)]_l=&\bar{\phi}_{l3}(t,\bar{k})\sum_{j=1}^4[\mu_2(0,t,k)]_j\check{s}_{j3}(t,k)
   +\bar{\phi}_{l4}(t,\bar{k})\sum_{j=1}^4[\mu_2(0,t,k)]_j\check{s}_{j4}(t,k) \vspace{0.1in} \\
  &  -\alpha e^{2ikL}\bar{\phi}_{l1}(t,\bar{k})\sum_{j=1}^4\mu_2(0,t,k)]_j\check{s}_{j1}(t,k) \vspace{0.1in} \\
   & -\alpha e^{2ikL}\bar{\phi}_{l2}(t,\bar{k})\sum_{j=1}^4[\mu_2(0,t,k)]_j\check{s}_{j2}(t,k), \quad l=3,4,
  \end{array}
   \ene\ees
Thus, the column vectors $[c(t,k)]_l,\, l=1,2$ are analytic and bounded in $D_4$ away from the possible zeros of $n_{11,22}(\mathbb{S})(k)$ and of order $O(\frac{1+e^{-2ikL}}{k})$ as $k\to \infty$, and the column vectors $[c(t,k)]_l,\, l=3,4$ are analytic and bounded in $D_1$ away from the possible zeros of $n_{33,44}(\mathbb{S})(k)$ and of order $O(\frac{1+e^{2ikL}}{k})$ as $k\to \infty$,

\subsection*{\it 4.2.\,  The asymptotic behaviors of eigenfunctions and global relation}

\quad It follows from the Lax pair (\ref{mulax}) that the eigenfunctions $\{\mu_j(x,t,k)\}_1^4$ possess the following asymptotics as $k\to\infty$ (see Appendix)
\bee\label{mua}
\begin{array}{rl}
\mu_j=&\!\!\!\!\!\d\mathbb{I}+\sum_{l=1}^2
\frac{1}{k^l}\left(\begin{array}{cccc} \mu_{j,11}^{(l)} & \mu_{j,12}^{(l)} & \mu_{j,13}^{(l)} & \mu_{j,14}^{(l)} \vspace{0.1in} \\
                                      \mu_{j,21}^{(l)} & \mu_{j,22}^{(l)} & \mu_{j,23}^{(l)} & \mu_{j,24}^{(l)} \vspace{0.1in}\\
                                      \mu_{j,31}^{(l)} & \mu_{j,32}^{(l)} & \mu_{j,33}^{(l)} & \mu_{j,34}^{(l)} \vspace{0.1in}\\
                                      \mu_{j,41}^{(l)} & \mu_{j,42}^{(l)} & \mu_{j,43}^{(l)} & \mu_{j,44}^{(l)}
  \end{array}\right) +O\left(\frac{1}{k^3}\right) \vspace{0.2in}\\
 =&\!\!\!\!\! \d\mathbb{I}+\frac{1}{k}\left(\!\!\!\!\begin{array}{cccc} \d\int_{(x_j, t_j)}^{(x,t)}\Delta_{11} & \!\!\d\int_{(x_j, t_j)}^{(x,t)}\Delta_{12}
  &\!\!\!\!\! \d-\frac{i}{2}q_1 & -\frac{i}{2}q_0 \vspace{0.1in}\\
  \d\int_{(x_j, t_j)}^{(x,t)}\Delta_{21} & \d\int_{(x_j, t_j)}^{(x,t)}\Delta_{22} & \d-\frac{i\beta}{2}q_0 & \d-\frac{i}{2}q_{-1} \vspace{0.1in}\\
                                    \d \frac{i\alpha}{2}\bar{q}_1 & \d\frac{i\alpha\beta}{2}\bar{q}_0 & \d\int_{(x_j, t_j)}^{(x,t)}\Delta_{33} & \d\int_{(x_j, t_j)}^{(x,t)}\Delta_{34} \vspace{0.1in}\\
                                     \d \frac{i\alpha}{2}\bar{q}_0 & \d\frac{i\alpha}{2}\bar{q}_{-1} & \d\int_{(x_j, t_j)}^{(x,t)}\Delta_{43} & \d\int_{(x_j, t_j)}^{(x,t)}\Delta_{44}
  \end{array}\!\!\!\!\right)\!
  +\!\frac{1}{k^2}\left(\!\!\!\!\begin{array}{cccc} \mu_{j,11}^{(2)} & \mu_{j,12}^{(2)} & \mu_{j,13}^{(2)} & \mu_{j,14}^{(2)} \vspace{0.1in}\\
                                      \mu_{j,21}^{(2)} & \mu_{j,22}^{(2)} & \mu_{j,23}^{(2)} & \mu_{j,24}^{(2)} \vspace{0.1in}\\
                                      \mu_{j,31}^{(2)} & \mu_{j,32}^{(2)} & \mu_{j,33}^{(2)} & \mu_{j,34}^{(2)} \vspace{0.1in}\\
                                      \mu_{j,41}^{(2)} & \mu_{j,42}^{(2)} & \mu_{j,43}^{(2)} & \mu_{j,44}^{(2)}
  \end{array}\!\!\!\right) \vspace{0.2in}\\
 &\!\!\!\!\! \d +O\left(\frac{1}{k^3}\right),
  \end{array}
\ene
where we have introduced the following functions
\bee\no
\left\{\!\!\!\begin{array}{rl}
\Delta_{11}=&\!\!\!\!-\Delta_{33}
=\d\frac{i\alpha}{2}(|q_1|^2+|q_0|^2)dx+\frac{\alpha}{2}\sum_{j=0,1}(q_j\bar{q}_{jx}-q_{jx}\bar{q}_j)dt, \vspace{0.1in} \\
\Delta_{22}=&\!\!\!\!-\Delta_{44}^{(1)}=\d\frac{i\alpha}{2}(|q_{-1}|^2+|q_0|^2)dx+\frac{\alpha}{2}\sum_{j=-1,0}(q_j\bar{q}_{jx}-q_{jx}\bar{q}_j)dt, \vspace{0.1in} \\
\Delta_{12}=&\!\!\!\! -\bar{\Delta}_{21}=-\Delta_{34}=\bar{\Delta}_{43} \vspace{0.1in} \\
=&\!\!\!\!\dfrac{i\alpha}{2}(\beta q_1\bar{q}_0+q_0\bar{q}_{-1})dx+\dfrac{\alpha}{2}
 (\beta q_1\bar{q}_{0x}-\beta q_{1x}\bar{q}_0+q_0\bar{q}_{-1x}-q_{0x}\bar{q}_{-1})dt,
\end{array}\right.
\ene
and
\bee\no
\left\{\begin{array}{rl}
\mu_{j,13}^{(2)}=&\!\!\d\frac{1}{4}q_{1x}+\frac{1}{2i}\left(q_1\mu_{j,33}^{(1)}+q_0\mu_{j,43}^{(1)}\right) \vspace{0.1in} \\ =&\d\frac{1}{4}q_{1x}+\frac{1}{2i}\left[q_1\int_{(x_j,t_j)}^{(x,t)}\Delta_{33}+q_0\int_{(x_j,t_j)}^{(x,t)}\Delta_{43}\right], \vspace{0.1in} \\
\mu_{j,14}^{(2)}=&\!\!\d\frac{1}{4}q_{0x}+\frac{1}{2i}\left(q_1\mu_{j,34}^{(1)}+q_0\mu_{j,44}^{(1)}\right) \vspace{0.1in} \\ =&\!\!\d\frac{1}{4}q_{0x}+\frac{1}{2i}\left[q_1\int_{(x_j,t_j)}^{(x,t)}\Delta_{34}+q_0\int_{(x_j,t_j)}^{(x,t)}\Delta_{44}\right], \vspace{0.1in} \\
\mu_{j,23}^{(2)}=&\!\!\d\frac{\beta}{4}q_{0x}+\frac{1}{2i}\left(\beta q_0\mu_{j,33}^{(1)}+q_{-1}\mu_{j,43}^{(1)}\right)\vspace{0.1in} \\
=&\!\!\d\frac{\beta}{4}q_{0x}+\frac{1}{2i}\left[\beta q_0\int_{(x_j,t_j)}^{(x,t)}\Delta_{33}+q_{-1}\int_{(x_j,t_j)}^{(x,t)}\Delta_{43}\right], \vspace{0.1in} \\
\mu_{j,24}^{(2)}=&\!\!\d\frac{1}{4}q_{-1x}+\frac{1}{2i}\left(\beta q_0\mu_{j,34}^{(1)}+q_{-1}\mu_{j,44}^{(1)}\right)\vspace{0.1in} \\
=&\!\!\d\frac{1}{4}q_{-1x}+\frac{1}{2i}\left[\beta q_0\int_{(x_j,t_j)}^{(x,t)}\Delta_{34}+q_{-1}\int_{(x_j,t_j)}^{(x,t)}\Delta_{44}\right],
 \end{array}\right.
\ene
\bee\no
\left\{\begin{array}{rl}
\mu_{j,31}^{(2)}=&\!\!\d\frac{\alpha}{4}\bar{q}_{1x}+\frac{i\alpha }{2}\left(\bar{q}_1\mu_{j,11}^{(1)}+\beta\bar{q}_0\mu_{j,21}^{(1)}\right) \vspace{0.1in} \\ =&\!\!\d\frac{\alpha}{4}\bar{q}_{1x}+\frac{i\alpha}{2}\left[\bar{q}_1\int_{(x_j,t_j)}^{(x,t)}\Delta_{11}+\beta\bar{q}_0
    \int_{(x_j,t_j)}^{(x,t)}\Delta_{21}\right], \vspace{0.1in} \\
\mu_{j,32}^{(2)}=&\!\!\d\frac{\alpha\beta}{4}\bar{q}_{0x}+\frac{i\alpha }{2}\left(\bar{q}_1\mu_{j,12}^{(1)}+\beta\bar{q}_0\mu_{j,22}^{(1)}\right) \vspace{0.1in} \\ =&\!\!\d\frac{\alpha\beta}{4}\bar{q}_{0x}+\frac{i\alpha}{2}\left[\bar{q}_1\int_{(x_j,t_j)}^{(x,t)}\Delta_{12}+\beta\bar{q}_0
    \int_{(x_j,t_j)}^{(x,t)}\Delta_{22}\right], \vspace{0.1in} \\
\mu_{j,41}^{(2)}=&\!\!\d\frac{\alpha}{4}\bar{q}_{0x}+\frac{i\alpha }{2}\left(\bar{q}_0\mu_{j,11}^{(1)}+\bar{q}_{-1}\mu_{j,21}^{(1)}\right) \vspace{0.1in} \\ =&\!\!\d\frac{\alpha}{4}\bar{q}_{0x}+\frac{i\alpha}{2}\left[\bar{q}_0\int_{(x_j,t_j)}^{(x,t)}\Delta_{11}+\beta\bar{q}_{-1}
    \int_{(x_j,t_j)}^{(x,t)}\Delta_{21}\right], \vspace{0.1in} \\
\mu_{j,42}^{(2)}=&\!\!\d\frac{\alpha}{4}\bar{q}_{-1x}+\frac{i\alpha }{2}\left(\bar{q}_0\mu_{j,12}^{(1)}+\bar{q}_{-1}\mu_{j,22}^{(1)}\right) \vspace{0.1in} \\ =&\!\!\d\frac{\alpha}{4}\bar{q}_{-1x}+\frac{i\alpha}{2}\left[\bar{q}_0\int_{(x_j,t_j)}^{(x,t)}\Delta_{12}+\beta\bar{q}_{-1}
    \int_{(x_j,t_j)}^{(x,t)}\Delta_{22}\right],
\end{array}\right.
\ene
The functions $\{\mu^{(i)}_{jl}=\mu^{(i)}_{jl}(x,t)\}_1^4,\, i=1, 2$ are independent of $k$.

We introduce the matrix-valued function $\Psi(t,k)=(\Psi_{ij}(t,k))_{4\times 4}$ as
\bee\label{mu2asy}
\begin{array}{rl}
\mu_2(0, t,k)=\Psi(t,k)=&\left(\begin{array}{cccc}
 \Psi_{11}(t, k) & \Psi_{12}(t, k) & \Psi_{13}(t, k) & \Psi_{14}(t, k) \vspace{0.1in} \\
\Psi_{21}(t, k) & \Psi_{22}(t, k) & \Psi_{23}(t, k) & \Psi_{24}(t, k) \vspace{0.1in} \\
\Psi_{31}(t, k) & \Psi_{32}(t, k) & \Psi_{33}(t, k) & \Psi_{34}(t, k) \vspace{0.1in} \\
\Psi_{41}(t, k) & \Psi_{42}(t, k) & \Psi_{43}(t, k) & \Psi_{44}(t, k)
 \end{array}
\right) \vspace{0.1in}\\
=&\d \mathbb{I}+\sum_{s=1}^2\frac{1}{k^s}\left(\begin{array}{cccc}
 \Psi_{11}^{(s)}(t) & \Psi_{12}^{(s)}(t) & \Psi_{13}^{(s)}(t) & \Psi_{14}^{(s)}(t) \vspace{0.1in} \\
 \Psi_{21}^{(s)}(t) & \Psi_{22}^{(s)}(t) & \Psi_{23}^{(s)}(t) & \Psi_{24}^{(s)}(t) \vspace{0.1in} \\
 \Psi_{31}^{(s)}(t) & \Psi_{32}^{(s)}(t) & \Psi_{33}^{(s)}(t) & \Psi_{34}^{(s)}(t) \vspace{0.1in} \\
 \Psi_{41}^{(s)}(t) & \Psi_{42}^{(s)}(t) & \Psi_{43}^{(s)}(t) & \Psi_{44}^{(s)}(t)
 \end{array}
\right) +O(\frac{1}{k^3}),
\end{array}
\ene

Based on the asymptotic of Eq.~(\ref{mua}) and the boundary data (\ref{ibv}) at $x=0$, we find
\bee\left\{\begin{array}{l}
\Psi_{13}^{(1)}(t)=-\d\frac{i}{2}u_{01}(t), \quad \Psi_{14}^{(1)}(t)=\beta\Psi_{23}^{(1)}(t)=-\frac{i}{2}u_{00}(t),\quad
\Psi_{24}^{(1)}(t)=-\d\frac{i}{2}u_{0-1}(t),
 \vspace{0.1in} \\
\Psi_{13}^{(2)}(t)=\d\frac{1}{4}u_{11}(t)-\frac{i}{2}\left[u_{01}(t)\Psi_{33}^{(1)}+u_{00}(t)\Psi_{43}^{(1)}\right], \vspace{0.1in} \\
\Psi_{14}^{(2)}(t)=\d\frac{1}{4}u_{10}(t)-\frac{i}{2}\left[u_{01}(t)\Psi_{34}^{(1)}+u_{00}(t)\Psi_{44}^{(1)}\right], \vspace{0.1in} \\
\Psi_{23}^{(2)}(t)=\d\frac{\beta}{4}u_{10}(t)-\frac{i}{2}\left[\beta u_{00}(t)\Psi_{33}^{(1)}+u_{0-1}(t)\Psi_{43}^{(1)}\right], \vspace{0.1in} \\
\Psi_{24}^{(2)}(t)=\d\frac{1}{4}u_{1-1}(t)-\frac{i}{2}\left[\beta u_{00}(t)\Psi_{34}^{(1)}+u_{0-1}(t)\Psi_{44}^{(1)}\right], \vspace{0.1in} \\
\Psi_{33}^{(1)}(t)=\d\frac{\alpha}{2}\int^t_0\sum_{j=0,1}\left[\bar{u}_{0j}(t)u_{1j}(t)-u_{0j}(t)\bar{u}_{1j}(t)\right]dt, \vspace{0.1in} \\
\Psi_{44}^{(1)}(t)=\d\frac{\alpha}{2}\int^t_0\sum_{j=-1, 0}\left[\bar{u}_{0j}(t)u_{1j}(t)-u_{0j}(t)\bar{u}_{1j}(t)\right]dt, \vspace{0.1in} \\
\Psi_{34}^{(1)}(t)=\displaystyle\frac{\alpha}{2}\int^t_0\left[\beta u_{11}(t)\bar{u}_{00}(t)-\beta u_{01}(t)\bar{u}_{10}(t)+u_{10}(t)\bar{u}_{0-1}(t)-u_{00}(t)\bar{u}_{1-1}(t)\right]dt,\vspace{0.1in}  \\
\Psi_{43}^{(1)}(t)=\displaystyle\frac{\alpha}{2}\int^t_0\left[\beta u_{10}(t)\bar{u}_{01}(t)-\beta u_{00}(t)\bar{u}_{11}(t)+u_{1-1}(t)\bar{u}_{00}(t)-u_{0-1}(t)\bar{u}_{10}(t)\right]dt,\vspace{0.1in} \\
\end{array}\right.
\ene

Thus we obtain the Dirichlet-Neumann boundary conditions at $x=0$ by using the spectral function:
\bee\label{ud}
\left\{\begin{array}{rl}
u_{01}(t)=&2i\Psi_{13}^{(1)}(t), \quad
u_{00}(t)=2i\Psi_{14}^{(1)}(t)=2i\beta\Psi_{23}^{(1)}(t),   \quad
u_{0-1}(t)= 2i\Psi_{24}^{(1)}(t),  \vspace{0.1in} \\
u_{11}(t)=&4\Psi_{13}^{(2)}(t)+2i\l[u_{01}(t)\Psi_{33}^{(1)}(t)+u_{00}(t)\Psi_{43}^{(1)}(t)\r] \vspace{0.1in} \\
u_{10}(t)=&4\Psi_{14}^{(2)}(t)+2i\l[u_{01}(t)\Psi_{34}^{(1)}(t)+u_{00}(t)\Psi_{44}^{(1)}(t)\r] \vspace{0.1in} \\
 =&4\beta\Psi_{23}^{(2)}(t)+2i\beta\l[\beta u_{00}(t)\Psi_{33}^{(1)}(t)+u_{0-1}(t)\Psi_{43}^{(1)}(t)\r],\vspace{0.1in} \\
u_{1-1}(t)=&4\Psi_{24}^{(2)}(t)+2i\l[\beta u_{00}(t)\Psi_{34}^{(1)}(t)+u_{0-1}(t)\Psi_{44}^{(1)}(t)\r],
\end{array}\right.
\ene

Similarly, we assume that the asymptotic formula of $\mu_3(L, t, k)=\phi(t,k)=\{\phi_{ij}(t,k)\}_{i,j=1}^4$ is of the from
\bee\label{mu3asy}
\begin{array}{rl}
\mu_3(L, t,k)=\phi(t,k)=&\left(\begin{array}{cccc}
 \phi_{11}(t, k) & \phi_{12}(t, k) & \phi_{13}(t, k) & \phi_{14}(t, k) \vspace{0.05in}\\
 \phi_{21}(t, k) & \phi_{22}(t, k) & \phi_{23}(t, k) & \phi_{24}(t, k) \vspace{0.05in}\\
  \phi_{31}(t, k) & \phi_{32}(t, k) & \phi_{33}(t, k) & \phi_{34}(t, k) \vspace{0.05in}\\
   \phi_{41}(t, k) & \phi_{42}(t, k) & \phi_{43}(t, k) & \phi_{44}(t, k)
 \end{array}
\right) \vspace{0.1in}\\
=&\d \mathbb{I}+\sum_{s=1}^2\frac{1}{k^s}\left(\begin{array}{cccc}
 \phi_{11}^{(s)}(t) & \phi_{12}^{(s)}(t) & \phi_{13}^{(s)}(t) & \phi_{14}^{(s)}(t) \vspace{0.05in}\\
 \phi_{21}^{(s)}(t) & \phi_{22}^{(s)}(t) & \phi_{23}^{(s)}(t) & \phi_{24}^{(s)}(t) \vspace{0.05in}\\
  \phi_{31}^{(s)}(t) & \phi_{32}^{(s)}(t) & \phi_{33}^{(s)}(t) & \phi_{34}^{(s)}(t) \vspace{0.05in}\\
   \phi_{41}^{(s)}(t) & \phi_{42}^{(s)}(t) & \phi_{43}^{(s)}(t) & \phi_{44}^{(s)}(t)
 \end{array}
\right) +O(\frac{1}{k^3}),
\end{array}
\ene

By using the asymptotic of Eq.~(\ref{mua}) and the boundary data (\ref{ibv}) at $x=L$, we find
\bee\left\{\begin{array}{l}
\phi_{13}^{(1)}(t)=-\d\frac{i}{2}v_{01}(t), \quad
\phi_{14}^{(1)}(t)=\beta\phi_{23}^{(1)}(t)=-\frac{i}{2}v_{00}(t),\quad
\phi_{24}^{(1)}(t)=-\d\frac{i}{2}v_{0-1}(t),
 \vspace{0.1in} \\
\phi_{13}^{(2)}(t)=\d\frac{1}{4}v_{11}(t)-\frac{i}{2}\left[v_{01}(t)\phi_{33}^{(1)}+v_{00}(t)\phi_{43}^{(1)}\right], \vspace{0.1in} \\
\phi_{14}^{(2)}(t)=\d\frac{1}{4}v_{10}(t)-\frac{i}{2}\left[v_{01}(t)\phi_{34}^{(1)}+v_{00}(t)\phi_{44}^{(1)}\right], \vspace{0.1in} \\
\phi_{23}^{(2)}(t)=\d\frac{\beta}{4}v_{10}(t)-\frac{i}{2}\left[\beta v_{00}(t)\phi_{33}^{(1)}+v_{0-1}(t)\phi_{43}^{(1)}\right], \vspace{0.1in} \\
\phi_{24}^{(2)}(t)=\d\frac{1}{4}v_{1-1}(t)-\frac{i}{2}\left[\beta v_{00}(t)\phi_{34}^{(1)}+v_{0-1}(t)\phi_{44}^{(1)}\right], \vspace{0.1in} \\
\phi_{33}^{(1)}(t)=\d\frac{\alpha}{2}\int^t_0\sum_{j=0,1}\left[\bar{v}_{0j}(t)v_{1j}(t)-v_{0j}(t)\bar{v}_{1j}(t)\right]dt, \vspace{0.1in} \\
\phi_{44}^{(1)}(t)=\d\frac{\alpha}{2}\int^t_0\sum_{j=-1, 0}\left[\bar{v}_{0j}(t)v_{1j}(t)-v_{0j}(t)\bar{v}_{1j}(t)\right]dt, \vspace{0.1in} \\
\phi_{34}^{(1)}(t)=\displaystyle\frac{\alpha}{2}\int^t_0\left[\beta v_{11}(t)\bar{v}_{00}(t)-\beta v_{01}(t)\bar{u}_{10}(t)+v_{10}(t)\bar{v}_{0-1}(t)-v_{00}(t)\bar{v}_{1-1}(t)\right]dt,\vspace{0.1in}  \\
\phi_{43}^{(1)}(t)=\displaystyle\frac{\alpha}{2}\int^t_0\left[\beta v_{10}(t)\bar{v}_{01}(t)-\beta v_{00}(t)\bar{v}_{11}(t)+v_{1-1}(t)\bar{v}_{00}(t)-v_{0-1}(t)\bar{v}_{10}(t)\right]dt,\vspace{0.1in} \\
\end{array}\right.
\ene
which generates the Dirichlet-Neumann boundary data at $x=L$ using the spectral function
\bee\label{vd}
\left\{\begin{array}{rl}
v_{01}(t)=&2i\phi_{13}^{(1)}(t), \quad
v_{00}(t)=2i\phi_{14}^{(1)}(t)=2i\beta\phi_{23}^{(1)}(t),   \quad
v_{0-1}(t)= 2i\phi_{24}^{(1)}(t),  \vspace{0.1in} \\
v_{11}(t)=&4\phi_{13}^{(2)}(t)+2i\l[v_{01}(t)\phi_{33}^{(1)}(t)+v_{00}(t)\phi_{43}^{(1)}(t)\r] \vspace{0.1in} \\
v_{10}(t)=&4\phi_{14}^{(2)}(t)+2i\l[v_{01}(t)\phi_{34}^{(1)}(t)+v_{00}(t)\phi_{44}^{(1)}(t)\r] \vspace{0.1in} \\
 =&4\beta\phi_{23}^{(2)}(t)+2i\beta\l[\beta v_{00}(t)\phi_{33}^{(1)}(t)+v_{0-1}(t)\phi_{43}^{(1)}(t)\r],\vspace{0.1in} \\
v_{1-1}(t)=&4\phi_{24}^{(2)}(t)+2i\l[\beta v_{00}(t)\phi_{34}^{(1)}(t)+v_{0-1}(t)\phi_{44}^{(1)}(t)\r],
\end{array}\right.
\ene

For the vanishing initial values, it follows from Eqs.~(\ref{c12o}) and (\ref{c34o}) that we have the following asymptotic of $c_{1j}(t,k),\,c_{j1}(t,k), j=3,4$, $c_{24}(k)$ and $c_{42}(k)$.

\vspace{0.1in}
\noindent {\bf Proposition 4.1.} {\it Let the initial and Dirichlet boundary data be compatible at points $x=0, L$ (i.e.,
$q_{0j}(0)=u_{0j}(0)$ at $x=0$ and  $q_{0j}(L)=v_{0j}(0)$ at $x=L$,\, $j=1,0,-1$), respectively. Then,
the global relation (\ref{gr}) in the vanishing initial data case implies that the large $k$ behaviors of
 $c_{1j}(t,k),\, c_{j1}(t,k),\, j=3,4$, $c_{24}(t,k)$, and $c_{42}(t,k)$ can be given as follows: }
\bes\bee
&\label{c13} \begin{array}{rl}
c_{13}(t,k)=&\!\!\d\frac{\Psi_{13}^{(1)}}{k}+\frac{\Psi_{13}^{(2)}+\Psi_{13}^{(1)}\bar{\phi}_{33}^{(1)}
+\Psi_{14}^{(1)}\bar{\phi}_{34}^{(1)}}{k^2}
+O\left(\frac{1}{k^3}\right) \vspace{0.08in}\\
&\qquad\displaystyle -\alpha\left[\frac{\bar{\phi}_{31}^{(1)}}{k}+\frac{\bar{\phi}_{31}^{(2)}+\Psi_{11}^{(1)}\bar{\phi}_{31}^{(1)}
+\Psi_{12}^{(1)}\bar{\phi}_{32}^{(1)}}{k^2}+O\left(\frac{1}{k^3}\right)\right]e^{2ikL},\quad {\rm as}\,\,\,k\to\infty,
\end{array} \vspace{0.1in}\\
&\label{c14} \begin{array}{rl}
c_{14}(t,k)=&\!\!\displaystyle\frac{\Psi_{14}^{(1)}}{k}+\frac{\Psi_{14}^{(2)}+\Psi_{14}^{(1)}\bar{\phi}_{44}^{(1)}
+\Psi_{13}^{(1)}\bar{\phi}_{43}^{(1)}}{k^2}
+O\left(\frac{1}{k^3}\right) \vspace{0.08in}\\
&\qquad \displaystyle-\alpha\left[\frac{\bar{\phi}_{41}^{(1)}}{k}+\frac{\bar{\phi}_{41}^{(2)}+\Psi_{11}^{(1)}\bar{\phi}_{41}^{(1)}
+\Psi_{12}^{(1)}\bar{\phi}_{42}^{(1)}}{k^2}+O\left(\frac{1}{k^3}\right)\right]e^{2ikL},\quad k\to\infty,
\end{array}\vspace{0.1in}\\
&\label{c24} \begin{array}{rl}
c_{24}(t,k)=&\!\!\displaystyle\frac{\Psi_{24}^{(1)}}{k}+\frac{\Psi_{24}^{(2)}+\Psi_{24}^{(1)}
\bar{\phi}_{44}^{(1)}+\Psi_{23}^{(1)}\bar{\phi}_{43}^{(1)}}{k^2}
+O\left(\frac{1}{k^3}\right) \vspace{0.08in}\\
&\qquad\displaystyle-\alpha\left[\frac{\bar{\phi}_{42}^{(1)}}{k}+\frac{\bar{\phi}_{42}^{(2)}
+\Psi_{21}^{(1)}\bar{\phi}_{41}^{(1)}
+\Psi_{22}^{(1)}\bar{\phi}_{42}^{(1)}}{k^2}+O\left(\frac{1}{k^3}\right)\right]e^{2ikL},\quad {\rm as}\,\,\,k\to\infty,
\end{array}
\ene
\ees
\bes\bee
&\label{co31} \begin{array}{rl}
c_{31}(t,k)=&\!\!\displaystyle\frac{\Psi_{31}^{(1)}}{k}+\frac{\Psi_{31}^{(2)}+\Psi_{31}^{(1)}\bar{\phi}_{11}^{(1)}
+\Psi_{32}^{(1)}\bar{\phi}_{12}^{(1)}}{k^2}
+O\left(\frac{1}{k^3}\right) \vspace{0.08in}\\
&\qquad\displaystyle -\alpha\left[\frac{\bar{\phi}_{13}^{(1)}}{k}+\frac{\bar{\phi}_{13}^{(2)}+\Psi_{33}^{(1)}\bar{\phi}_{13}^{(1)}
+\Psi_{34}^{(1)}\bar{\phi}_{14}^{(1)}}{k^2}+O\left(\frac{1}{k^3}\right)\right]e^{-2ikL},\quad {\rm as}\,\,\, k\to\infty,
\end{array} \vspace{0.1in}\\
&\label{co41} \begin{array}{rl}
c_{41}(t,k)=&\!\!\displaystyle\frac{\Psi_{41}^{(1)}}{k}+\frac{\Psi_{41}^{(2)}+\Psi_{41}^{(1)}\bar{\phi}_{11}^{(1)}
+\Psi_{42}^{(1)}\bar{\phi}_{12}^{(1)}}{k^2}
+O\left(\frac{1}{k^3}\right) \vspace{0.08in}\\
&\qquad\displaystyle -\alpha\left[\frac{\bar{\phi}_{14}^{(1)}}{k}+\frac{\bar{\phi}_{14}^{(2)}+\Psi_{43}^{(1)}\bar{\phi}_{13}^{(1)}
+\Psi_{44}^{(1)}\bar{\phi}_{14}^{(1)}}{k^2}+O\left(\frac{1}{k^3}\right)\right]e^{-2ikL},\quad {\rm as}\,\,\, k\to\infty,
\end{array}\vspace{0.1in}\\
&\label{co42} \begin{array}{rl}
c_{42}(t,k)=&\!\!\displaystyle\frac{\Psi_{42}^{(1)}}{k}+\frac{\Psi_{42}^{(2)}+\Psi_{41}^{(1)}\bar{\phi}_{21}^{(1)}
+\Psi_{42}^{(1)}\bar{\phi}_{22}^{(1)}}{k^2}
+O\left(\frac{1}{k^3}\right) \vspace{0.08in}\\
&\qquad\displaystyle -\alpha\left[\frac{\bar{\phi}_{24}^{(1)}}{k}+\frac{\bar{\phi}_{24}^{(2)}+\Psi_{43}^{(1)}\bar{\phi}_{23}^{(1)}
+\Psi_{44}^{(1)}\bar{\phi}_{24}^{(1)}}{k^2}+O\left(\frac{1}{k^3}\right)\right]e^{-2ikL},\quad {\rm as}\,\,\, k\to\infty,
\end{array}
\ene
\ees

\vspace{0.1in}
\noindent {\bf Proof.} It follows from the global relation (\ref{gr}) under the vanishing initial data that we get
\bes\bee
&\label{ca} c_{13}(t,k)=-\alpha e^{2ikL}[\Psi_{11}(t,k)\bar{\phi}_{31}(t,\bar{k})+\Psi_{12}(t,k)\bar{\phi}_{32}(t,\bar{k})] +\Psi_{13}(t,k)\bar{\phi}_{33}(t,\bar{k})+\Psi_{14}(t,k)\bar{\phi}_{34}(t,\bar{k}), \vspace{0.1in}\\
& \label{cb} c_{14}(t,k)=-\alpha e^{2ikL}[\Psi_{11}(t,k)\bar{\phi}_{41}(t,\bar{k})+\Psi_{12}(t,k)\bar{\phi}_{42}(t,\bar{k})] +\Psi_{13}(t,k)\bar{\phi}_{43}(t,\bar{k})+\Psi_{14}(t,k)\bar{\phi}_{44}(t,\bar{k}),
 \vspace{0.1in}\\
& \label{cc} c_{24}(t,k)=-\alpha e^{2ikL}[\Psi_{21}(t,k)\bar{\phi}_{41}(t,\bar{k})+\Psi_{22}(t,k)\bar{\phi}_{42}(t,\bar{k})] +\Psi_{23}(t,k)\bar{\phi}_{43}(t,\bar{k})+\Psi_{24}(t,k)\bar{\phi}_{44}(t,\bar{k}),
\ene\ees

Recalling the $t$-part of the Lax pair (\ref{mulax})
\bee\label{tlax}
 \mu_t+2ik^2[\sigma_4, \mu]=V(x,t,k)\mu,
 \ene
It follows from Eq.~(\ref{tlax}) that the first column of Eq.~(\ref{tlax}) with $\mu=\mu_2(0,t,k)=\Psi(t,k)$ is
\bee\label{psit11}
\left\{ \begin{array}{rl}
 \Psi_{11,t}(t,k)=&\!\!\!\! 2k(u_{01}\Psi_{31}+u_{00}\Psi_{41})+i(u_{11}\Psi_{31}+u_{10}\Psi_{41}) \vspace{0.08in}\\
  &-i\alpha[(|u_{01}|^2+|u_{00}|^2)\Psi_{11}+(\beta u_{01}\bar{u}_{00}+u_{00}\bar{u}_{0-1})\Psi_{21}], \vspace{0.08in}\\
 \Psi_{21,t}(t,k)=&\!\!\!\! 2k(\beta u_{00}\Psi_{31}+u_{0-1}\Psi_{41})+i(\beta u_{10}\Psi_{31}+u_{1-1}\Psi_{41}) \vspace{0.08in}\\
     & -i\alpha[(\beta u_{00}\bar{u}_{01}+u_{0-1}\bar{u}_{00})\Psi_{11}+(|u_{0-1}|^2+|u_{00}|^2)\Psi_{21}],\vspace{0.08in}\\
 \Psi_{31,t}(t,k)=&\!\!\!\! 4ik^2\Psi_{31}+2\alpha k(\bar{u}_{01}\Psi_{11}+\beta \bar{u}_{00}\Psi_{21})
      -i\alpha(\bar{u}_{11}\Psi_{11}+\beta \bar{u}_{10}\Psi_{21}) \vspace{0.08in}\\
    & +i\alpha[(|u_{01}|^2+|u_{00}|^2)\Psi_{31}+(\beta u_{0-1}\bar{u}_{00}+u_{00}\bar{u}_{01})\Psi_{41}] \vspace{0.08in}\\
 \Psi_{41,t}(t,k)=&\!\!\!\! 4ik^2\Psi_{41}+2\alpha k(\bar{u}_{00}\Psi_{11}+\bar{u}_{0-1}\Psi_{21})
   -i\alpha(\bar{u}_{10}\Psi_{11}+\bar{u}_{1-1}\Psi_{21})\vspace{0.08in}\\
& +i\alpha[(\beta u_{00}\bar{u}_{0-1}+u_{01}\bar{u}_{00})\Psi_{31}+(|u_{0-1}|^2+|u_{00}|^2)\Psi_{41}],
\end{array}\right.
\ene
the second column of Eq.~(\ref{tlax}) with $\mu=\mu_2(0,t,k)=\Psi(t,k)$ yields
\bee\label{psit12}
\left\{ \begin{array}{rl}
 \Psi_{12,t}(t,k)=&\!\!\!\! 2k(u_{01}\Psi_{32}+u_{00}\Psi_{42})+i(u_{11}\Psi_{32}+u_{10}\Psi_{42}) \vspace{0.08in}\\
  &-i\alpha[(|u_{01}|^2+|u_{00}|^2)\Psi_{12}+(\beta u_{01}\bar{u}_{00}+u_{00}\bar{u}_{0-1})\Psi_{22}], \vspace{0.08in}\\
 \Psi_{22,t}(t,k)=&\!\!\!\! 2k(\beta u_{00}\Psi_{32}+u_{0-1}\Psi_{42})+i(\beta u_{10}\Psi_{32}+u_{1-1}\Psi_{42}) \vspace{0.08in}\\
     & -i\alpha[(\beta u_{00}\bar{u}_{01}+u_{0-1}\bar{u}_{00})\Psi_{12}+(|u_{0-1}|^2+|u_{00}|^2)\Psi_{22}],\vspace{0.08in}\\
 \Psi_{32,t}(t,k)=&\!\!\!\! 4ik^2\Psi_{32}+2\alpha k(\bar{u}_{01}\Psi_{12}+\beta \bar{u}_{00}\Psi_{22})
      -i\alpha(\bar{u}_{11}\Psi_{12}+\beta \bar{u}_{10}\Psi_{22}) \vspace{0.08in}\\
    & +i\alpha[(|u_{01}|^2+|u_{00}|^2)\Psi_{32}+(\beta u_{0-1}\bar{u}_{00}+u_{00}\bar{u}_{01})\Psi_{42}] \vspace{0.08in}\\
 \Psi_{42,t}(t,k)=&\!\!\!\! 4ik^2\Psi_{42}+2\alpha k(\bar{u}_{00}\Psi_{12}+\bar{u}_{0-1}\Psi_{22})
   -i\alpha(\bar{u}_{10}\Psi_{12}+\bar{u}_{1-1}\Psi_{22})\vspace{0.08in}\\
& +i\alpha[(\beta u_{00}\bar{u}_{0-1}+u_{01}\bar{u}_{00})\Psi_{32}+(|u_{0-1}|^2+|u_{00}|^2)\Psi_{42}],
\end{array}\right.
\ene
the third column of Eq.~(\ref{tlax}) with $\mu=\mu_2(0,t,k)=\Psi(t,k)$ is of
\bee\label{psit13}
\left\{ \begin{array}{rl}
 \Psi_{13,t}(t,k)=&\!\!\!\! -4ik^2\Psi_{13}+2k(u_{01}\Psi_{33}+u_{00}\Psi_{43})+i(u_{11}\Psi_{33}+u_{10}\Psi_{43}) \vspace{0.08in}\\
  &-i\alpha[(|u_{01}|^2+|u_{00}|^2)\Psi_{13}+(\beta u_{01}\bar{u}_{00}+u_{00}\bar{u}_{0-1})\Psi_{23}], \vspace{0.08in}\\
 \Psi_{23,t}(t,k)=&\!\!\!\! -4ik^2\Psi_{23}+2k(\beta u_{00}\Psi_{33}+u_{0-1}\Psi_{43})+i(\beta u_{10}\Psi_{33}+u_{1-1}\Psi_{43}) \vspace{0.08in}\\
     & -i\alpha[(\beta u_{00}\bar{u}_{01}+u_{0-1}\bar{u}_{00})\Psi_{13}+(|u_{0-1}|^2+|u_{00}|^2)\Psi_{23}],\vspace{0.08in}\\
 \Psi_{33,t}(t,k)=&\!\!\!\! 2\alpha k(\bar{u}_{01}\Psi_{13}+\beta \bar{u}_{00}\Psi_{23})
      -i\alpha(\bar{u}_{11}\Psi_{13}+\beta \bar{u}_{10}\Psi_{23}) \vspace{0.08in}\\
    & +i\alpha[(|u_{01}|^2+|u_{00}|^2)\Psi_{33}+(\beta u_{0-1}\bar{u}_{00}+u_{00}\bar{u}_{01})\Psi_{43}] \vspace{0.08in}\\
 \Psi_{43,t}(t,k)=&\!\!\!\! 2\alpha k(\bar{u}_{00}\Psi_{13}+\bar{u}_{0-1}\Psi_{23})
   -i\alpha(\bar{u}_{10}\Psi_{13}+\bar{u}_{1-1}\Psi_{23})\vspace{0.08in}\\
& +i\alpha[(\beta u_{00}\bar{u}_{0-1}+u_{01}\bar{u}_{00})\Psi_{33}+(|u_{0-1}|^2+|u_{00}|^2)\Psi_{43}],
\end{array}\right.
\ene
and the fourth column of Eq.~(\ref{tlax}) with $\mu=\mu_2(0,t,k)=\Psi(t,k)$ is
\bee\label{psit14}
\left\{ \begin{array}{rl}
 \Psi_{14,t}(t,k)=&\!\!\!\! -4ik^2\Psi_{14}+2k(u_{01}\Psi_{34}+u_{00}\Psi_{44})+i(u_{11}\Psi_{34}+u_{10}\Psi_{44}) \vspace{0.08in}\\
  &-i\alpha[(|u_{01}|^2+|u_{00}|^2)\Psi_{14}+(\beta u_{01}\bar{u}_{00}+u_{00}\bar{u}_{0-1})\Psi_{24}], \vspace{0.08in}\\
 \Psi_{24,t}(t,k)=&\!\!\!\! -4ik^2\Psi_{24}+2k(\beta u_{00}\Psi_{34}+u_{0-1}\Psi_{44})+i(\beta u_{10}\Psi_{34}+u_{1-1}\Psi_{44}) \vspace{0.08in}\\
     & -i\alpha[(\beta u_{00}\bar{u}_{01}+u_{0-1}\bar{u}_{00})\Psi_{14}+(|u_{0-1}|^2+|u_{00}|^2)\Psi_{24}],\vspace{0.08in}\\
 \Psi_{34,t}(t,k)=&\!\!\!\! 2\alpha k(\bar{u}_{01}\Psi_{14}+\beta \bar{u}_{00}\Psi_{24})
      -i\alpha(\bar{u}_{11}\Psi_{14}+\beta \bar{u}_{10}\Psi_{24}) \vspace{0.08in}\\
    & +i\alpha[(|u_{01}|^2+|u_{00}|^2)\Psi_{34}+(\beta u_{0-1}\bar{u}_{00}+u_{00}\bar{u}_{01})\Psi_{44}] \vspace{0.08in}\\
 \Psi_{44,t}(t,k)=&\!\!\!\! 2\alpha k(\bar{u}_{00}\Psi_{14}+\bar{u}_{0-1}\Psi_{24})
   -i\alpha(\bar{u}_{10}\Psi_{14}+\bar{u}_{1-1}\Psi_{24})\vspace{0.08in}\\
& +i\alpha[(\beta u_{00}\bar{u}_{0-1}+u_{01}\bar{u}_{00})\Psi_{34}+(|u_{0-1}|^2+|u_{00}|^2)\Psi_{44}],
\end{array}\right.
\ene

Suppose that $\Psi_{j1}(t,k)$'s,\, $j=1,2,3,4$ are of the form
\bee\label{psi1}
\left(\begin{array}{c} \Psi_{11}(t,k) \\ \Psi_{21}(t,k) \\ \Psi_{31}(t,k) \\ \Psi_{41}(t,k) \end{array}\right)
=\left(a_{10}(t)+\frac{a_{11}(t)}{k}+\frac{a_{12}(t)}{k^2}+\cdots\right)+\left(b_{10}(t)+\frac{b_{11}(t)}{k}
+\frac{b_{12}(t)}{k^2}+\cdots\right)e^{4ik^2t},
\ene
where the $4\times 1$ column vector functions $a_{1j}(t),b_{1j}(t)\, (j=0,1,...,)$ are independent of $k$.

By substituting Eq.~(\ref{psi1}) into Eq.(\ref{psit11}) and using the initial conditions
$a_{10}(0)+b_{10}(0)=(1, 0,0,0)^T, \, a_{11}(0)+b_{11}(0)=(0, 0, 0, 0)^T,$ we have
\bee\label{psi11g}
\begin{array}{rl}
\left(\begin{array}{c} \Psi_{11} \\ \Psi_{21} \\ \Psi_{31} \\ \Psi_{41} \end{array}\right)
=&\d\left(\begin{array}{c} 1 \\ 0 \\ 0 \\ 0 \end{array}\right)
+\frac{1}{k}\left(\begin{array}{c} \Psi_{11}^{(1)} \vspace{0.05in}\\ \Psi_{21}^{(1)} \vspace{0.05in}\\ \Psi_{31}^{(1)} \vspace{0.05in}\\ \Psi_{41}^{(1)} \end{array}\right)
+\frac{1}{k^2}\left(\begin{array}{c} \Psi_{11}^{(2)} \vspace{0.05in}\\ \Psi_{21}^{(2)} \vspace{0.05in}\\ \Psi_{31}^{(2)} \vspace{0.05in}\\ \Psi_{41}^{(2)} \end{array}\right)
+O\left(\frac{1}{k^3}\right) \vspace{0.1in}\\
&+\d \left[\frac{1}{k}\left(\begin{array}{c} 0 \\ 0 \vspace{0.05in}\\ -\frac{i\alpha}{2}\bar{u}_{01}(0) \vspace{0.05in}\\ -\frac{i\alpha}{2}\bar{u}_{00}(0) \end{array}\right)+O\left(\frac{1}{k^2}\right)  \right]e^{4ik^2t},
\end{array}
\ene

Similarly, it follows from Eqs.~(\ref{psit12})-(\ref{psit14}) that we have the asymptotic formulae for $\Psi_{ij},\, i=1,2,3,4; j=2,3,4$ in the forms
\bee\begin{array}{rl}
\left(\begin{array}{c} \Psi_{12} \\ \Psi_{22} \\ \Psi_{32} \\ \Psi_{42} \end{array}\right)
=&\d \left(\begin{array}{c} 0 \\ 1 \\ 0 \\ 0 \end{array}\right)
+\frac{1}{k}\left(\begin{array}{c} \Psi_{12}^{(1)} \vspace{0.05in}\\ \Psi_{22}^{(1)} \vspace{0.05in}\\ \Psi_{32}^{(1)} \vspace{0.05in}\\ \Psi_{42}^{(1)} \end{array}\right)
+\frac{1}{k^2}\left(\begin{array}{c} \Psi_{12}^{(2)} \vspace{0.05in}\\ \Psi_{22}^{(2)} \vspace{0.05in}\\ \Psi_{32}^{(2)} \vspace{0.05in}\\ \Psi_{42}^{(2)} \end{array}\right)
+O\left(\frac{1}{k^3}\right)\vspace{0.1in}\\
&+\d\left[\frac{1}{k}\left(\begin{array}{c} 0 \\ 0 \vspace{0.05in}\\ -\frac{i\alpha\beta}{2}\bar{u}_{00}(0) \vspace{0.05in}\\ -\frac{i\alpha}{2}\bar{u}_{0-1}(0) \end{array}\right)+O\left(\frac{1}{k^2}\right)\right]e^{4ik^2t},
\label{psi12g}
\end{array}\ene
\bee\begin{array}{rl}
\left(\begin{array}{c} \Psi_{13} \\ \Psi_{23} \\ \Psi_{33} \\ \Psi_{43} \end{array}\right)
=&\d \left(\begin{array}{c} 0 \\ 0 \\ 1 \\ 0 \end{array}\right)
+\frac{1}{k}\left(\begin{array}{c} \Psi_{13}^{(1)} \vspace{0.05in}\\ \Psi_{23}^{(1)} \vspace{0.05in}\\ \Psi_{33}^{(1)} \vspace{0.05in}\\ \Psi_{43}^{(1)} \end{array}\right)
+\frac{1}{k^2}\left(\begin{array}{c} \Psi_{13}^{(2)} \vspace{0.05in}\\ \Psi_{23}^{(2)} \vspace{0.05in}\\ \Psi_{33}^{(2)} \vspace{0.05in}\\ \Psi_{43}^{(2)} \end{array}\right)
+O\left(\frac{1}{k^3}\right)\vspace{0.1in}\\
&+\d \left[\frac{1}{k}\left(\begin{array}{c} \frac{i}{2}u_{01}(0) \vspace{0.05in}\\  \frac{i\beta}{2}u_{00}(0) \vspace{0.05in}\\ 0 \vspace{0.05in}\\ 0\end{array}\right)+O\left(\frac{1}{k^2}\right)  \right]e^{-4ik^2t},
\label{psi13g}
\end{array}\ene
and
\bee\begin{array}{rl}
\left(\begin{array}{c} \Psi_{14} \\ \Psi_{24} \\ \Psi_{34} \\ \Psi_{44} \end{array}\right)
=&\d \left(\begin{array}{c} 0 \\ 0 \\ 0 \\ 1 \end{array}\right)
+\frac{1}{k}\left(\begin{array}{c} \Psi_{14}^{(1)} \vspace{0.05in}\\ \Psi_{24}^{(1)} \vspace{0.05in}\\ \Psi_{34}^{(1)} \vspace{0.05in}\\ \Psi_{44}^{(1)} \end{array}\right)
+\frac{1}{k^2}\left(\begin{array}{c} \Psi_{14}^{(2)} \vspace{0.05in}\\ \Psi_{24}^{(2)} \vspace{0.05in}\\ \Psi_{34}^{(2)} \vspace{0.05in}\\ \Psi_{44}^{(2)} \end{array}\right)
+O\left(\frac{1}{k^3}\right)\vspace{0.1in}\\
&+\d\left[\frac{1}{k}\left(\begin{array}{c} \frac{i}{2}u_{00}(0) \vspace{0.05in}\\  \frac{i}{2}u_{0-1}(0) \vspace{0.05in}\\ 0 \vspace{0.05in}\\ 0\end{array}\right)+O\left(\frac{1}{k^2}\right) \right]e^{-4ik^2t},
\label{psi14gg}
\end{array}\ene

The substitution of Eqs.~(\ref{psi11g})-(\ref{psi14gg}) into Eq.~(\ref{ca}) yields Eq.~(\ref{c13}). Similarly, we can also get Eqs.~(\ref{c14}) and (\ref{c24}).

Similar to Eqs.~(\ref{psit11})-(\ref{psit14}) for $\mu_2(0,t,k)$, we also know that the function $\mu(x,t,k)=\mu_3(L, t,k)$ at $x=L$ satisfy the $t$-part of Lax pair (\ref{tlax}) such that we have the first column of Eq.~(\ref{tlax}) with $\mu=\mu_3(L,t,k)=\phi(t,k)$
\bee\label{psit21}
\left\{ \begin{array}{rl}
 \phi_{11,t}(t,k)=&\!\!\!\! 2k(v_{01}\phi_{31}+v_{00}\phi_{41})+i(v_{11}\phi_{31}+v_{10}\phi_{41}) \vspace{0.08in}\\
  &-i\alpha[(|v_{01}|^2+|v_{00}|^2)\phi_{11}+(\beta v_{01}\bar{v}_{00}+v_{00}\bar{v}_{0-1})\phi_{21}], \vspace{0.08in}\\
 \phi_{21,t}(t,k)=&\!\!\!\! 2k(\beta v_{00}\phi_{31}+v_{0-1}\phi_{41})+i(\beta v_{10}\phi_{31}+v_{1-1}\phi_{41}) \vspace{0.08in}\\
     & -i\alpha[(\beta v_{00}\bar{v}_{01}+v_{0-1}\bar{u}_{00})\phi_{11}+(|v_{0-1}|^2+|v_{00}|^2)\phi_{21}],\vspace{0.08in}\\
 \phi_{31,t}(t,k)=&\!\!\!\! 4ik^2\phi_{31}+2\alpha k(\bar{v}_{01}\phi_{11}+\beta \bar{v}_{00}\phi_{21})
      -i\alpha(\bar{v}_{11}\phi_{11}+\beta \bar{v}_{10}\phi_{21}) \vspace{0.08in}\\
    & +i\alpha[(|v_{01}|^2+|v_{00}|^2)\phi_{31}+(\beta v_{0-1}\bar{v}_{00}+v_{00}\bar{v}_{01})\phi_{41}] \vspace{0.08in}\\
 \phi_{41,t}(t,k)=&\!\!\!\!4ik^2\phi_{41}+2\alpha k(\bar{v}_{00}\phi_{11}+\bar{v}_{0-1}\phi_{21})
   -i\alpha(\bar{v}_{10}\phi_{11}+\bar{v}_{1-1}\phi_{21})\vspace{0.08in}\\
& +i\alpha[(\beta v_{00}\bar{v}_{0-1}+v_{01}\bar{v}_{00})\phi_{31}+(|v_{0-1}|^2+|v_{00}|^2)\phi_{41}],
\end{array}\right.
\ene
the second column of Eq.~(\ref{tlax}) with $\mu=\mu_3(L,t,k)=\phi(t,k)$
\bee\label{psit22}
\left\{ \begin{array}{rl}
 \phi_{12,t}(t,k)=&\!\!\!\! 2k(v_{01}\phi_{32}+v_{00}\phi_{42})+i(v_{11}\phi_{32}+v_{10}\phi_{42}) \vspace{0.08in}\\
  &-i\alpha[(|v_{01}|^2+|v_{00}|^2)\phi_{12}+(\beta v_{01}\bar{v}_{00}+v_{00}\bar{v}_{0-1})\phi_{22}], \vspace{0.08in}\\
 \phi_{22,t}(t,k)=&\!\!\!\! 2k(\beta v_{00}\phi_{32}+v_{0-1}\phi_{42})+i(\beta v_{10}\phi_{32}+v_{1-1}\phi_{42}) \vspace{0.08in}\\
     & -i\alpha[(\beta v_{00}\bar{v}_{01}+v_{0-1}\bar{v}_{00})\phi_{12}+(|v_{0-1}|^2+|v_{00}|^2)\phi_{22}],\vspace{0.08in}\\
 \phi_{32,t}(t,k)=&\!\!\!\!4ik^2\phi_{32}+2\alpha k(\bar{v}_{01}\phi_{12}+\beta \bar{v}_{00}\phi_{22})
      -i\alpha(\bar{v}_{11}\phi_{12}+\beta \bar{v}_{10}\phi_{22}) \vspace{0.08in}\\
    & +i\alpha[(|v_{01}|^2+|u_{00}|^2)\phi_{32}+(\beta v_{0-1}\bar{v}_{00}+v_{00}\bar{v}_{01})\phi_{42}] \vspace{0.08in}\\
 \phi_{42,t}(t,k)=&\!\!\!\!4ik^2\phi_{42}+2\alpha k(\bar{v}_{00}\phi_{12}+\bar{v}_{0-1}\phi_{22})
   -i\alpha(\bar{v}_{10}\phi_{12}+\bar{v}_{1-1}\phi_{22})\vspace{0.08in}\\
& +i\alpha[(\beta v_{00}\bar{v}_{0-1}+v_{01}\bar{v}_{00})\phi_{32}+(|v_{0-1}|^2+|v_{00}|^2)\phi_{42}],
\end{array}\right.
\ene
the third column of Eq.~(\ref{tlax}) with $\mu=\mu_3(L,t,k)=\phi(t,k)$
\bee\label{psit23}
\left\{ \begin{array}{rl}
 \phi_{13,t}(t,k)=&\!\!\!\!-4ik^2\phi_{13}+2k(v_{01}\phi_{33}+v_{00}\phi_{43})+i(v_{11}\phi_{33}+v_{10}\phi_{43}) \vspace{0.08in}\\
  &-i\alpha[(|v_{01}|^2+|v_{00}|^2)\phi_{13}+(\beta v_{01}\bar{v}_{00}+v_{00}\bar{v}_{0-1})\phi_{23}], \vspace{0.08in}\\
 \phi_{23,t}(t,k)=&\!\!\!\!-4ik^2\phi_{23}+2k(\beta v_{00}\phi_{33}+v_{0-1}\phi_{43})+i(\beta v_{10}\phi_{33}+v_{1-1}\phi_{43}) \vspace{0.08in}\\
     & -i\alpha[(\beta v_{00}\bar{v}_{01}+v_{0-1}\bar{v}_{00})\phi_{13}+(|v_{0-1}|^2+|v_{00}|^2)\phi_{23}],\vspace{0.08in}\\
 \phi_{33,t}(t,k)=&\!\!\!\!2\alpha k(\bar{v}_{01}\phi_{13}+\beta \bar{v}_{00}\phi_{23})
      -i\alpha(\bar{v}_{11}\phi_{13}+\beta \bar{v}_{10}\phi_{23}) \vspace{0.08in}\\
    & +i\alpha[(|v_{01}|^2+|v_{00}|^2)\phi_{33}+(\beta v_{0-1}\bar{v}_{00}+v_{00}\bar{v}_{01})\phi_{43}] \vspace{0.08in}\\
 \phi_{43,t}(t,k)=&\!\!\!\!2\alpha k(\bar{v}_{00}\phi_{13}+\bar{v}_{0-1}\phi_{23})
   -i\alpha(\bar{v}_{10}\phi_{13}+\bar{v}_{1-1}\phi_{23})\vspace{0.08in}\\
& +i\alpha[(\beta v_{00}\bar{v}_{0-1}+v_{01}\bar{v}_{00})\phi_{33}+(|v_{0-1}|^2+|v_{00}|^2)\phi_{43}],
\end{array}\right.
\ene
and the fourth column of Eq.~(\ref{tlax}) with $\mu=\mu_3(L,t,k)=\phi(t,k)$
\bee\label{psit24}
\left\{ \begin{array}{rl}
 \phi_{14,t}(t,k)=&\!\!\!\!-4ik^2\phi_{14}+2k(v_{01}\phi_{34}+v_{00}\phi_{44})+i(v_{11}\phi_{34}+v_{10}\phi_{44}) \vspace{0.08in}\\
  &-i\alpha[(|v_{01}|^2+|v_{00}|^2)\phi_{14}+(\beta v_{01}\bar{v}_{00}+v_{00}\bar{v}_{0-1})\phi_{24}], \vspace{0.08in}\\
 \phi_{24,t}(t,k)=&\!\!\!\!-4ik^2\phi_{24}+2k(\beta v_{00}\phi_{34}+v_{0-1}\phi_{44})+i(\beta v_{10}\phi_{34}+v_{1-1}\phi_{44}) \vspace{0.08in}\\
     & -i\alpha[(\beta v_{00}\bar{v}_{01}+v_{0-1}\bar{v}_{00})\phi_{14}+(|v_{0-1}|^2+|v_{00}|^2)\phi_{24}],\vspace{0.08in}\\
 \phi_{34,t}(t,k)=&\!\!\!\!2\alpha k(\bar{v}_{01}\phi_{14}+\beta \bar{v}_{00}\phi_{24})
      -i\alpha(\bar{v}_{11}\phi_{14}+\beta \bar{v}_{10}\phi_{24}) \vspace{0.08in}\\
    & +i\alpha[(|v_{01}|^2+|v_{00}|^2)\phi_{34}+(\beta v_{0-1}\bar{v}_{00}+v_{00}\bar{v}_{01})\phi_{44}] \vspace{0.08in}\\
 \phi_{44,t}(t,k)=&\!\!\!\!2\alpha k(\bar{v}_{00}\phi_{14}+\bar{v}_{0-1}\phi_{24})-i\alpha(\bar{v}_{10}\phi_{14}+\bar{v}_{1-1}\phi_{24})\vspace{0.08in}\\
& +i\alpha[(\beta v_{00}\bar{v}_{0-1}+u_{01}\bar{v}_{00})\phi_{34}+(|v_{0-1}|^2+|v_{00}|^2)\phi_{44}],
\end{array}\right.
\ene

Similarly, we can also obtain the asymptotic formulae for $\phi_{ij},\, i,j=1,2,3,4$. The substitution of these formulae into Eq.~(\ref{ca}) and using the assumption the initial and boundary data are compatible at $x=0$ and $x=L$, we find the asymptotic result (\ref{c13}) of $c_{13}(t,k)$ for $k\to \infty$. Similarly we can also show Eqs.~(\ref{c14}) and (\ref{c24}) for $c_{j4}(t,k),\, j=1,2$ as $k\to \infty$.

 Similarly, it follows from the global relation (\ref{gr}) under the vanishing initial data that we have
\bes\bee
&\label{cag1}
c_{31}(t,k)=\Psi_{31}(t,k)\bar{\phi}_{11}(t,\bar{k})+\Psi_{32}(t,k)\bar{\phi}_{12}(t,\bar{k})-\alpha e^{-2ikL}[\Psi_{33}(t,k)\bar{\phi}_{13}(t,\bar{k})+\Psi_{34}(t,k)\bar{\phi}_{14}(t,\bar{k})],\\
& \label{cbg2} c_{41}(t,k)=\Psi_{41}(t,k)\bar{\phi}_{11}(t,\bar{k})+\Psi_{42}(t,k)\bar{\phi}_{12}(t,\bar{k})-\alpha e^{-2ikL}[\Psi_{43}(t,k)\bar{\phi}_{13}(t,\bar{k})+\Psi_{44}(t,k)\bar{\phi}_{14}(t,\bar{k}),\\
& \label{ccg3} c_{42}(t,k)=\Psi_{41}(t,k)\bar{\phi}_{21}(t,\bar{k})+\Psi_{42}(t,k)\bar{\phi}_{22}(t,\bar{k})-\alpha e^{-2ikL}[\Psi_{43}(t,k)\bar{\phi}_{23}(t,\bar{k})+\Psi_{44}(t,k)\bar{\phi}_{24}(t,\bar{k})],
\ene\ees
such that we can show Eqs.~(\ref{co31})-(\ref{co42}) by means of $c_{j1}(t,k),\, j=3,4$ and $c_{42}(t,k)$ as $k\to \infty$.  $\square$

\subsection*{\it 4.3.\, The map between Dirichlet and Neumann problems}

\quad In what follows we mainly show that the spectral functions $S(k)$ and $S_L(k)$ can be expressed in terms of the prescribed Dirichlet and Neumann boundary data and the initial data using the solution of a system of integral equations.

For simplicity, we define the notations as
\bee
 F_{\pm} (t,k)=F(t,k)\pm F(t, -k), \qquad \Sigma_{\pm}=\Sigma_{\pm}(k)=e^{2ikL}\pm e^{-2ikL}.
\ene
The sign $\partial D_j,\ j=1,...,4$ stands for the boundary of the $j$th quadrant $D_j$, oriented so that $D_j$ lies to the left of $\partial D_j$.
$\partial D_3^0$ denotes the boundary contour which has not contain the zeros of $\Sigma_-(k)$ and $\partial D_3^0=-\partial D_1^0$.

\vspace{0.1in}
\noindent{\bf Theorem 4.2.} {\it Let $q_{0j}(x)=q_j(x,t=0)\in C^{\infty}[0, L],\, j=1,0,-1$ be an initial data of Eq.~(\ref{pnls}) on the finite interval $x\in [0, L]$ and  $T<\infty$.

For the Dirichlet problem, the smooth boundary data $u_{0j}(t)$ and $v_{0j}(t)\, (j=1,0,-1)$ on the interval $t\in [0, T)$ are sufficiently smooth and compatible with the initial data $q_{0j}(x),\, (j=1,0,-1)$ at points $(x_2, t_2)=(0, 0)$ and $(x_3, t_3)=(L, 0)$, respectively, i.e., $u_{0j}(0)=q_{0j}(0),\, v_{0j}(0)=q_{0j}(L), \, j=1,0,-1$.

For the Neumann problem, the smooth boundary data $u_{1j}(t)$ and $v_{1j}(t)\, (j=1,0,-1)$ on the interval $t\in [0, T)$ are sufficiently smooth and compatible with the initial data $q_{0j}(x),\, (j=1,0,-1)$ at the origin $(x_2, t_2)=(0, 0)$ and $(x_3, t_3)=(L, 0)$, respectively, i.e., $u_{1j}(0)=\partial_x q_{0j}(0),\, v_{1j}(0)=\partial_x q_{0j}(L), \, j=1,0,-1$.}

{\it For simplicity, let $n_{33,44}(\mathbb{S})(k)$ have no zeros in the domain $D_1$. Then the spectral functions $S(k)$ and $S_L(k)$ are defined by
\bee\label{skm}
S(k)\!\!=\!\!\left(\!\!\!\begin{array}{cccc}
 \overline{\Psi_{11}(T,\bar{k})} & \overline{\Psi_{21}(T,\bar{k})} & -\alpha\overline{\Psi_{31}(T,\bar{k})}e^{4ik^2T} & -\alpha\overline{\Psi_{41}(T,\bar{k})}e^{4ik^2T} \vspace{0.1in}\\
\overline{\Psi_{12}(T,\bar{k})} & \overline{\Psi_{22}(T,\bar{k})} & -\alpha\overline{\Psi_{32}(T,\bar{k})}e^{4ik^2T} & -\alpha\overline{\Psi_{42}(T,\bar{k})}e^{4ik^2T} \vspace{0.1in}\\
-\alpha\overline{\Psi_{13}(T,\bar{k})}e^{-4ik^2T} & -\alpha\overline{\Psi_{23}(T,\bar{k})}e^{-4ik^2T} & \overline{\Psi_{33}(T,\bar{k})} & \overline{\Psi_{43}(T,\bar{k})} \vspace{0.1in}\\
-\alpha\overline{\Psi_{14}(T,\bar{k})}e^{-4ik^2T} & -\alpha\overline{\Psi_{24}(T,\bar{k})}e^{-4ik^2T} & \overline{\Psi_{34}(T,\bar{k})} & \overline{\Psi_{44}(T,\bar{k})}
 \end{array}
\!\!\!\right),\,\,
\ene

\bee\label{slm}
S_L(k)\!=\!\left(\!\!\!\begin{array}{cccc}
 \overline{\phi_{11}(T,\bar{k})} & \overline{\phi_{21}(T,\bar{k})} & -\alpha\overline{\phi_{31}(T,\bar{k})}e^{4ik^2T} & -\alpha\overline{\phi_{41}(T,\bar{k})}e^{4ik^2T} \vspace{0.1in}\\
\overline{\phi_{12}(T,\bar{k})} & \overline{\phi_{22}(T,\bar{k})} & -\alpha\overline{\phi_{32}(T,\bar{k})}e^{4ik^2T} & -\alpha\overline{\phi_{42}(T,\bar{k})}e^{4ik^2T} \vspace{0.1in}\\
-\alpha\overline{\phi_{13}(T,\bar{k})}e^{-4ik^2T} & -\alpha\overline{\phi_{23}(T,\bar{k})}e^{-4ik^2T} & \overline{\phi_{33}(T,\bar{k})} & \overline{\phi_{43}(T,\bar{k})} \vspace{0.1in}\\
-\alpha\overline{\phi_{14}(T,\bar{k})}e^{-4ik^2T} & -\alpha\overline{\phi_{24}(T,\bar{k})}e^{-4ik^2T} & \overline{\phi_{34}(T,\bar{k})} & \overline{\phi_{44}(T,\bar{k})}
 \end{array}
\!\!\!\right),\,\,
\ene
and the complex-valued functions $\{\Psi_{ij}(t,k)\}_{i,j=1}^4$ can be expressed using the following system of integral equations
\bee\label{psit10}
\left\{ \begin{array}{rl}
 \Psi_{11}(t,k)=&\!\!\! 1+\displaystyle\int_0^t \left\{-i\alpha[(|u_{01}|^2+|u_{00}|^2)\Psi_{11}+(\beta u_{01}\bar{u}_{00}+u_{00}\bar{u}_{0-1})\Psi_{21}]\right. \vspace{0.08in}\\
 &\displaystyle \qquad+\left. (2ku_{01}+iu_{11})\Psi_{31}+(2ku_{00}+iu_{10})\Psi_{41} \right\}(t',k)dt', \vspace{0.08in}\\
\Psi_{21}(t,k)=&\!\!\! \displaystyle\int_0^t
   \left\{-i\alpha[(\beta u_{00}\bar{u}_{01}+u_{0-1}\bar{u}_{00})\Psi_{11}+(|u_{0-1}|^2+|u_{00}|^2)\Psi_{21}]\right. \vspace{0.08in}\\
    &\qquad+\left. (2k(\beta u_{00}+i\beta u_{10})\Psi_{31}+ (2ku_{0-1}+iu_{1-1})\Psi_{41}\right\}(t',k)dt', \vspace{0.08in}\\
 \Psi_{31}(t,k)=&\!\!\!\displaystyle \int_0^te^{4ik^2(t-t')}\left\{2\alpha k(\bar{u}_{01}\Psi_{11}+\beta \bar{u}_{00}\Psi_{21})
      -i\alpha(\bar{u}_{11}\Psi_{11}+\beta \bar{u}_{10}\Psi_{21})\right. \vspace{0.08in}\\
  & \qquad\left.+i\alpha[(|u_{01}|^2+|u_{00}|^2)\Psi_{31}+(\beta u_{0-1}\bar{u}_{00}+u_{00}\bar{u}_{01})\Psi_{41}]\right\}
  (t',k)dt', \vspace{0.08in}\\
 \Psi_{41}(t,k)=&\!\!\! \displaystyle\int_0^te^{4ik^2(t-t')}\left[
  2\alpha k(\bar{u}_{00}\Psi_{11}+\bar{u}_{0-1}\Psi_{21})
   -i\alpha(\bar{u}_{10}\Psi_{11}+\bar{u}_{1-1}\Psi_{21})\right.\vspace{0.08in}\\
  & \qquad \displaystyle\left.+i\alpha[(\beta u_{00}\bar{u}_{0-1}+u_{01}\bar{u}_{00})\Psi_{31}+(|u_{0-1}|^2+|u_{00}|^2)\Psi_{41}]\right](t',k)dt',
\end{array}\right.
\ene

\bee\label{psit20}
\left\{ \begin{array}{rl}
 \Psi_{12}(t,k)=&\!\!\! \displaystyle\int_0^t
 \left\{-i\alpha[(|u_{01}|^2+|u_{00}|^2)\Psi_{12}+(\beta u_{01}\bar{u}_{00}+u_{00}\bar{u}_{0-1})\Psi_{22}]\right. \vspace{0.08in}\\
 & \qquad\displaystyle\left.+2k(u_{01}\Psi_{32}+u_{00}\Psi_{42})+i(u_{11}\Psi_{32}+u_{10}\Psi_{42})\right\}(t',k)dt',  \vspace{0.08in}\\
\Psi_{22}(t,k)=&\!\!\! \displaystyle 1+\int_0^t\left\{-i\alpha[(\beta u_{00}\bar{u}_{01}+u_{0-1}\bar{u}_{00})\Psi_{12}+(|u_{0-1}|^2+|u_{00}|^2)\Psi_{22}]\right.\vspace{0.08in}\\
& \qquad\displaystyle\left. +2k(\beta u_{00}\Psi_{32}+u_{0-1}\Psi_{42})+i(\beta u_{10}\Psi_{32}+u_{1-1}\Psi_{42})\right\}(t',k)dt', \vspace{0.08in}\\
 \Psi_{32}(t,k)=&\!\!\! \displaystyle\int_0^t e^{4ik^2(t-t')}\left\{2\alpha k(\bar{u}_{01}\Psi_{12}+\beta \bar{u}_{00}\Psi_{22})
      -i\alpha(\bar{u}_{11}\Psi_{12}+\beta \bar{u}_{10}\Psi_{22})\right. \vspace{0.08in}\\
& \qquad\left. \displaystyle+i\alpha[(|u_{01}|^2+|u_{00}|^2)\Psi_{32}+(\beta u_{0-1}\bar{u}_{00}+u_{00}\bar{u}_{01})\Psi_{42}]\right\}(t',k)dt', \vspace{0.08in}\\
 \Psi_{42}(t,k)=&\!\!\! \displaystyle\int_0^t e^{4ik^2(t-t')}\left\{
 2\alpha k(\bar{u}_{00}\Psi_{12}+\bar{u}_{0-1}\Psi_{22})
   -i\alpha(\bar{u}_{10}\Psi_{12}+\bar{u}_{1-1}\Psi_{22})\right.\vspace{0.08in}\\
&\qquad \displaystyle\left.+i\alpha[(\beta u_{00}\bar{u}_{0-1}+u_{01}\bar{u}_{00})\Psi_{32}+(|u_{0-1}|^2+|u_{00}|^2)\Psi_{42}]\right\}(t',k)dt',
\end{array}\right.
\ene

\bee\label{psit30}
\left\{ \begin{array}{rl}
 \Psi_{13}(t,k)=&\!\!\! \displaystyle\int_0^t
 e^{-4ik^2(t-t')}\left\{-i\alpha[(|u_{01}|^2+|u_{00}|^2)\Psi_{13}+(\beta u_{01}\bar{u}_{00}+u_{00}\bar{u}_{0-1})\Psi_{23}]\right., \vspace{0.08in}\\
 &\qquad\displaystyle\left.+2k(u_{01}\Psi_{33}+u_{00}\Psi_{43})+i(u_{11}\Psi_{33}+u_{10}\Psi_{43})\right\}(t',k)dt', \vspace{0.08in}\\
\Psi_{23}(t,k)=&\!\!\! \displaystyle\int_0^t
 e^{-4ik^2(t-t')}\left\{-i\alpha[(\beta u_{00}\bar{u}_{01}+u_{0-1}\bar{u}_{00})\Psi_{13}+(|u_{0-1}|^2+|u_{00}|^2)\Psi_{23}]\right.\vspace{0.08in}\\
&\qquad\displaystyle\left. +2k(\beta u_{00}\Psi_{33}+u_{0-1}\Psi_{43})+i(\beta u_{10}\Psi_{33}+u_{1-1}\Psi_{43})\right\}(t',k)dt', \vspace{0.08in}\\
 \Psi_{33}(t,k)=&\!\!\! 1+\displaystyle\int_0^t
 \left\{2\alpha k(\bar{u}_{01}\Psi_{13}+\beta \bar{u}_{00}\Psi_{23})
      -i\alpha(\bar{u}_{11}\Psi_{13}+\beta \bar{u}_{10}\Psi_{23})\right. \vspace{0.08in}\\
& \qquad\displaystyle\left.+i\alpha[(|u_{01}|^2+|u_{00}|^2)\Psi_{33}+(\beta u_{0-1}\bar{u}_{00}+u_{00}\bar{u}_{01})\Psi_{43}]\right\}(t',k)dt', \vspace{0.08in}\\
 \Psi_{43}(t,k)=&\!\!\! \displaystyle\int_0^t\left\{2\alpha k(\bar{u}_{00}\Psi_{13}+\bar{u}_{0-1}\Psi_{23})
   -i\alpha(\bar{u}_{10}\Psi_{13}+\bar{u}_{1-1}\Psi_{23})\right. \vspace{0.08in}\\
 &\qquad \displaystyle\left.+i\alpha[(\beta u_{00}\bar{u}_{0-1}+u_{01}\bar{u}_{00})\Psi_{33}+(|u_{0-1}|^2+|u_{00}|^2)\Psi_{43}]\right\}(t',k)dt',
\end{array}\right.
\ene
and
\bee\label{psit40}
\left\{ \begin{array}{rl}
 \Psi_{14}(t,k)=&\!\!\! \displaystyle\int_0^t
 e^{-4ik^2(t-t')}\left\{-i\alpha[(|u_{01}|^2+|u_{00}|^2)\Psi_{14}+(\beta u_{01}\bar{u}_{00}+u_{00}\bar{u}_{0-1})\Psi_{24}]\right. \vspace{0.08in}\\
 &\qquad\displaystyle \left.+2k(u_{01}\Psi_{34}+u_{00}\Psi_{44})+i(u_{11}\Psi_{34}+u_{10}\Psi_{44})\right\}(t',k)dt',  \vspace{0.08in}\\
   \Psi_{24}(t,k)=&\!\!\! \displaystyle\int_0^te^{-4ik^2(t-t')}\left\{-i\alpha[(\beta u_{00}\bar{u}_{01}+u_{0-1}\bar{u}_{00})\Psi_{14}+(|u_{0-1}|^2+|u_{00}|^2)\Psi_{24}]\right.\vspace{0.08in}\\
 &\qquad\displaystyle\left.+ 2k(\beta u_{00}\Psi_{34}+u_{0-1}\Psi_{44})+i(\beta u_{10}\Psi_{34}+u_{1-1}\Psi_{44})\right\}(t',k)dt', \vspace{0.08in}\\
 \Psi_{34}(t,k)=&\!\!\! \displaystyle\int_0^t
 \left[2\alpha k(\bar{u}_{01}\Psi_{14}+\beta \bar{u}_{00}\Psi_{24})
      -i\alpha(\bar{u}_{11}\Psi_{14}+\beta \bar{u}_{10}\Psi_{24})\right. \vspace{0.08in}\\
&\qquad\displaystyle \left.+i\alpha[(|u_{01}|^2+|u_{00}|^2)\Psi_{34}+(\beta u_{0-1}\bar{u}_{00}+u_{00}\bar{u}_{01})\Psi_{44}]\right\}(t',k)dt', \vspace{0.08in}\\
 \Psi_{44}(t,k)=&\!\!\! 1+\displaystyle\int_0^t\left\{
 2\alpha k(\bar{u}_{00}\Psi_{14}+\bar{u}_{0-1}\Psi_{24})
   -i\alpha(\bar{u}_{10}\Psi_{14}+\bar{u}_{1-1}\Psi_{24})\right.\vspace{0.08in}\\
&\qquad\displaystyle\left. +i\alpha[(\beta u_{00}\bar{u}_{0-1}+u_{01}\bar{u}_{00})\Psi_{34}+(|u_{0-1}|^2+|u_{00}|^2)\Psi_{44}]\right\}(t',k)dt',
\end{array}\right.
\ene }
{\it The functions $\{\phi_{ij}(t,k)\}_{i,j=1}^4$ are of the same integral equations (\ref{psit10})-(\ref{psit40}) by replacing
the functions $\{u_{0j}(t),\, u_{1j}(t)\}$ with $\{v_{0j}(t),\, v_{1j}(t)\},\, (j=1,0,-1)$, respectively, that is,
\bee\no
\phi_{ij}(t,k)=\Phi_{ij}(t,k)\big|_{\{u_{0l}(t)\to v_{0l}(t),\, u_{1l}(t)\to v_{1l}(t),\, \Psi_{ij}\to \phi_{ij}\}},\, (i,j=1,2,3,4;\, l=1,0,-1).
\ene

(i) For the given Dirichlet boundary data, the unknown Neumann boundary data $\{u_{1j}(t),\, j=1,0,-1\}$ at $x=0$ and $\{v_{1j}(t),\, j=1,0,-1\}$ at $x=L$,\, $0<t<T$ can be found as
\bes\bee\label{u11}
&\begin{array}{rl}
 u_{11}(t)=&\!\!\!\!\d\frac{1}{\pi}\int_{\partial D_3^0}\!\!
\left[\frac{2\Sigma_+(k)}{i\Sigma_-(k)}(k\Psi_{13-}\!+iu_{01})
-(u_{01}\bar{\phi}_{33-}\!+u_{00}\bar{\phi}_{34-})
  +\frac{4i}{\Sigma_-(k)}(\alpha k\bar{\phi}_{31-}\!+i v_{01})\right]dk \vspace{0.08in}\\
&+\d\int_{\partial D_3^0}\frac{4k}{i\pi\Sigma_-(k)}
\Big\{[\Psi_{13}(\bar{\phi}_{33}-1)\!+\!\Psi_{14}\bar{\phi}_{34}]e^{-2ikL}\!-\!\alpha \l[(\Psi_{11}-1)\bar{\phi}_{31}+\Psi_{12}\bar{\phi}_{32}\r]\Big\}_-dk \qquad \vspace{0.08in}\\
&+\d\frac{2}{\pi}\int_{\partial D_3^0}(u_{01}\Psi_{33-}+u_{00}\Psi_{43-})dk,
\end{array} \\
\label{u10}
&\begin{array}{rl}
 u_{10}(t)=&\!\!\!\!\d\frac{1}{\pi}\int_{\partial D_3^0}\!\!
\left[\frac{2\Sigma_+(k)}{i\Sigma_-(k)}(k\Psi_{14-}\!+iu_{00})
-(u_{01}\bar{\phi}_{43-}\!+u_{00}\bar{\phi}_{44-})
  +\frac{4i}{\Sigma_-(k)}(\alpha k\bar{\phi}_{41-}\!+i v_{00})\right]dk \vspace{0.08in}\\
&+\d\int_{\partial D_3^0}\frac{4k}{i\pi\Sigma_-(k)}
\Big\{[\Psi_{13}\bar{\phi}_{43}\!+\!\Psi_{14}(\bar{\phi}_{44}-1)]e^{-2ikL}\!-\!\alpha [(\Psi_{11}-1)\bar{\phi}_{41}+\Psi_{12}\bar{\phi}_{42}]\Big\}_-dk  \qquad \vspace{0.08in}\\
&+\d\frac{2}{\pi}\int_{\partial D_3^0}(u_{01}\Psi_{34-}+u_{00}\Psi_{44-})dk,
\end{array} \\
\label{u1-1}
&\begin{array}{rl}
 u_{1-1}(t)=&\!\!\!\!\d\frac{1}{\pi}\int_{\partial D_3^0}\left[\frac{2\Sigma_+(k)}{i\Sigma_-(k)}(k\Psi_{24-}\!+iu_{0-1})
-(u_{00}\bar{\phi}_{43-}\!+u_{0-1}\bar{\phi}_{44-})
  +\frac{4i}{\Sigma_-(k)}(\alpha k\bar{\phi}_{42-}\!+i v_{0-1})\right]dk \vspace{0.08in}\\
&+\d\int_{\partial D_3^0}\frac{4k}{i\pi\Sigma_-(k)}
\Big\{[\Psi_{23}\bar{\phi}_{43}\!+\!\Psi_{24}(\bar{\phi}_{44}-1)]e^{-2ikL}\!-\!\alpha [\Psi_{21}\bar{\phi}_{41}+(\Psi_{22}-1)\bar{\phi}_{42}]\Big\}_-dk  \qquad \vspace{0.08in}\\
&+\d\frac{2}{\pi}\int_{\partial D_3^0}(\beta u_{00}\Psi_{34-}+u_{0-1}\Psi_{44-})dk,
\end{array}
\ene\ees
and
\bes\bee\label{v11}
&\begin{array}{rl}
 v_{11}(t)=&\!\!\!\!\d\frac{1}{\pi}\int_{\partial D_3^0}\left[\frac{2i\Sigma_+(k)}{\Sigma_-(k)}(k\phi_{13-}\!+iv_{01})
   -(v_{01}\bar{\Psi}_{33-}+v_{00}\bar{\Psi}_{34-})
    -\frac{4i}{\Sigma_-(k)}(\alpha k\bar{\Psi}_{31-}+i u_{01})\right]dk \vspace{0.08in}\\
&+\d\int_{\partial D_3^0}\frac{4k}{i\pi\Sigma_-(k)}
\Big\{\alpha [(\phi_{11}-1)\bar{\Psi}_{31}+\phi_{12}\bar{\Psi}_{32}]
-[\phi_{13}(\bar{\Psi}_{33}-1)+\phi_{14}\bar{\Psi}_{34}]e^{2ikL}\Big\}_-dk \qquad \vspace{0.08in}\\
&+\d\frac{2}{\pi}\int_{\partial D_3^0}(v_{01}\phi_{33-}+v_{00}\phi_{43-})dk,
\end{array} \vspace{0.1in}\\
&\label{v10}
\begin{array}{rl}
 v_{10}(t)=&\!\!\!\!\d\frac{1}{\pi}\int_{\partial D_3^0}\!\!
\left[\frac{2i\Sigma_+(k)}{\Sigma_-(k)}(k\phi_{14-}\!+iv_{00})-(v_{01}\bar{\Psi}_{43-}+v_{00}\bar{\Psi}_{44-})
  -\frac{4i}{\Sigma_-(k)}(\alpha k\bar{\Psi}_{41-}+i u_{00})\right]dk \vspace{0.08in}\\
&+\d\int_{\partial D_3^0}\frac{4k}{i\pi\Sigma_-(k)}
\Big\{\alpha [(\phi_{11}-1)\bar{\Psi}_{41}+\phi_{12}\bar{\Psi}_{42}]
-[\phi_{13}\bar{\Psi}_{43}+\phi_{14}(\bar{\Psi}_{44}-1)]e^{2ikL}\Big\}_-dk \qquad \vspace{0.08in}\\
&+\d\frac{2}{\pi}\int_{\partial D_3^0}(v_{01}\phi_{34-}+v_{00}\phi_{44-})dk,
\end{array} \vspace{0.08in}\\
&\label{v1-1}
\begin{array}{rl}
 v_{1-1}(t)=&\!\!\!\!\d\frac{1}{\pi}\int_{\partial D_3^0}\!\!
\left[\frac{2i\Sigma_+(k)}{\Sigma_-(k)}(k\phi_{24-}\!+iv_{0-1})-(v_{00}\bar{\Psi}_{43-}+v_{0-1}\bar{\Psi}_{44-})
  -\frac{4i}{\Sigma_-(k)}(\alpha k\bar{\Psi}_{42-}+i u_{0-1})\right]dk \vspace{0.08in}\\
&+\d\int_{\partial D_3^0}\frac{4k}{i\pi\Sigma_-(k)}
\Big\{\alpha [\phi_{21}\bar{\Psi}_{41}+(\phi_{22}-1)\bar{\Psi}_{42}]
-[\phi_{23}\bar{\Psi}_{43}+\phi_{24}(\bar{\Psi}_{44}-1)]e^{2ikL}\Big\}_-dk \qquad \vspace{0.08in}\\
&+\d\frac{2}{\pi}\int_{\partial D_3^0}(\beta v_{00}\phi_{34-}+v_{0-1}\phi_{44-})dk,
\end{array}
\ene\ees

(ii) For the given Neumann  boundary data, the unknown Dirichlet boundary data $\{u_{0j}(t),\, j=1,0,-1\}$ at $x=0$ and $\{v_{0j}(t),\, j=1,0,-1\}$ at $x=L$, $0<t<T$ can be obtained as
\bes\bee \label{u01}
&\begin{array}{rl}
u_{01}(t)=&\!\!\!\! \d \int_{\partial D_3^0}\frac{\Sigma_+(k)\Psi_{13+}-2\alpha\bar{\phi}_{31+}}{\pi\Sigma_-(k)}dk
 \vspace{0.08in}\\
& +\d\int_{\partial D_3^0}\frac{2}{\pi\Sigma_-(k)}\Big\{\left[\Psi_{13}(\bar{\phi}_{33}-1)+\Psi_{14}\bar{\phi}_{34}\right]e^{-2ikL}
\!-\!\alpha [(\Psi_{11}-1)\bar{\phi}_{31}+\Psi_{12}\bar{\phi}_{32}]\Big\}_+dk,\quad
\end{array}\vspace{0.08in}\\
& \label{u00} \begin{array}{rl}
u_{00}(t)=&\!\!\!\! \d \int_{\partial D_3^0}\frac{\Sigma_+(k)\Psi_{14+}-2\alpha\bar{\phi}_{41+}}{\pi\Sigma_-(k)}dk \vspace{0.08in}\\
&+\d\int_{\partial D_3^0}\frac{2}{\pi\Sigma_-(k)}\Big\{\left[\Psi_{13}\bar{\phi}_{43}+\Psi_{14}(\bar{\phi}_{44}-1)\right]e^{-2ikL}
\!-\!\alpha [(\Psi_{11}-1)\bar{\phi}_{41}+\Psi_{12}\bar{\phi}_{42}]\Big\}_+dk,\quad
\end{array}\vspace{0.08in}\\
& \label{u0-1} \begin{array}{rl}
u_{0-1}(t)=&\!\!\!\! \d \int_{\partial D_3^0}\frac{\Sigma_+(k)\Psi_{24+}-2\alpha\bar{\phi}_{42+}}{\pi\Sigma_-(k)}dk \vspace{0.08in}\\
&+\d\int_{\partial D_3^0}\frac{2}{\pi\Sigma_-(k)}\Big\{\left[\Psi_{23}\bar{\phi}_{43}+\Psi_{24}(\bar{\phi}_{44}-1)\right]e^{-2ikL}
\!-\!\alpha [\Psi_{21}\bar{\phi}_{41}+(\Psi_{22}-1)\bar{\phi}_{42}]\Big\}_+dk,\quad
\end{array}
\ene\ees
and
\bes\bee \label{v01}
&\begin{array}{rl}
v_{01}(t)=&\!\!\!\!
 \d \int_{\partial D_3^0}\frac{-\Sigma_+(k)\phi_{13+}+2\alpha\bar{\Psi}_{31+}}{\pi\Sigma_-(k)}dk \vspace{0.08in}\\
&+\d\int_{\partial D_3^0}\frac{2}{\pi\Sigma_-(k)}\Big\{\alpha [(\phi_{11}-1)\bar{\Psi}_{31}+\phi_{12}\bar{\Psi}_{32}]\!-\!\left[\phi_{13}(\bar{\Psi}_{33}-1)
+\phi_{14}\bar{\Psi}_{34}\right]e^{2ikL}\Big\}_+dk,\quad
\end{array} \vspace{0.08in}\\
& \label{v00} \begin{array}{rl}
v_{00}(t)=&\!\!\!\!  \d \int_{\partial D_3^0}\frac{-\Sigma_+(k)\phi_{14+}+2\alpha\bar{\Psi}_{41+}}{\pi\Sigma_-(k)}dk \vspace{0.08in}\\
&+\d\int_{\partial D_3^0}\frac{2}{\pi\Sigma_-(k)}\Big\{\alpha [(\phi_{11}-1)\bar{\Psi}_{41}+\phi_{12}\bar{\Psi}_{42}]\!-\!\left[\phi_{13}\bar{\Psi}_{43}
+\phi_{14}(\bar{\Psi}_{44}-1)\right]e^{2ikL}\Big\}_+dk,\quad
\end{array}\vspace{0.08in}\\
& \label{v0-1} \begin{array}{rl}
v_{0-1}(t)=&\!\!\!\!  \d \int_{\partial D_3^0}\frac{-\Sigma_+(k)\phi_{24+}+2\alpha\bar{\Psi}_{42+}}{\pi\Sigma_-(k)}dk \vspace{0.08in}\\
&+\d\int_{\partial D_3^0}\frac{2}{\pi\Sigma_-(k)}\Big\{\alpha [\phi_{21}\bar{\Psi}_{41}+(\phi_{22}-1)\bar{\Psi}_{42}]\!-\!\left[\phi_{23}\bar{\Psi}_{43}
+\phi_{24}(\bar{\Psi}_{44}-1)\right]e^{2ikL}\Big\}_+dk,\quad
\end{array}
\ene\ees}
where $\Psi_{13}=\Psi_{13}(t,k),\, \bar{\phi}_{33}=\overline{\phi_{33}(t, \bar{k})}=\bar{\phi}_{33}(t, \bar{k})$, and other functions have the similar expressions.

\vspace{0.1in}

\noindent {\bf Proof.} We can  prove  Eqs.~(\ref{skm}) and (\ref{slm}) by means of Eqs.~(\ref{sss}) and (\ref{sl}) with replacing $T$ by $t$, that is,  $S(k)=e^{-2ik^2t\hat{\sigma}_4}\mu_2^{-1}(0,t,k)$ and $S_L(k)=e^{-2ik^2t\hat{\sigma}_4}\mu_3^{-1}(L,t,k)$ and the symmetry relation (\ref{symmetry}).
Moreover, Eqs.~(\ref{psit10})-(\ref{psit40}) for $\Psi_{ij}(t,k),\, i,j=1,2,3,4$ can be given in terms of the Volteral integral equations of $\mu_2(0,t,k)$. Similarly, the expressions of $\phi_{ij}(t,k),\, i,j=1,2,3,4$ can be obtained via the Volteral integral equations of $\mu_3(L,t,k)$.

\vspace{0.1in}
In the following we will certify Eqs.~(\ref{u11})-(\ref{v0-1}).

(i) Applying the Cauchy's theorem to Eq.~(\ref{mu2asy}), we have
\bee\begin{array}{l}
-\dfrac{i\pi }{2}\Psi_{33}^{(1)}(t)=\displaystyle\int_{\partial D_2}[\Psi_{33}(t,k)-1]dk=\int_{\partial D_4}[\Psi_{33}(t,k)-1]dk, \vspace{0.08in}\\
-\dfrac{i\pi }{2}\Psi_{43}^{(1)}(t)=\displaystyle\int_{\partial D_2}\Psi_{43}(t,k)dk=\int_{\partial D_4}\Psi_{43}(t,k)dk, \vspace{0.08in}\\
-\dfrac{i\pi }{2}\Psi_{13}^{(2)}(t)=\displaystyle\int_{\partial D_2}\left[k\Psi_{13}(t,k)+\frac{i}{2}u_{01}(t)\right]dk=\int_{\partial D_4}\left[k\Psi_{13}(t,k)+\frac{i}{2}u_{01}(t)\right]dk,
\end{array}\ene

We further find
\bes\bee
&\label{psi33}
\begin{array}{rl}
i\pi \Psi_{33}^{(1)}(t)=& \d-\left(\int_{\partial D_2}+\int_{\partial D_4}\right)[\Psi_{33}(t,k)-1]dk = \left(\int_{\partial D_1}+\int_{\partial D_3}\right)[\Psi_{33}(t,k)-1]dk \vspace{0.08in}\\
=&  \d\int_{\partial D_3}[\Psi_{33}(t,k)-1]dk-\int_{\partial D_3}[\Psi_{33}(t,-k)-1]dk =\int_{\partial D_3}\Psi_{33-}(t,k)dk,
\end{array} \\
&\label{psi43}\begin{array}{rl}
i\pi \Psi_{43}^{(1)}(t)=&  \d -\left(\int_{\partial D_2}+\int_{\partial D_4}\right)\Psi_{43}(t,k)dk =\d \int_{\partial D_3}\Psi_{43-}(t,k)dk,
\end{array} \qquad\quad\quad
\ene\ees
and
\bee\label{psi13}
\begin{array}{rl}
i\pi \Psi_{13}^{(2)}(t)=&\!\!\! \d \left(\int_{\partial D_1}+\int_{\partial D_3}\right)\left[k\Psi_{13}(t,k)+\frac{i}{2}u_{01}(t)\right]dk\vspace{0.08in}\\
=&\!\!\! \d
\int_{\partial D_3}\left[k\Psi_{13}(t,k)+\frac{i}{2}u_{01}(t)\right]_-dk \vspace{0.08in}\\
=&\!\!\! \d\int_{\partial D_3^0}\left\{k\Psi_{13}(t,k)+\frac{i}{2}u_{01}(t)+\frac{2e^{-2ikL}}{\Sigma_-(k)}\left[k\Psi_{13}(t,k)+\frac{i}{2}u_{01}(t)\right]\right\}_-dk+C_1(t) \vspace{0.08in}\\
=&\!\!\! \d\int_{\partial D_3^0}\frac{\Sigma_+(k)}{\Sigma_-(k)}(k\Psi_{13-}+iu_{01})dk+C_1(t),
\end{array}\ene
where we have introduced the function $C_1(t)$ as
\bee\no
C_1(t)=-\d\int_{\partial D_3^0}\left\{\frac{2e^{-2ikL}}{\Sigma_-}\left[k\Psi_{13}(t,k)+\frac{i}{2}u_{01}(t)\right]\right\}_-dk,
\ene

We use the global relation (\ref{ca}) to further reduce $C_1(t)$ in the form
\bee \label{it} \begin{array}{rl}
C_1(t)=&\!\!\!-\d\int_{\partial D_3^0}\left\{\frac{2e^{-2ikL}}{\Sigma_-}\left[k\Psi_{13}(t,k)+\frac{i}{2}u_{01}(t)\right]\right\}_-dk \vspace{0.08in}\\
=&\!\!\! \d\int_{\partial D_3^0}\left\{\frac{2e^{-2ikL}}{\Sigma_-}\left[\Psi_{13}^{(1)}-kc_{13}(t,k)+
\frac{\Psi_{13}^{(1)}\bar{\phi}_{33}^{(1)}+\Psi_{14}^{(1)}\bar{\phi}_{34}^{(1)}}{k}-\alpha \bar{\phi}_{31}^{(1)}e^{2ikL}\right]\right\}_-dk \vspace{0.08in}\\
&\!\!\!-\d\int_{\partial D_3^0}\left\{\frac{2e^{-2ikL}}{\Sigma_-}\left[
\frac{\Psi_{13}^{(1)}\bar{\phi}_{33}^{(1)}+\Psi_{14}^{(1)}\bar{\phi}_{34}^{(1)}}{k}
 +\alpha(k\bar{\phi}_{31}-\bar{\phi}_{31}^{(1)})e^{2ikL}\right]\right\}_-dk \vspace{0.08in}\\
&\!\!\! +\d\int_{\partial D_3^0}\left\{\frac{2ke^{-2ikL}}{\Sigma_-}\left[
\Psi_{13}(\bar{\phi}_{33}-1)+\Psi_{14}\bar{\phi}_{34}-\alpha ((\Psi_{11}-1)\bar{\phi}_{31}+\Psi_{12}\bar{\phi}_{32})e^{2ikL}\right]\right\}_-dk,
\end{array}
\ene

By applying the Cauchy's theorem and asymptotic (\ref{c13}) to Eq.~(\ref{it}), we find that the terms on the right-hand side of
Eq.~(\ref{it}) are of the form
\bee\label{it2}
\begin{array}{rl}
C_1(t)=&-i\pi\Psi_{13}^{(2)}-\d\int_{\partial D_3^0}\left[\frac{i}{2}(u_{01}\bar{\phi}_{33-}+u_{00}\bar{\phi}_{34-})
  +\frac{2\alpha}{\Sigma_-(k)}(k\bar{\phi}_{31-}+i\alpha v_{01})\right]dk \vspace{0.08in}\\
&+\d\int_{\partial D_3^0}\frac{2k}{\Sigma_-}
\left\{[\Psi_{13}(\bar{\phi}_{33}-1)\!+\!\Psi_{14}\bar{\phi}_{34}]e^{-2ikL}\!-\!\alpha [(\Psi_{11}-1)\bar{\phi}_{31}+\Psi_{12}\bar{\phi}_{32}]\right\}_-dk,
\end{array}
\ene

It follows from Eqs.~(\ref{psi13}) and (\ref{it2}) that we have
\bee\label{psi13g}
\begin{array}{rl}
2i\pi \Psi_{13}^{(2)}(t)=&\!\!\!\!\! \d\int_{\partial D_3^0}\!\!
\left[\frac{\Sigma_+}{\Sigma_-}(k\Psi_{13-}\!+iu_{01})
-\frac{i}{2}(u_{01}\bar{\phi}_{33-}\!+u_{00}\bar{\phi}_{34-})
  -\frac{2}{\Sigma_-}(\alpha k\bar{\phi}_{31-}\!+i v_{01})\right]dk \vspace{0.08in}\\
&\!\!\!\!+\d\int_{\partial D_3^0}\frac{2k}{\Sigma_-}
\left\{[\Psi_{13}(\bar{\phi}_{33}-1)\!+\!\Psi_{14}\bar{\phi}_{34}]e^{-2ikL}\!-\!\alpha [(\Psi_{11}-1)\bar{\phi}_{31}+\Psi_{12}\bar{\phi}_{32}]\right\}_-dk,
\end{array}\ene
Thus substituting Eqs.~(\ref{psi33}), (\ref{psi43}) and (\ref{psi13g}) into the third one of system (\ref{ud}), we can get Eq.~(\ref{u11}).

Applying the Cauchy's theorem to Eq.~(\ref{mu2asy}), we have
\bee\label{psi14}
\begin{array}{rl}
i\pi \Psi_{14}^{(2)}(t)=&\!\!\! \d \left(\int_{\partial D_1}+\int_{\partial D_3}\right)\left[k\Psi_{14}(t,k)+\frac{i}{2}u_{00}(t)\right]dk\vspace{0.08in}\\
=&\!\!\! \d\int_{\partial D_3}\left[k\Psi_{14}(t,k)+\frac{i}{2}u_{00}(t)\right]_-dk \vspace{0.08in}\\
=&\!\!\! \d\int_{\partial D_3^0}\left\{k\Psi_{14}(t,k)+\frac{i}{2}u_{00}(t)
+\frac{2e^{-2ikL}}{\Sigma_-(k)}\left[k\Psi_{14}(t,k)+\frac{i}{2}u_{00}(t)\right]\right\}_-dk+C_2(t) \vspace{0.08in}\\
=&\!\!\! \d\int_{\partial D_3^0}\frac{\Sigma_+(k)}{\Sigma_-(k)}(k\Psi_{14-}+iu_{00})dk+C_2(t),
\end{array}\ene
where we have introduced the function $C_2(t)$ as
\bee\no
C_2(t)=-\d\int_{\partial D_3^0}\left\{\frac{2e^{-2ikL}}{\Sigma_-}\left[k\Psi_{14}(t,k)+\frac{i}{2}u_{00}(t)\right]\right\}_-dk,
\ene

We apply the Cauchy's theorem, asymptotic (\ref{c14}), and the global relation (\ref{cb}) to $C_2(t)$ to have
\bee \label{it3} \begin{array}{rl}
C_2(t)=&\!\!\! -\d\int_{\partial D_3^0}\left\{\frac{2e^{-2ikL}}{\Sigma_-}\left[k\Psi_{14}(t,k)+\frac{i}{2}u_{00}(t)\right]\right\}_-dk \vspace{0.08in}\\
=&\!\!\!\d\int_{\partial D_3^0}\left\{\frac{2e^{-2ikL}}{\Sigma_-}\left[\Psi_{14}^{(1)}-kc_{14}(t,k)+
\frac{\Psi_{13}^{(1)}\bar{\phi}_{43}^{(1)}+\Psi_{14}^{(1)}\bar{\phi}_{44}^{(1)}}{k}-\alpha \bar{\phi}_{41}^{(1)}e^{2ikL}\right]\right\}_-dk \vspace{0.08in}\\
&\!\!\! -\d\int_{\partial D_3^0}\left\{\frac{2e^{-2ikL}}{\Sigma_-}\left[
\frac{\Psi_{13}^{(1)}\bar{\phi}_{43}^{(1)}+\Psi_{14}^{(1)}\bar{\phi}_{44}^{(1)}}{k}
 +\alpha(k\bar{\phi}_{41}-\bar{\phi}_{41}^{(1)})e^{2ikL}\right]\right\}_-dk \vspace{0.08in}\\
&\!\!\!+\d\int_{\partial D_3^0}\left\{\frac{2ke^{-2ikL}}{\Sigma_-}\left[
\Psi_{13}\bar{\phi}_{43}+\Psi_{14}(\bar{\phi}_{44}-1)-\alpha ((\Psi_{11}-1)\bar{\phi}_{41}+\Psi_{12}\bar{\phi}_{42})e^{2ikL}\right]\right\}_-dk\vspace{0.08in}\\
=&\!\!\! -i\pi\Psi_{14}^{(2)}-\d\int_{\partial D_3^0}\left[\frac{i}{2}(u_{01}\bar{\phi}_{43-}+u_{00}\bar{\phi}_{44-})
  +\frac{2}{\Sigma_-(k)}(\alpha k\bar{\phi}_{41-}+i v_{00})\right]dk \vspace{0.08in}\\
&+\d\int_{\partial D_3^0}\frac{2k}{\Sigma_-}
\left\{[\Psi_{13}\bar{\phi}_{43}\!+\!\Psi_{14}(\bar{\phi}_{44}-1)]e^{-2ikL}\!-\!\alpha [(\Psi_{11}-1)\bar{\phi}_{41}+\Psi_{12}\bar{\phi}_{42}]\right\}_-dk,
\end{array}
\ene

It follows from Eqs.~(\ref{psi14}) and (\ref{it3}) that we have
\bee\label{psi14g}
\begin{array}{rl}
2i\pi \Psi_{14}^{(2)}(t)=&\!\!\!\!\! \d\int_{\partial D_3^0}\!\!
\left[\frac{\Sigma_+}{\Sigma_-}(k\Psi_{14-}\!+iu_{00})
-\frac{i}{2}(u_{01}\bar{\phi}_{43-}\!+u_{00}\bar{\phi}_{44-})
  -\frac{2}{\Sigma_-}(\alpha k\bar{\phi}_{41-}\!+i v_{00})\right]dk \vspace{0.08in}\\
&\!\!\!\!+\d\int_{\partial D_3^0}\frac{2k}{\Sigma_-}
\left\{[\Psi_{13}\bar{\phi}_{43}\!+\!\Psi_{14}(\bar{\phi}_{44}-1)]e^{-2ikL}\!-\!\alpha [(\Psi_{11}-1)\bar{\phi}_{41}+\Psi_{12}\bar{\phi}_{42}]\right\}_-dk,
\end{array}\ene

We further find
\bee
\label{psi34}i\pi \Psi_{j4}^{(1)}(t)=\d \int_{\partial D_3}\Psi_{j4-}(t,k)dk,\quad j=3,4,
\ene
from Eq.~(\ref{mu2asy}). Thus substituting Eqs.~(\ref{psi14g}) and (\ref{psi34}) into the fourth one of system (\ref{ud}), we can deduce Eq.~(\ref{u10}). Similarly, we can also show Eq.~(\ref{u1-1}).

To use Eq.~(\ref{vd}) to show Eq.~(\ref{v11}) for $v_{11}(t)$ we need to find these functions $\phi_{33}^{(1)}(t,k),\, \phi_{43}^{(1)}(t,k)$, and $\phi_{13}^{(2)}(t,k)$.  Applying the Cauchy's theorem to Eq.~(\ref{mu3asy}), we have
\bee\label{phi13}
\begin{array}{rl}
i\pi \phi_{13}^{(2)}(t)=&\!\!\! \d\int_{\partial D_3}\left[k\phi_{13}(t,k)-\phi_{13}^{(1)}\right]_-dk \vspace{0.08in}\\
=&\!\!\! \d\int_{\partial D_3^0}\left[k\phi_{13}(t,k)-\phi_{13}^{(1)}-\frac{2e^{2ikL}}{\Sigma_-(k)}\l(k\phi_{13}-\phi_{13}^{(1)}\r)\right]_-dk+C_3(t) \vspace{0.08in}\\
=&\!\!\! \d\int_{\partial D_3^0}-\frac{\Sigma_+(k)}{\Sigma_-(k)}\left[k\phi_{13-}-\phi_{13}^{(1)}\right]dk+C_3(t),
\end{array}\ene
where the function $C_3(t)$ is defined as
\bee\no
C_3(t)=\d\int_{\partial D_3^0}\left\{\frac{2e^{2ikL}}{\Sigma_-}\left[k\phi_{13}(t,k)-\phi_{13}^{(1)}\right]\right\}_-dk,
\ene

We need to further reduce $C_3(t)$ by using the asymptotic (\ref{co31}), the global relation (\ref{cag1}) and the Cauchy's theorem such that we find that $C_3(t)$ can be reduced to
\bee \label{jt} \begin{array}{rl}
C_3(t)=&\!\!\!\d\int_{\partial D_3^0}\left\{-\frac{2}{\Sigma_-}\left[\alpha k\bar{c}_{31}(t,\bar{k})+\phi_{13}^{(1)}e^{2ikL}+
\frac{\phi_{13}^{(1)}\bar{\Psi}_{33}^{(1)}+\phi_{14}^{(1)}\bar{\Psi}_{34}^{(1)}}{k}e^{2ikL}-\alpha \bar{\Psi}_{31}^{(1)}\right]\right\}_-dk \vspace{0.08in}\\
&\!\!\!+\d\int_{\partial D_3^0}\left\{\frac{2}{\Sigma_-}\left[\frac{\phi_{13}^{(1)}\bar{\Psi}_{33}^{(1)}+\phi_{14}^{(1)}\bar{\Psi}_{34}^{(1)}}{k}e^{2ikL}
+\alpha(k\bar{\Psi}_{31}-\bar{\Psi}_{31}^{(1)})\right]\right\}_-dk \vspace{0.08in}\\
&\!\!\!+\d\int_{\partial D_3^0}\left\{\frac{2k}{\Sigma_-}\left[\alpha ((\phi_{11}-1)\bar{\Psi}_{31}+\phi_{12}\bar{\Psi}_{32})
-[\phi_{13}(\bar{\Psi}_{33}-1)+\phi_{14}\bar{\Psi}_{34}]e^{2ikL}\right]\right\}_-dk\vspace{0.08in}\\
=&\!\!\! -i\pi\phi_{13}^{(1)}+\d\int_{\partial D_3^0}\left[\frac{1}{2i}(v_{01}\bar{\Psi}_{33-}+v_{00}\bar{\Psi}_{34-})
+\frac{2}{\Sigma_-}(\alpha k\bar{\Psi}_{31-}+i u_{01})\right]dk \vspace{0.08in}\\
&\!\!\!+\d\int_{\partial D_3^0}\frac{2k}{\Sigma_-}\left\{\alpha [(\phi_{11}-1)\bar{\Psi}_{31}+\phi_{12}\bar{\Psi}_{32}]
-[\phi_{13}(\bar{\Psi}_{33}-1)+\phi_{14}\bar{\Psi}_{34}]e^{2ikL}\right\}_-dk,\\
\end{array}
\ene

Eqs.~(\ref{phi13}) and (\ref{jt}) imply that
\bee\label{phi13g}
\begin{array}{rl}
2i\pi \phi_{13}^{(2)}(t)=&\!\!\!\!\!\! \d\int_{\partial D_3^0}\left[\frac{\Sigma_+}{\Sigma_-}(\phi_{13}^{(1)}-k\phi_{13-})
\!+\!\frac{1}{2i}(v_{01}\bar{\Psi}_{33-}\!+\!v_{00}\bar{\Psi}_{34-})\!+\!\frac{2}{\Sigma_-}(\alpha k\bar{\Psi}_{31-}\!+\!i u_{01})\right]dk \vspace{0.08in}\\
&\!\!\!\!\!\!\!+\d\int_{\partial D_3^0}\frac{2k}{\Sigma_-}\left\{\alpha [(\phi_{11}\!-\!1)\bar{\Psi}_{31}+\phi_{12}\bar{\Psi}_{32}]
\!-\![\phi_{13}(\bar{\Psi}_{33}-1)\!+\!\phi_{14}\bar{\Psi}_{34}]e^{2ikL}\right\}_-dk,\\
\end{array}\ene

We also have
\bee
\label{phi33}
 i\pi \phi_{j3}^{(1)}(t)=\d \int_{\partial D_3}\phi_{j3-}(t,k)dk,\quad j=3,4,
\ene
from Eq.~(\ref{mu3asy}).

Thus substituting Eqs.~(\ref{phi13g}) and (\ref{phi33}) into the third one of system (\ref{vd}) yields Eq.~(\ref{v11}). Similarly, we can also show  Eqs.~(\ref{v10}) and (\ref{v1-1}).

(ii) We now deduce the Dirichlet boundary value problems given by Eqs.~(\ref{u01})-(\ref{u0-1}) at $x=0$ from the known Neumann boundary value problems. It follows from the first one of Eq.~(\ref{ud}) that $u_{01}(t)$ can be expressed by means of $\Psi_{13}^{(1)}$.

Applying the Cauchy's theorem to Eq.~(\ref{mu2asy}) yields
\bee\label{psi13g2}
\begin{array}{rl}
i\pi \Psi_{13}^{(1)}(t)=&\!\!\! \d \left(\int_{\partial D_1}+\int_{\partial D_3}\right)\Psi_{13}(t,k)dk =\int_{\partial D_3}\Psi_{13-}(t,k)dk \vspace{0.08in}\\
=&\!\!\! \d\int_{\partial D_3^0}\left[\Psi_{13-}(t,k)+\frac{2}{\Sigma_-(k)}(e^{-2ikL}\Psi_{13})_+\right]dk+C_4(t) \vspace{0.08in}\\
=&\!\!\! \d \int_{\partial D_3^0}\frac{\Sigma_+(k)}{\Sigma_-(k)}\Psi_{13+}dk+C_4(t),
\end{array}\ene
where we have introduced the function $C_4(t)$ as
\bee\label{k1}
C_4(t)=-\d\int_{\partial D_3^0}\frac{2}{\Sigma_-(k)}(e^{-2ikL}\Psi_{13})_+dk,
\ene

By applying the global relation (\ref{ca}), the Cauchy's theorem and asymptotics (\ref{c13}) to Eq.~(\ref{k1}), we find
\bee \label{k1g} \begin{array}{rl}
C_4(t)=&-\d\int_{\partial D_3^0}\frac{2}{\Sigma_-}(e^{-2ikL}\Psi_{13})_+dk \vspace{0.08in}\\
=&\d\int_{\partial D_3^0}\frac{2}{\Sigma_-}\left[-c_{13}(t,k)e^{-2ikL}-\alpha\bar{\phi}_{31}\right]_+dk \vspace{0.08in}\\
&+\d\int_{\partial D_3^0}\frac{2}{\Sigma_-}\Big\{\left[\Psi_{13}(\bar{\phi}_{33}-1)+\Psi_{14}\bar{\phi}_{34}\right]e^{-2ikL}-\alpha [(\Psi_{11}-1)\bar{\phi}_{31}+\Psi_{12}\bar{\phi}_{32}]\Big\}_+dk \vspace{0.08in}\\
=& -i\pi\Psi_{13}^{(1)}-\d\int_{\partial D_3^0}\frac{2\alpha}{\Sigma_-}\bar{\phi}_{31+}dk \vspace{0.08in}\\
&+\d\int_{\partial D_3^0}\frac{2}{\Sigma_-}\Big\{-\alpha [(\Psi_{11}-1)\bar{\phi}_{31}+\Psi_{12}\bar{\phi}_{32}]+\left[\Psi_{13}(\bar{\phi}_{33}-1)
+\Psi_{14}\bar{\phi}_{34}\right]e^{-2ikL}\Big\}_+dk,
\end{array}
\ene

Eqs.~(\ref{psi13g2}) and (\ref{k1g}) imply that
\bee \label{psi13g2g}
\begin{array}{rl}
2i\pi \Psi_{13}^{(1)}(t)=&\!\!\!\!\!\! \d \int_{\partial D_3^0}\left[\frac{\Sigma_+(k)}{\Sigma_-(k)}\Psi_{13+}-\frac{2\alpha}{\Sigma_-}\bar{\phi}_{31+}\right]dk \vspace{0.08in}\\
&\!\!\!\!\!\!+\d\int_{\partial D_3^0}\frac{2}{\Sigma_-}\Big\{-\alpha [(\Psi_{11}\!-\!1)\bar{\phi}_{31}\!+\!\Psi_{12}\bar{\phi}_{32}]\!+\!\left[\Psi_{13}(\bar{\phi}_{33}-1)
+\Psi_{14}\bar{\phi}_{34}\right]e^{-2ikL}\Big\}_+dk,
\end{array}
\ene
Thus, substituting Eq.~(\ref{psi13g2g}) into the first one of Eq.~(\ref{ud}) yields Eq.~(\ref{u01}). Similarly, by applying the expressions of $\Psi_{j4}^{(1)}(t),\, j=1,2$ to the second and third ones of Eq.~(\ref{ud}), we can obtain Eqs.~(\ref{u00}) and (\ref{u0-1}).

We now derive the Dirichlet boundary value problems (\ref{v01})-(\ref{v0-1}) at $x=L$ from the known Neumann boundary value problems. It follows from the first one of Eq.~(\ref{vd}) that $v_{01}(t)$ can be expressed by means of $\phi_{13}^{(1)}$.
Applying the Cauchy's theorem to Eq.~(\ref{mu3asy}) yields
\bee\label{phi13g2}
\begin{array}{rl}
i\pi \phi_{13}^{(1)}(t)=&\!\!\! \d \left(\int_{\partial D_1}+\int_{\partial D_3}\right)\phi_{13}(t,k)dk =\int_{\partial D_3}\phi_{13-}(t,k)dk \vspace{0.08in}\\
=&\!\!\! \d\int_{\partial D_3^0}\left[\phi_{13-}(t,k)-\frac{2}{\Sigma_-(k)}(e^{2ikL}\phi_{13})_+\right]dk+C_5(t) \vspace{0.08in}\\
=&\!\!\! \d \int_{\partial D_3^0}-\frac{\Sigma_+(k)}{\Sigma_-(k)}\phi_{13+}dk+C_5(t),
\end{array}\ene
where we have introduced the function $C_5(t)$ as
\bee\label{k2}
C_5(t)=\d\int_{\partial D_3^0}\frac{2}{\Sigma_-(k)}(e^{2ikL}\phi_{13})_+dk,
\ene

By applying the global relation (\ref{cag1}), the Cauchy's theorem and asymptotics (\ref{co31}) to Eq.~(\ref{k2}), we find
\bee \label{k2g} \begin{array}{rl}
C_5(t)=&\d\int_{\partial D_3^0}\frac{2}{\Sigma_-}(e^{2ikL}\phi_{13})_+dk \vspace{0.08in}\\
=&\d\int_{\partial D_3^0}\frac{2\alpha}{\Sigma_-}\left[-\bar{c}_{31}(t,\bar{k})+\bar{\Psi}_{31}\right]_+dk \vspace{0.08in}\\
&+\d\int_{\partial D_3^0}\frac{2}{\Sigma_-}\left\{\alpha[(\phi_{11}-1)\bar{\Psi}_{31}+\phi_{12}\bar{\Psi}_{32}]-\left[\phi_{13}(\bar{\Psi}_{33}-1)
+\phi_{14}\bar{\Psi}_{34}\right]e^{2ikL}\right\}_+dk \vspace{0.08in}\\
=& -i\pi\phi_{13}^{(1)}+\d\int_{\partial D_3^0}\frac{2\alpha}{\Sigma_-}\bar{\Psi}_{31+}dk \vspace{0.08in}\\
&+\d\int_{\partial D_3^0}\frac{2}{\Sigma_-}\left\{\alpha[(\phi_{11}-1)\bar{\Psi}_{31}+\phi_{12}\bar{\Psi}_{32}]-\left[\phi_{13}(\bar{\Psi}_{33}-1)
+\phi_{14}\bar{\Psi}_{34}\right]e^{2ikL}\right\}_+dk,
\end{array}
\ene

Eqs.~(\ref{phi13g2}) and (\ref{k2g}) imply that
\bee \label{phi13g2g}
\begin{array}{rl}
2i\pi \phi_{13}^{(1)}(t)=&\!\!\! \d \int_{\partial D_3^0}\left[-\frac{\Sigma_+(k)}{\Sigma_-(k)}\phi_{13+}+\frac{2\alpha}{\Sigma_-}\bar{\Psi}_{31+}\right]dk \vspace{0.08in}\\
&+\d\int_{\partial D_3^0}\frac{2}{\Sigma_-}\left\{\alpha[(\phi_{11}-1)\bar{\Psi}_{31}+\phi_{12}\bar{\Psi}_{32}]-\left[\phi_{13}(\bar{\Psi}_{33}-1)
+\phi_{14}\bar{\Psi}_{34}\right]e^{2ikL}\right\}_+dk ,
\end{array}
\ene
Thus, substituting of Eq.~(\ref{phi13g2g}) into the first one of Eq.~(\ref{vd}) yields Eq.~(\ref{v01}). Similarly, by applying the expression of $\phi_{j4}^{(1)}(t),\, j=1,2$ to the second and third ones of Eq.~(\ref{vd}), we can obtain Eqs.~(\ref{v00}) and (\ref{v0-1}). $\square$

\subsection*{\it 4.3.\, The effective characterizations }

\quad For the given Dirichlet boundary data $\{u_{0j}(t),\, v_{0j}(t),\, j=1,0,-1\}$, substituting Eqs. (\ref{u11})-(\ref{v1-1}) into Eqs.~(\ref{psit10})-(\ref{psit40}) and the similar expressions for $\{\phi_{ij}(t,k)\}_{i,j=1}^4$ yields a system of
quadratic nonlinear integral equations for $\{\Psi_{ij}(t,k),\, \phi_{ij}(t,k)\}_{i,j=1}^4$. The nonlinear integral system
gives an effective characterization of the spectral functions $\{\Psi(t,k),\, \phi(t,k)\}$ for the given Dirichlet problem.
In what follows we use the perturbation expression approach to exhibit them in detail.

Substituting these perturbated expressions
\bee\label{solue}
\left\{\begin{array}{rl}
\Psi_{ij}(t,k)=&\!\!\Psi_{ij}^{[0]}(t,k)+\epsilon\Psi_{ij}^{[1]}(t,k)+\epsilon^2\Psi_{ij}^{[2]}(t,k)+\cdots, \quad  i,j=1,2,3,4, \vspace{0.08in}\\
\phi_{ij}(t,k)=&\!\!\phi_{ij}^{[0]}(t,k)+\epsilon\phi_{ij}^{[1]}(t,k)+\epsilon^2\phi_{ij}^{[2]}(t,k)+\cdots, \quad  i,j=1,2,3,4, \vspace{0.08in}\\
u_{0j}(t)=&\!\!\epsilon u_{0j}^{[1]}(t)+\epsilon^2 u_{0j}^{[2]}(t)+\epsilon^3 u_{0j}^{[3]}(t)+\cdots, \quad  j=1,0,-1, \vspace{0.08in}\\
v_{0j}(t)=&\!\!\epsilon v_{0j}^{[1]}(t)+\epsilon^2 v_{0j}^{[2]}(t)+\epsilon^3 v_{0j}^{[3]}(t)+\cdots, \quad  j=1,0,-1, \vspace{0.08in}\\
u_{1j}(t)=&\!\!\epsilon u_{1j}^{[1]}(t)+\epsilon^2 u_{1j}^{[2]}(t)+\epsilon^3 u_{1j}^{[3]}(t)+\cdots, \quad  j=1,0,-1, \vspace{0.08in}\\
v_{1j}(t)=&\!\! \epsilon v_{1j}^{[1]}(t)+\epsilon^2 v_{1j}^{[2]}(t)+\epsilon^3 v_{1j}^{[3]}(t)+\cdots, \quad  j=1,0,-1,
\end{array}\right.
\ene
into Eqs.~(\ref{psit11})-(\ref{psit14}), where $\epsilon>0$ is a small parameter, we find these terms of $O(1)$ and
$O(\epsilon)$ of $\Psi_{ij}(t,k)$ as
\bee
O(1): \left\{\begin{array}{l}
 \Psi_{jj}^{[0]}=1,\quad j=1,2,3,4, \vspace{0.1in}\\
 \Psi_{ij}^{[0]}=0,\quad i,j=1,2,3,4,\, i\not=j,
\end{array}\right. \qquad\qquad\qquad\qquad\qquad\quad
\ene
\bee \label{epsilon}
O(\epsilon): \left\{\begin{array}{l}
 \Psi_{11}^{[1]}=\Psi_{12}^{[1]}=\Psi_{21}^{[1]}=\Psi_{22}^{[1]}=
 \Psi_{33}^{[1]}=\Psi_{34}^{[1]}=\Psi_{43}^{[1]}=\Psi_{44}^{[1]}=0, \vspace{0.1in}\\
 \Psi_{13}^{[1]}(t,k)=\d\int_0^te^{-4ik^2(t-t')}\left(2ku_{01}^{[1]}+iu_{11}^{[1]}\right)(t')dt',\vspace{0.1in}\\
 \Psi_{14}^{[1]}(t,k)=\d\int_0^te^{-4ik^2(t-t')}\left(2ku_{00}^{[1]}+iu_{10}^{[1]}\right)(t')dt',\vspace{0.1in}\\
 \Psi_{23}^{[1]}(t,k)=\d\beta\int_0^te^{-4ik^2(t-t')}\left(2ku_{00}^{[1]}+iu_{10}^{[1]}\right)(t')dt',\vspace{0.1in}\\
 \Psi_{24}^{[1]}(t,k)=\d\int_0^te^{-4ik^2(t-t')}\left(2ku_{0-1}^{[1]}+iu_{1-1}^{[1]}\right)(t')dt',\vspace{0.1in}\\
 \Psi_{31}^{[1]}(t,k)=\d\alpha\int_0^te^{4ik^2(t-t')}\left(2k\bar{u}_{01}^{[1]}-i\bar{u}_{11}^{[1]}\right)(t')dt',\vspace{0.1in}\\
  \Psi_{32}^{[1]}(t,k)=\d\alpha\beta\int_0^te^{4ik^2(t-t')}\left(2k\bar{u}_{00}^{[1]}-i\bar{u}_{10}^{[1]}\right)(t')dt',\vspace{0.1in}\\
 \Psi_{41}^{[1]}(t,k)=\d\alpha\int_0^te^{4ik^2(t-t')}\left(2k\bar{u}_{00}^{[1]}-i\bar{u}_{10}^{[1]}\right)(t')dt',\vspace{0.1in}\\
 \Psi_{42}^{[1]}(t,k)=\d\alpha\int_0^te^{4ik^2(t-t')}\left(2k\bar{u}_{0-1}^{[1]}-i\bar{u}_{1-1}^{[1]}\right)(t')dt',
 \end{array}\right.
\ene
and other terms $O(\epsilon^s),\, s>1$ of $\Psi_{ij}(t,k)$, which are omitted here.

Similarly, we can also obtain the analogous expressions for $\{\phi_{ij}^{[l]}\}_{i,j=1}^4, l=0,1$
by means of the boundary values at $x=L$, that is, $\{v_{ij}^{[l]}\}, i=0,1; j=1,2; l=0,1$.

If we assume that $n_{33,44}(\mathbb{S})$ has no zero points, then we expand Eqs.~(\ref{u11})-(\ref{v1-1}) to have
\bee \label{u0111}\left\{\begin{array}{l}
 u_{11}^{[1]}(t)=\d\int_{\partial D_3^0}\left[\frac{2\Sigma_+}{i\pi\Sigma_-}\left(k\Psi_{13-}^{[1]}\!+iu_{01}^{[1]}\right)
  +\frac{4i}{\pi\Sigma_-}\left(\alpha k\bar{\phi}_{31-}^{[1]}\!+i v_{01}^{[1]}\right)\right]dk, \vspace{0.08in}\\
 u_{10}^{[1]}(t)=\d\int_{\partial D_3^0}\left[\frac{2\Sigma_+}{i\pi\Sigma_-}\left(k\Psi_{14-}^{[1]}\!+iu_{00}^{[1]}\right)
  +\frac{4i}{\pi\Sigma_-}\left(\alpha k\bar{\phi}_{41-}^{[1]}\!+i v_{00}^{[1]}\right)\right]dk, \vspace{0.08in}\\
  u_{1-1}^{[1]}(t)=\d\int_{\partial D_3^0}\left[\frac{2\Sigma_+}{i\pi\Sigma_-}\left(k\Psi_{24-}^{[1]}\!+iu_{0-1}^{[1]}\right)
  +\frac{4i}{\pi\Sigma_-}\left(\alpha k\bar{\phi}_{42-}^{[1]}\!+i v_{0-1}^{[1]}\right)\right]dk, \vspace{0.08in}\\
 v_{11}^{[1]}(t)=\d\int_{\partial D_3^0}\left[\frac{2i\Sigma_+}{\pi\Sigma_-}\left(k\phi_{13-}^{[1]}\!+iv_{01}^{[1]}\right)
  -\frac{4i}{\pi\Sigma_-}\left(\alpha k\bar{\Psi}_{31-}^{[1]}+i u_{01}^{[1]}\right)\right]dk, \vspace{0.08in}\\
 v_{10}^{[1]}(t)=\d\int_{\partial D_3^0}\left[\frac{2i\Sigma_+}{\pi\Sigma_-}\left(k\phi_{14-}^{[1]}\!+iv_{00}^{[1]}\right)
  -\frac{4i}{\pi\Sigma_-}\left(\alpha k\bar{\Psi}_{41-}^{[1]}+i u_{00}^{[1]}\right)\right]dk,\vspace{0.08in}\\
 v_{1-1}^{[1]}(t)=\d\int_{\partial D_3^0}\left[\frac{2i\Sigma_+}{\pi\Sigma_-}\left(k\phi_{24-}^{[1]}\!+iv_{0-1}^{[1]}\right)
  -\frac{4i}{\pi\Sigma_-}\left(\alpha k\bar{\Psi}_{42-}^{[1]}+i u_{0-1}^{[1]}\right)\right]dk,
\end{array}\right.
\ene

It further follows from Eq.~(\ref{epsilon}) that we have
\bee\label{psi-}
 \left\{\begin{array}{l}
  \Psi_{13-}^{[1]}(t,k)=\d 4k\int_0^te^{-4ik^2(t-t')}u_{01}^{[1]}(t')dt',\vspace{0.1in}\\
 \Psi_{14-}^{[1]}(t,k)=\d 4k\int_0^te^{-4ik^2(t-t')}u_{00}^{[1]}(t')dt',\vspace{0.1in}\\
 \Psi_{24-}^{[1]}(t,k)=\d 4k\int_0^te^{-4ik^2(t-t')}u_{0-1}^{[1]}(t')dt',\vspace{0.1in}\\
  \Psi_{31-}^{[1]}(t,k)=\d 4\alpha k\int_0^te^{4ik^2(t-t')}\bar{u}_{01}^{[1]}(t')dt',\vspace{0.1in}\\
 \Psi_{41-}^{[1]}(t,k)=\d 4\alpha k\int_0^te^{4ik^2(t-t')}\bar{u}_{00}^{[1]}(t')dt',\vspace{0.1in}\\
 \Psi_{42-}^{[1]}(t,k)=\d 4\alpha k\int_0^te^{4ik^2(t-t')}\bar{u}_{0-1}^{[1]}(t')dt',
\end{array}\right.
\ene
Similarly, we have
\bee\label{phi-}
 \left\{\begin{array}{l}
 \phi_{13-}^{[1]}(t,k)=\d 4k \int_0^te^{-4ik^2(t-t')}v_{01}^{[1]}(t')dt',\vspace{0.1in}\\
 \phi_{14-}^{[1]}(t,k)=\d 4k \int_0^te^{-4ik^2(t-t')}v_{00}^{[1]}(t')dt',\vspace{0.1in}\\
 \phi_{24-}^{[1]}(t,k)=\d 4k \int_0^te^{-4ik^2(t-t')}v_{0-1}^{[1]}(t')dt',\vspace{0.1in}\\
 \phi_{31-}^{[1]}(t,k)=\d 4 \alpha k\int_0^te^{4ik^2(t-t')}\bar{v}_{01}^{[1]}(t')dt',\vspace{0.1in}\\
 \phi_{41-}^{[1]}(t,k)=\d 4 \alpha k\int_0^te^{4ik^2(t-t')}\bar{v}_{00}^{[1]}(t')dt',\vspace{0.1in}\\
 \phi_{42-}^{[1]}(t,k)=\d 4 \alpha k\int_0^te^{4ik^2(t-t')}\bar{v}_{0-1}^{[1]}(t')dt',
 \end{array}\right.
\ene

Therefore, the unknown Nuemann  boundary problem can now be solved perturbatively as follows: for the given $n_{33,44}(\mathbb{S})$ without zero points and Dirichlet  boundary data $u_{0j}^{[1]}$ and $v_{0j}^{[1]},\, j=1,0,-1$ at $x=0$, we can find these functions $\{\Psi_{13-}^{[1]},\, \Psi_{14-}^{[1]},\, \Psi_{24-}^{[1]},\, \Psi_{31-}^{[1]},\, \Psi_{41-}^{[1]},\, \Psi_{42-}^{[1]},\,\phi_{13-}^{[1]},\, \phi_{14-}^{[1]},\, \phi_{24-}^{[1]},\, \phi_{31-}^{[1]},\, \phi_{41-}^{[1]},\, \phi_{42-}^{[1]}\}$ in terms of Eqs.~(\ref{psi-}) and (\ref{phi-}). And then we can obtain $u_{1j}^{[1]}(t)$ and $v_{1j}^{[1]}(t),\, j=1,0,-1$ from Eq.~(\ref{u0111}). Finally, we have $\Psi_{ij}^{[1]}$ from Eq.~(\ref{epsilon}). Similarly, we can also find  $\phi_{ij}^{[1]}$.

Similarly, it follows from Eqs.~(\ref{u01})-(\ref{v0-1}) that we have
\bee \label{u1101}
\left\{\begin{array}{l}
u_{01}^{[1]}(t)=\d \int_{\partial D_3^0}\left[\frac{\Sigma_+(k)}{\pi\Sigma_-(k)}\Psi_{13+}^{[1]}
-\frac{2\alpha}{\pi\Sigma_-}\bar{\phi}_{31+}^{[1]}\right]dk, \vspace{0.08in}\\
u_{00}^{[1]}(t)=\d \int_{\partial D_3^0}\left[\frac{\Sigma_+(k)}{\pi\Sigma_-(k)}\Psi_{14+}^{[1]}-\frac{2\alpha}{\pi\Sigma_-}\bar{\phi}_{41+}^{[1]}\right]dk, \vspace{0.08in}\\
u_{0-1}^{[1]}(t)=\d \int_{\partial D_3^0}\left[\frac{\Sigma_+(k)}{\pi\Sigma_-(k)}\Psi_{24+}^{[1]}-\frac{2\alpha}{\pi\Sigma_-}\bar{\phi}_{42+}^{[1]}\right]dk, \vspace{0.08in}\\
v_{01}^{[1]}(t)=\d \int_{\partial D_3^0}\left[-\frac{\Sigma_+(k)}{\pi\Sigma_-(k)}\phi_{13+}^{[1]}+\frac{2\alpha}{\pi\Sigma_-}\bar{\Psi}_{31+}^{[1]}\right]dk, \vspace{0.08in}\\
v_{00}^{[1]}(t)=\d \int_{\partial D_3^0}\left[-\frac{\Sigma_+(k)}{\pi\Sigma_-(k)}\phi_{14+}^{[1]}+\frac{2\alpha}{\pi\Sigma_-}\bar{\Psi}_{41+}^{[1]}\right]dk, \vspace{0.08in}\\
v_{0-1}^{[1]}(t)=\d \int_{\partial D_3^0}\left[-\frac{\Sigma_+(k)}{\pi\Sigma_-(k)}\phi_{24+}^{[1]}+\frac{2\alpha}{\pi\Sigma_-}\bar{\Psi}_{42+}^{[1]}\right]dk,
\end{array}\right.
\ene

It further follows from Eq.~(\ref{epsilon}) that we have
\bee \label{psi+1}
\left\{\begin{array}{l}
  \Psi_{13+}^{[1]}(t,k)=\d 2i\int_0^te^{-4ik^2(t-t')}u_{11}^{[1]}(t')dt',\vspace{0.1in}\\
 \Psi_{14+}^{[1]}(t,k)=\d  2i\int_0^te^{-4ik^2(t-t')}u_{10}^{[1]}(t')dt',\vspace{0.1in}\\
  \Psi_{24+}^{[1]}(t,k)=\d  2i\int_0^te^{-4ik^2(t-t')}u_{1-1}^{[1]}(t')dt',\vspace{0.1in}\\
  \Psi_{31+}^{[1]}(t,k)=\d -2i\alpha\int_0^te^{4ik^2(t-t')}\bar{u}_{11}^{[1]}(t')dt',\vspace{0.1in}\\
  \Psi_{41+}^{[1]}(t,k)=\d -2i\alpha\int_0^te^{4ik^2(t-t')}\bar{u}_{10}^{[1]}(t')dt',\vspace{0.1in}\\
  \Psi_{42+}^{[1]}(t,k)=\d -2i\alpha\int_0^te^{4ik^2(t-t')}\bar{u}_{1-1}^{[1]}(t')dt',
 \end{array} \right.
\ene

Similarly we get
 \bee \label{psi+2}
\left\{\begin{array}{l}
  \phi_{13+}^{[1]}(t,k)=\d 2i\int_0^te^{-4ik^2(t-t')}v_{11}^{[1]}(t')dt',\vspace{0.1in}\\
 \phi_{14+}^{[1]}(t,k)=\d  2i\int_0^te^{-4ik^2(t-t')}v_{10}^{[1]}(t')dt',\vspace{0.1in}\\
 \phi_{24+}^{[1]}(t,k)=\d  2i\int_0^te^{-4ik^2(t-t')}v_{1-1}^{[1]}(t')dt',\vspace{0.1in}\\
  \phi_{31+}^{[1]}(t,k)=\d -2i\alpha\int_0^te^{4ik^2(t-t')}\bar{v}_{11}^{[1]}(t')dt',\vspace{0.1in}\\
 \phi_{41+}^{[1]}(t,k)=\d  -2i\alpha\int_0^te^{4ik^2(t-t')}\bar{v}_{10}^{[1]}(t')dt',\vspace{0.1in}\\
 \phi_{42+}^{[1]}(t,k)=\d  -2i\alpha\int_0^te^{4ik^2(t-t')}\bar{v}_{1-1}^{[1]}(t')dt',
\end{array} \right.
\ene
Therefore, the unknown Dirichlet  boundary problem can now be solved perturbatively as follows: for the given $n_{33,44}(\mathbb{S})$ without zero points and Neumann  boundary data at $x=L$ $u_{1j}^{[1]}$ and $v_{1j}^{[1]}\, j=1,0,-1$, we can determine these functions $\{\Psi_{13+}^{[1]},\, \Psi_{14+}^{[1]},\, \Psi_{24+}^{[1]},\, \Psi_{31+}^{[1]},\, \Psi_{41+}^{[1]},\, \Psi_{42+}^{[1]},\,\phi_{13+}^{[1]},\, \phi_{14+}^{[1]},\, \phi_{24+}^{[1]},\, \phi_{31+}^{[1]},\, \phi_{41+}^{[1]},\, \phi_{42+}^{[1]}\}$  from Eqs.~(\ref{psi+1}) and (\ref{psi+2}). Moreover, we can further find $u_{0j}^{[1]}$ and $v_{0j}^{[1]}\, j=1,0,-1$ from Eq.~(\ref{u1101}). Finally, we can have $\Psi_{ij}^{[1]}$ from Eq.~(\ref{epsilon}). Similarly, we can also find  $\phi_{ij}^{[1]}$.

In fact, the above-obtained recursive formulae can be continued indefinitely. We assume that they hold for all $0\leq j\leq n-1$, then for $n>0$,
the substitution of Eq.~(\ref{solue}) into Eqs.~(\ref{u11})-(\ref{u1-1}) yields the terms of $O(\epsilon^n)$ as
\bes\bee\label{u11e}
u_{11}^{[n]}(t)=\d\int_{\partial D_3^0}\frac{2}{i\pi\Sigma_-}\left[\Sigma_+\left(k\Psi_{13-}^{[n]}+iu_{01}^{[n]}\right)
 -2\left(\alpha k\bar{\phi}_{31-}^{[n]}+i v_{01}^{[n]}\right)\right]dk+ {\rm lower \,\, order \,\, terms},
 \\ \label{u10e}
 u_{10}^{[n]}(t)=\d\int_{\partial D_3^0}\frac{2}{i\pi\Sigma_-}
\left[\Sigma_+\left(k\Psi_{14-}^{[n]}+iu_{00}^{[n]}\right)-2\left(\alpha k\bar{\phi}_{41-}^{[n]}+i v_{00}^{[n]}\right)\right]dk+ {\rm lower \,\, order \,\, terms}, \\ \label{u1-1e}
 u_{1-1}^{[n]}(t)=\d\int_{\partial D_3^0}\frac{2}{i\pi\Sigma_-}
\left[\Sigma_+\left(k\Psi_{24-}^{[n]}+iu_{0-1}^{[n]}\right)-2\left(\alpha k\bar{\phi}_{42-}^{[n]}+i v_{0-1}^{[n]}\right)\right]dk+ {\rm lower \,\, order \,\, terms},
\ene\ees
where `lower order terms' stands for the result involving known terms of lower order.

The terms of $O(\epsilon^n)$ for  $\Psi_{ij}$ in Eqs.~(\ref{psit10})-(\ref{psit40}) and the similar equations for $\phi_{ij}$ yield
\bee\label{pp0}
\left\{\begin{array}{l}
\Psi_{13}^{[n]}(t,k)=\displaystyle\int_0^t e^{-4ik^2(t-t')}\left(2ku_{01}^{[n]}+iu_{11}^{[n]}\right)(t')dt'
   + {\rm lower \,\, order \,\, terms},  \vspace{0.1in}\\
\bar{\phi}_{31}^{[n]}(t,\bar{k})=\d\alpha\int_0^t e^{-4ik^2(t-t')}\left(2kv_{01}^{[n]}+iv_{11}^{[n]}\right)(t')dt'+ {\rm lower \,\, order \,\, terms}, \vspace{0.1in}\\
\Psi_{14}^{[n]}(t,k)=\d \int_0^t e^{-4ik^2(t-t')}\left(2ku_{00}^{[n]}+iu_{10}^{[n]}\right)(t')dt' + {\rm lower \,\, order \,\, terms},  \vspace{0.1in}\\
 \bar{\phi}_{41}^{[n]}(t,\bar{k})=\d \alpha\int_0^te^{-4ik^2(t-t')}\left(2kv_{00}^{[n]}+iv_{10}^{[n]}\right)(t')dt' + {\rm lower \,\, order \,\, terms},\vspace{0.1in}\\
\Psi_{24}^{[n]}(t,k)=\d \int_0^t e^{-4ik^2(t-t')}\left(2ku_{0-1}^{[n]}+iu_{1-1}^{[n]}\right)(t')dt' + {\rm lower \,\, order \,\, terms},  \vspace{0.1in}\\
 \bar{\phi}_{42}^{[n]}(t,\bar{k})=\d \alpha\int_0^te^{-4ik^2(t-t')}\left(2kv_{0-1}^{[n]}+iv_{1-1}^{[n]}\right)(t')dt' + {\rm lower \,\, order \,\, terms},
\end{array}\right.
\ene
which leads to
\bee \label{pp-}
\left\{\begin{array}{l}
\Psi_{13-}^{[n]}(t,k)=\d 4k\int_0^t e^{-4ik^2(t-t')}u_{01}^{[n]}(t')dt'+ {\rm lower \,\, order \,\, terms},   \vspace{0.1in}\\
\bar{\phi}_{31-}^{[n]}(t,\bar{k})=\d 4\alpha k\int_0^t e^{-4ik^2(t-t')}v_{01}^{[n]}(t')dt'+ {\rm lower \,\, order \,\, terms}, \vspace{0.1in}\\
\Psi_{14-}^{[n]}(t,k)=\d 4k\int_0^t e^{-4ik^2(t-t')}u_{00}^{[n]}(t')dt'+ {\rm lower \,\, order \,\, terms},   \vspace{0.1in}\\
\bar{\phi}_{41-}^{[n]}(t,\bar{k})=\d 4\alpha k\int_0^t e^{-4ik^2(t-t')}v_{00}^{[n]}(t')dt'+ {\rm lower \,\, order \,\, terms},\vspace{0.1in}\\
\Psi_{24-}^{[n]}(t,k)=\d 4k\int_0^t e^{-4ik^2(t-t')}u_{0-1}^{[n]}(t')dt'+ {\rm lower \,\, order \,\, terms},   \vspace{0.1in}\\
\bar{\phi}_{42-}^{[n]}(t,\bar{k})=\d 4\alpha k\int_0^t e^{-4ik^2(t-t')}v_{0-1}^{[n]}(t')dt'+ {\rm lower \,\, order \,\, terms},
\end{array}\right.
\ene
It follows from system (\ref{pp-}) that $\Psi_{1j-}^{[n]},\, \Psi_{24-}^{[n]},\, \bar{\phi}_{j1-}^{[n]},\, \bar{\phi}_{42-}^{[n]},\, j=3,4$ can be obtained at each step from the known Dirichlet
boundary data $u_{0j}^{[n]}$ and $v_{0j}^{[n]},\, j=1,0,-1$ such that we know that the Neumann boundary data at $x=0$ $u_{1j}^{[n]},\, j=1,0,-1$  can then be found by Eqs.~(\ref{u11e})-(\ref{u1-1e}). Similarly, we also show that the Neumann boundary data at $x=L$ $v_{1j}^{[n]},\, j=1,0,-1$ can then be determined by the known Dirichlet boundary data $u_{0j}^{[n]}$ and $v_{0j}^{[n]},\, j=1,0,-1$ .

Similarly, the substitution of Eq.~(\ref{solue}) into Eqs.~(\ref{u01})-(\ref{u0-1}) yields the terms of $O(\epsilon^n)$ as
\bes\bee \label{u01e}
& u_{01}^{[n]}(t)=\d \int_{\partial D_3^0}\frac{1}{\pi\Sigma_-(k)}
\left[\Sigma_+(k)\Psi_{13+}^{[n]}-2\alpha\bar{\phi}_{31+}^{[n]}\right]dk+ {\rm lower \,\, order \,\, terms},\\
& \label{u00e}
u_{00}^{[n]}(t)=\d \int_{\partial D_3^0}\frac{1}{\pi\Sigma_-(k)}
\left[\Sigma_+(k)\Psi_{14+}^{[n]}-2\alpha\bar{\phi}_{41+}^{[n]}\right]dk+ {\rm lower \,\, order \,\, terms},\\
& \label{u0-1e}
u_{0-1}^{[n]}(t)=\d \int_{\partial D_3^0}\frac{1}{\pi\Sigma_-(k)}
\left[\Sigma_+(k)\Psi_{24+}^{[n]}-2\alpha\bar{\phi}_{42+}^{[n]}\right]dk+ {\rm lower \,\, order \,\, terms},
\ene\ees

Eq.~(\ref{pp0}) implies that
\bee\label{pp+}
\left\{\begin{array}{l}
\Psi_{13+}^{[n]}(t,k)=\d 2i\int_0^t e^{-4ik^2(t-t')}u_{11}^{[n]}(t')dt' + {\rm lower \,\, order \,\, terms},  \vspace{0.1in}\\
\bar{\phi}_{31+}^{[n]}(t,\bar{k})=\d 2i\alpha\int_0^t e^{-4ik^2(t-t')}v_{11}^{[n]}(t')dt'+ {\rm lower \,\, order \,\, terms}, \vspace{0.1in}\\
\Psi_{14+}^{[n]}(t,k)=\d 2i\int_0^t e^{-4ik^2(t-t')}u_{10}^{[n]}(t')dt' + {\rm lower \,\, order \,\, terms},  \vspace{0.1in}\\
 \bar{\phi}_{41+}^{[n]}(t,\bar{k})=\d 2i\alpha\int_0^te^{-4ik^2(t-t')}v_{10}^{[n]}(t')dt' + {\rm lower \,\, order \,\, terms},\vspace{0.1in}\\
\Psi_{24+}^{[n]}(t,k)=\d 2i\int_0^t e^{-4ik^2(t-t')}u_{1-1}^{[n]}(t')dt' + {\rm lower \,\, order \,\, terms},  \vspace{0.1in}\\
 \bar{\phi}_{42+}^{[n]}(t,\bar{k})=\d 2i\alpha\int_0^te^{-4ik^2(t-t')}v_{1-1}^{[n]}(t')dt' + {\rm lower \,\, order \,\, terms},
\end{array}\right.
\ene
It follows from system (\ref{pp+}) that $\Psi_{1j+}^{[n]},\, \Psi_{24+}^{[n]},\, \bar{\phi}_{j1+}^{[n]},\, \bar{\phi}_{42+}^{[n]},\, j=3,4$ can be determined at each step from the known Neumann
boundary data at $x=0$ $u_{1j}^{[n]}$ and $v_{1j}^{[n]},\, j=1,0,-1$ such that we know that the Dirichlet boundary data $u_{0j}^{[n]},\, 1,0,-1$ can then be given by Eqs.~(\ref{u01e})-(\ref{u0-1e}). Similarly, we also show that the Dirichlet boundary data at $x=L$ $v_{0j}^{[n]},\,j=1,0,-1$ can then be determined
by the known Neumann boundary data at $x=L$ $u_{1j}^{[n]}$ and $v_{1j}^{[n]},\, j=1,0,-1$.

\subsection*{\it 4.4. \, The large $L$ limit}

\quad The formulae for $u_{0j}(t)$ and $u_{1j}(t),\, j=1,0,-1$ of Theorem 4.2 in the limit $L\to \infty$ can reduce to the corresponding
ones on the half-line. Since when $L\to \infty$,
\bee \label{limit}
\begin{array}{l} v_{0j}\to 0,\quad v_{1j}\to 0,\quad j=1,0,-1, \quad
 \phi_{ij}\to \delta_{ij}, \quad \d\frac{\Sigma_+(k)}{\Sigma_-(k)}\to 1 \quad {\rm as} \quad k\to \infty \quad {\rm in} \quad D_3,
\end{array}
\ene

Thus, according to Eq.~(\ref{limit}), the $L\to \infty$ limits of Eqs.~(\ref{u11})-(\ref{u1-1}), (\ref{u01})-(\ref{u0-1}) yield the unknown Neumann boundary data
\bee
\left\{\begin{array}{rl}
 u_{11}(t)=&\!\!\!\!\d -\frac{1}{\pi}\int_{\partial D_3^0}
\left[2i(k\Psi_{13-}\!+iu_{01})+u_{01}\bar{\phi}_{33-}-2(u_{01}\Psi_{33-}+u_{00}\Psi_{43-})\right]dk,\vspace{0.1in} \\
 u_{10}(t)=&\!\!\!\!\d -\frac{1}{\pi}\int_{\partial D_3^0}
\left[2i(k\Psi_{14-}\!+iu_{00})+u_{00}\bar{\phi}_{44-}-2(u_{01}\Psi_{34-}+u_{00}\Psi_{44-})\right]dk,\vspace{0.1in} \\
 u_{1-1}(t)=&\!\!\!\!\d -\frac{1}{\pi}\int_{\partial D_3^0}
\left[2i(k\Psi_{24-}\!+iu_{0-1})+u_{0-1}\bar{\phi}_{44-}-2(\beta u_{00}\Psi_{34-}+u_{0-1}\Psi_{44-})\right]dk,
\end{array}\right.
\ene
for the given Dirichlet boundary problem, and the unknown Dirichlet boundary data
\bee
 u_{01}(t)=\d \frac{1}{\pi}\int_{\partial D_3^0}\Psi_{13+}dk, \quad
u_{00}(t)=\d \frac{1}{\pi}\int_{\partial D_3^0}\Psi_{14+}dk,\quad
u_{0-1}(t)=\d \frac{1}{\pi}\int_{\partial D_3^0}\Psi_{24+}dk,
\ene
for the given Neumann boundary problem.

\section{The GLM representations and equivalence}

\quad Nowadays we rededuce the eigenfunctions $\Psi(t,k)$ and $\phi(t,k)$ in Sec. 4 by means of the Gel'fand-Levitan-Marchenko (GLM) approach~\cite{f4,glm1,glm2,glm3}. Moreover, the global relation can be used to find the unknown Neumann (Dirichlet) boundary values from the given
Dirichlet (Neumann) boundary values in terms of the GLM representations. Moreover, the GLM representations are shown to be equivalent to the
ones obtained in Sec. 4. Finally, the linearizable boundary conditions are presented for  the  GLM representations.

\vspace{0.1in}
\noindent \subsection*{\it 5.1. The GLM representations for $\Psi(t,k)$ and $\phi(t,k)$}
\vspace{0.1in}

\noindent {\bf Proposition 5.1.} {\it The eigenfunctions $\Psi(t,k)$ and $\phi(t,k)$ defined in Sec. 4 have the GLM representations
\bes\bee\label{psiglm}
&\Psi(t,k)=\mathbb{I}_{4\times 4}
+\d\int_{-t}^t\left[L(t,s)+\left(k+\frac{i}{2}U^{(0)}\sigma_4\right)G(t,s)\right]e^{-2ik^2(s-t)\sigma_4}ds,
\\
\label{phiglm}
&\phi(t,k)=\mathbb{I}_{4\times 4}+\d\int_{-t}^t\left[\mathcal{L}(t,s)+\left(k+\frac{i}{2}\mathcal{U}^{(L)}\sigma_4\right)\mathcal{G}(t,s)\right]e^{-2ik^2(s-t)\sigma_4}ds,
\ene\ees
where the $4\times 4$ matrix-valued functions $L(t, s)=(L_{ij})_{4\times 4}$ and $G(t, s)=(G_{ij})_{4\times 4},\, -t\leq s\leq t$ satisfy a
Goursat system
\bee\label{lge}\left\{
\begin{array}{rl}
L_t(t, s)+\sigma_4L_s(t,s)\sigma_4=&\!\!\! i\sigma_4U_x^{(0)}L(t,s) -\dfrac{1}{2}\left\{(U^{(0)})^3+i\dot{U}^{(0)}\sigma_4+\l[U_x^{(0)},  U^{(0)}\r]\right\}G(t,s), \vspace{0.1in}\\
G_t(t, s)+\sigma_4G_s(t,s)\sigma_4=&\!\!\! 2U^{(0)}L(t,s)+i\sigma_4U_x^{(0)} G(t,s),
\end{array}\right.
\ene
with the initial conditions
\bee\label{lgi}
\left\{
\begin{array}{rl}
L_{11}(t,-t)=&\!\!\!L_{12}(t,-t)=L_{21}(t,-t)=L_{22}(t,-t)=L_{33}(t,-t) \vspace{0.1in}\\
 =&\!\!\!L_{34}(t,-t)=L_{43}(t,-t)=L_{44}(t,-t)=0, \vspace{0.1in}\\
G_{11}(t,-t)=&\!\!\! G_{12}(t,-t)=G_{21}(t,-t)=G_{22}(t,-t)=G_{33}(t,-t) \vspace{0.1in}\\
  =&\!\!\! G_{34}(t,-t)=G_{43}(t,-t)=G_{44}(t,-t)=0, \vspace{0.1in}\\
G_{13}(t,t)=&\!\!\! u_{01}(t),\,\, G_{24}(t,t)=u_{0-1}(t),\,\, G_{14}(t,t)=u_{00}(t),\,\, G_{23}(t,t)=\beta u_{00}(t), \vspace{0.1in}\\
G_{31}(t,t)=&\!\!\!\alpha \bar{u}_{01}(t),\,\, G_{42}(t,t)=\alpha \bar{u}_{0-1}(t), \,\, G_{32}(t,t)=\alpha\beta \bar{u}_{00}(t),\,\, G_{41}(t,t)=\alpha\bar{u}_{00}(t), \vspace{0.1in}\\
L_{13}(t,t)=&\d\!\!\!\frac{i}{2}u_{11}(t),\,\, L_{24}(t,t)=\frac{i}{2}u_{1-1}(t),\,\,L_{14}(t,t)=\frac{i}{2}u_{10}(t), \,\,L_{23}(t,t)=\frac{i}{2}\beta u_{10}(t), \vspace{0.1in}\\
L_{31}(t,t)=&\d\!\!\! -\frac{i}{2}\alpha \bar{u}_{11}(t),\,\, L_{42}(t,t)=-\frac{i}{2}\alpha \bar{u}_{1-1}(t),\,\,L_{32}(t,t)=-\frac{i}{2}\alpha\beta \bar{u}_{10}(t), \,\, L_{41}(t,t)=-\frac{i}{2}\alpha\bar{u}_{10}(t),
\end{array}\right.
\ene
and
\bee \label{U0}
\begin{array}{c}
U^{(0)}=\left(\begin{array}{cccc}
            0 & 0 & u_{01}(t) & u_{00}(t) \vspace{0.1in}\\
            0&  0 & \beta u_{00}(t) & u_{0-1}(t) \vspace{0.1in}\\
            \alpha \bar{u}_{01}(t) & \alpha\beta \bar{u}_{00}(t) & 0 & 0\vspace{0.1in} \\
            \alpha \bar{u}_{00}(t) & \alpha \bar{u}_{0-1}(t) & 0 & 0
            \end{array}\right), \quad \dot{U}^{(0)}=\dfrac{d}{dt}U^{(0)},\vspace{0.1in}\\
U_x^{(0)}=\left(\begin{array}{cccc}
            0 & 0 & u_{11}(t) & u_{10}(t) \vspace{0.1in}\\
            0&  0 & \beta u_{10}(t) & u_{11}(t) \vspace{0.1in}\\
            \alpha \bar{u}_{11}(t) & \alpha\beta \bar{u}_{10}(t) & 0 & 0 \vspace{0.1in}\\
            \alpha \bar{u}_{10}(t) & \alpha \bar{u}_{1-1}(t) & 0 & 0
            \end{array}\right),
            \end{array}
                        \ene
Similarly, $\mathcal{L}(t,s),\, \mathcal{G}(t,s)$ satisfy the similar Eqs.~(\ref{lge}) and (\ref{lgi}) with $L(t,s)\to\mathcal{L}(t,s),\, G(t,s)\to \mathcal{G}(t,s),\, u_{0j}(t)\to v_{0j}(t),\, u_{1j}(t)\to v_{1j}(t),\,
 U^{(0)}\to \mathcal{U}^{(L)}=U^{(0)}\big|_{u_{0j}(t)\to v_{0j}(t)},\, U_x^{(0)}\to \mathcal{U}_x^{(L)}=U_x^{(0)}\big|_{u_{1j}(t)\to v_{1j}(t)},\, j=1,0,-1$.}

\vspace{0.1in}
\noindent {\bf Proof.} We assume that  the function
\bee \label{psi}
\psi(t,k)=e^{-2ik^2t\sigma_4}+\d\int_{-t}^t[L_0(t,s)+kG(t,s)]e^{-2ik^2s\sigma_4}ds,
\ene
satisfies the time-part of Lax pair (\ref{lax}) with the boundary data $\psi(0,k)=\mathbb{I}$ at $x=0$, where $L_0(t,s)$ and $G(t,s)$ are the unknown $4\times 4$ matrix-valued functions. We substitute Eq.~(\ref{psi}) into the time-part of Lax pair (\ref{lax}) with the boundary data (\ref{ibv}) and use the identity
\bee
\d\int_{-t}^tF(t,s)e^{-2ik^2s\sigma_4}ds=\frac{i}{2k^2}\left[F(t,t)e^{-2ik^2t\sigma_4}-F(t,-t)e^{2ik^2t\sigma_4}-
  \d\int_{-t}^tF_s(t,s)e^{-2ik^2s\sigma_4}ds\right]\sigma_4,
\ene
where the function $F(t,s)$ is a $4\times 4$ matrix-valued function. As a result, we have
\bee\label{lg}
\left\{
\begin{array}{l}
L_0(t, -t)+\sigma_4L_0(t,-t)\sigma_4=-iU^{(0)}G(t,-t)\sigma_4, \vspace{0.1in}\\
G(t, -t)+\sigma_4G(t,-t)\sigma_4=0, \vspace{0.1in}\\
L_0(t, t)-\sigma_4L_0(t,t)\sigma_4=iU^{(0)}G(t,t)\sigma_4+V_0^{(0)}, \vspace{0.1in}\\
G(t, t)-\sigma_4G(t,t)\sigma_4=2U^{(0)}, \vspace{0.1in}\\
L_{0t}(t, s)+\sigma_4L_{0s}(t,s)\sigma_4=-iU^{(0)}G_s(t,s)\sigma_4+V_0^{(0)}L_0(t,s), \vspace{0.1in}\\
G_{t}(t, s)+\sigma_4G_s(t,s)\sigma_4=2U^{(0)}L_0(t,s)+V_0^{(0)}G(t,s),
\end{array}\right.
\ene
where $U^{(0)}$ is given by Eq.~(\ref{U0}) and
\bee\no\begin{array}{rl}
 V_0^{(0)}=&\!\!\!\! i\sigma_4\left[U_x^{(0)}-(U^{(0)})^2\right] \vspace{0.1in}\\
  =&\!\!\!\! i\left(\!\!\!\!\!\begin{array}{cccc}
           -\alpha(|u_{01}|^2\!+\!|u_{00}|^2) &\!\! -\alpha(\beta u_{01}\bar{u}_{00}\!+\!u_{00}\bar{u}_{0-1}) &
           u_{11} & u_{10} \vspace{0.1in}\\
           -\alpha(\beta u_{00}\bar{u}_{01}\!+\!u_{0-1}\bar{u}_{00}) & \!\!  -\alpha(|u_{00}|^2\!+\!|u_{0-1}|^2) &            \beta u_{10} &  u_{1-1} \vspace{0.1in}\\
           -\alpha\bar{u}_{11} & -\alpha\beta\bar{u}_{10} & \!\! \alpha(|u_{01}|^2\!+\!|u_{00}|^2) & \!\!
           \alpha(\beta u_{0-1}\bar{u}_{00}\!+\! u_{00}\bar{u}_{01}) \vspace{0.1in}\\
            -\alpha\bar{u}_{10} & -\alpha \bar{u}_{1-1} & \!\!\!\!
            \alpha(\beta u_{00}\bar{u}_{0-1}\!+\!u_{01}\bar{u}_{00}) &  \alpha(|u_{0-1}|^2\!+\!|u_{00}|^2)
            \end{array}\!\!\!\!\right).
           \end{array}
            \ene

We further introduce the new matrix $L(t,s)$ by
\bee
L(t,s)=L_0(t,s)-\frac{i}{2}U^{(0)}\sigma_4G(t,s),
\ene
such that we can simplify the first four equations of system (\ref{lg}) as
\bee\no
\left\{
\begin{array}{l}
L(t, -t)+\sigma_4L(t,-t)\sigma_4=0, \vspace{0.1in}\\
G(t, -t)+\sigma_4G(t,-t)\sigma_4=0, \vspace{0.1in}\\
L(t, t)-\sigma_4L(t,t)\sigma_4=V_0^{(0)}, \vspace{0.1in}\\
G(t, t)-\sigma_4G(t,t)\sigma_4=2U^{(0)},
\end{array}\right.
\ene
which leads to Eq.~(\ref{lgi}), and from the last two equations of system (\ref{lg}) we have Eq.~(\ref{lge}). By using  the transformation
(\ref{mud}), that is, $\mu_2(0, t, k)=\Psi(t,k)=\psi(t,k)e^{2ik^2t\sigma_4}$, we know that $\Psi(t,k)$ is given by Eq.~(\ref{psiglm}).
Similarly, we can also show that Eq.~(\ref{phiglm}) holds. $\square$

\vspace{0.1in}

For convenience, we rewrite a $4\times 4$ matrix $C=(C_{ij})_{4\times 4}$ as
\bee\no
C=(C_{ij})_{4\times 4}=\left(\begin{array}{cc}
 \tilde{C}_{11} & \tilde{C}_{12} \vspace{0.05in}\cr
 \tilde{C}_{21} & \tilde{C}_{22}
\end{array}\right),
\ene
with
\bee \no
\tilde{C}_{11}=\left(\begin{array}{cc} C_{11} & C_{12} \vspace{0.05in}\cr C_{21} & C_{22}\end{array} \right),\,
\tilde{C}_{12}=\left(\begin{array}{cc} C_{13} & C_{14} \vspace{0.05in}\cr C_{23} & C_{24}\end{array} \right), \,
\tilde{C}_{21}=\left(\begin{array}{cc} C_{31} & C_{32} \vspace{0.05in}\cr C_{41} & C_{42}\end{array} \right),\,
\tilde{C}_{22}=\left(\begin{array}{cc} C_{33} & C_{34} \vspace{0.05in}\cr C_{43} & C_{44}\end{array} \right).
\ene

The Dirichlet and Neumann boundary values at $x=0, L$ can simply be rewritten as
\bee
 u_j(t)=\left(\begin{array}{cc}
 u_{j1}(t) & \beta u_{j0}(t) \vspace{0.05in}\cr
 u_{j0}(t) & u_{j-1}(t)
\end{array}
 \right),\quad
 v_j(t)=\left(\begin{array}{cc}
 v_{j1}(t) & \beta v_{j0}(t) \vspace{0.05in}\cr
 v_{j0}(t) & v_{j-1}(t)
\end{array}
 \right), \quad j=1,2,
\ene

For a matrix-valued function $F(t,s)$, we introduce the $\hat{F}(t,k)$ by
\bee\no
 \hat{F}(t,k)=\d\int_{-t}^{t}F(t,s)e^{2ik^2(s-t)}ds,
\ene
Thus, the GLM expressions (\ref{psiglm}) and (\ref{phiglm}) for $\{\Psi_{ij}(t,k),\, \phi_{ij}(t,k)\}$ can be rewritten as
\bes
\bee \label{psiglmg}
&\left\{\begin{array}{l}
\d\tilde{\Psi}_{11}(t,k)=\mathbb{I}_{2\times 2}+\hat{\tilde{L}}_{11}-\frac{i}{2}u_0^T(t)\hat{\tilde{G}}_{21}+k\hat{\tilde{G}}_{11},  \vspace{0.1in}\\
\d\tilde{\Psi}_{12}(t,k)=\hat{\tilde{L}}_{12}-\frac{i}{2}u_0^T(t)\hat{\tilde{G}}_{22}+k\hat{\tilde{G}}_{12}, \vspace{0.1in}\\
\d\tilde{\Psi}_{21}(t,k)=\hat{\tilde{L}}_{21}+\frac{i\alpha}{2}\bar{u}_0(t)\hat{\tilde{G}}_{11}+k\hat{\tilde{G}}_{21}, \vspace{0.1in}\\
\d\tilde{\Psi}_{22}(t,k)=\mathbb{I}_{2\times 2}+\hat{\tilde{L}}_{22}+\frac{i\alpha}{2}\bar{u}_0(t)\hat{\tilde{G}}_{12}+k\hat{\tilde{G}}_{22},
\end{array}\right.\\
\label{phiglmg}
&\left\{\begin{array}{l}
\d\tilde{\phi}_{11}(t,k)=\mathbb{I}_{2\times 2}+\hat{\tilde{\mathcal{L}}}_{11}-\frac{i}{2}v_0^T(t)\hat{\tilde{\mathcal{G}}}_{21}+k\hat{\tilde{\mathcal{G}}}_{11},  \vspace{0.1in}\\
\d\tilde{\phi}_{12}(t,k)=\hat{\tilde{\mathcal{L}}}_{12}-\frac{i}{2}v_0^T(t)\hat{\tilde{\mathcal{G}}}_{22}+k\hat{\tilde{\mathcal{G}}}_{12}, \vspace{0.1in}\\
\d\tilde{\phi}_{21}(t,k)=\hat{\tilde{\mathcal{L}}}_{21}+\frac{i\alpha}{2}\bar{v}_0(t)\hat{\tilde{\mathcal{G}}}_{11}+k\hat{\tilde{\mathcal{G}}}_{21}, \vspace{0.1in}\\
\d\tilde{\phi}_{22}(t,k)=\mathbb{I}_{2\times 2}+\hat{\tilde{\mathcal{L}}}_{22}+\frac{i\alpha}{2}\bar{v}_0(t)\hat{\tilde{\mathcal{G}}}_{12}+k\hat{\tilde{\mathcal{G}}}_{22},
\end{array}\right.
\ene\ees

\vspace{0.1in}
\noindent {\bf Proposition 5.2.}
\bes\bee
&\label{f12e} \d \lim_{t'\to t}\d\int_{\partial D_1^0}\frac{k}{\Sigma_-}e^{4ik^2(t-t')}\!\!\left(\tilde{F}_{12}(t,k)e^{-2ikL}\right)_{\!\!-}dk=\!\!
\d\int_{\partial D_1^0}\!\!\left[\frac{i k}{2}u_0^T\hat{\tilde{G}}_{22}\!+\!\frac{ k}{2i}u_0^T\bar{\hat{\tilde{\mathcal{G}}}}_{22}^T\!+\!\frac{k}{\Sigma_-}\left(\tilde{F}_{12}(t,k)e^{-2ikL}\right)_-\right]dk,
\qquad \vspace{0.1in}\\
&\label{f21e}  \d\lim_{t'\to t}\d\int_{\partial D_1^0}\frac{k}{\Sigma_-}e^{4ik^2(t-t')}\tilde{F}_{21-}(t,k)dk=
\d\int_{\partial D_1^0}\left[\frac{i\alpha k}{2}v_0^T\hat{\tilde{\mathcal{G}}}_{22}+\frac{\alpha k}{2i}v_0^T\bar{\hat{\tilde{G}}}_{22}^T+\frac{k}{\Sigma_-}\tilde{F}_{21-}(t,k)\right]dk, \vspace{0.1in}\\
&\label{f12e+}
\d \lim_{t'\to t}\d\int_{\partial D_1^0}\frac{1}{\Sigma_-}e^{4ik^2(t-t')}\left(\tilde{F}_{12}(t,k)e^{-2ikL}\right)_+dk=
\d\int_{\partial D_1^0}\frac{1}{\Sigma_-}\left(\tilde{F}_{12}(t,k)e^{-2ikL}\right)_+dk, \vspace{0.1in}\\
&\label{f21e+}
\d\lim_{t'\to t}\d\int_{\partial D_1^0}\frac{1}{\Sigma_-}e^{4ik^2(t-t')}\tilde{F}_{21+}(t,k)dk=
\d\int_{\partial D_1^0}\frac{1}{\Sigma_-}\tilde{F}_{21+}(t,k)dk, \qquad\qquad\qquad\qquad
\ene
\ees
{\it where the $2\times 2$ matrix-valued functions $\tilde{F}_{12}(t,k)$ and $\tilde{F}_{21}(t,k)$ are given by}
\bee
\label{f12eq}
\begin{array}{rl}
\tilde{F}_{12}(t,k)=&\!\!\!\! \d -\frac{i}{2}u_0^T(t)\hat{\tilde{G}}_{22}+\frac{i}{2}\bar{\hat{\tilde{\mathcal{G}}}}_{11}^Tv_0^T(t)e^{2ikL}
-\alpha\l(\tilde{\Psi}_{11}-\mathbb{I}\r)\bar{\tilde{\phi}}_{21}^Te^{2ikL}
+\tilde{\Psi}_{12}\l(\bar{\tilde{\phi}}_{22}(t,\bar{k})^T-\mathbb{I}\r)
\vspace{0.1in}\\
= &\!\!\!\!  \d-\frac{i}{2}u_0^T(t)\hat{\tilde{G}}_{22}+\frac{i}{2}\bar{\hat{\tilde{\mathcal{G}}}}_{11}^Tv_0^T(t)e^{2ikL}  \vspace{0.1in}\\
 &\!\!\!\! \d+ \left(\hat{\tilde{L}}_{12}-\frac{i}{2}u_0^T(t)\hat{\tilde{G}}_{22}+k\hat{\tilde{G}}_{12}\right)
   \left(\bar{\hat{\tilde{\mathcal{L}}}}_{22}^T-\frac{i\alpha}{2}\bar{\hat{\tilde{\mathcal{G}}}}_{12}^Tv_0^T(t)
   +k\bar{\hat{\tilde{\mathcal{G}}}}_{22}^T\right)  \vspace{0.1in}\\
  &\!\!\!\!  \d -\alpha e^{2ikL}\left(\hat{\tilde{L}}_{11}-\frac{i}{2}u_0^T(t)\hat{\tilde{G}}_{21}+k\hat{\tilde{G}}_{11}\right)
  \left(\bar{\hat{\tilde{\mathcal{L}}}}_{21}^T-\frac{i\alpha}{2}\bar{\hat{\tilde{\mathcal{G}}}}_{11}^Tv_0^T(t)
  +k\bar{\hat{\tilde{\mathcal{G}}}}_{21}^T\right),
\end{array}
\ene
\bee
\label{f21eq}
\begin{array}{rl}
\tilde{F}_{21}(t,k)= &\!\!\!\!  \d -\frac{i\alpha}{2}\bar{\hat{\tilde{G}}}^T_{11}u_0^T(t)+\frac{i\alpha}{2}v_0^T(t)\hat{\tilde{\mathcal{G}}}_{22}e^{2ikL}
+\l(\tilde{\phi}_{11}-\mathbb{I}\r)\bar{\tilde{\Psi}}^T_{21}(t,\bar{k})
 -\alpha \tilde{\phi}_{12}\l(\bar{\tilde{\Psi}}_{22}^T(t,\bar{k})-\mathbb{I}\r)e^{2ikL}
\vspace{0.1in}\\
= &\!\!\!\!  \d -\frac{i\alpha}{2}\bar{\hat{\tilde{G}}}^T_{11}u_0^T(t)+\frac{i\alpha}{2}v_0^T(t)\hat{\tilde{\mathcal{G}}}_{22}e^{2ikL} \vspace{0.1in}\\
 &\!\!\!\! \d+\left(\hat{\tilde{\mathcal{L}}}_{11}-\frac{i}{2}v_0^T(t)\hat{\tilde{\mathcal{G}}}_{21}+k\hat{\tilde{\mathcal{G}}}_{11}\right)
  \left(\bar{\hat{\tilde{L}}}^T_{21}-\frac{i\alpha}{2}\bar{\hat{\tilde{G}}}^T_{11}u_0^T(t)+k\bar{\hat{\tilde{G}}}^T_{21}\right) \qquad  \vspace{0.1in}\\
  &\!\!\!\! \d -\alpha e^{2ikL}\left(\hat{\tilde{\mathcal{L}}}_{12}-\frac{i}{2}v_0^T(t)\hat{\tilde{\mathcal{G}}}_{22}+k\hat{\tilde{\mathcal{G}}}_{12}\right)
   \left(\bar{\hat{\tilde{L}}}^T_{22}-\frac{i\alpha}{2}\bar{\hat{\tilde{G}}}^T_{12}u_0^T(t)+k\bar{\hat{\tilde{G}}}^T_{22}\right),
   \end{array}
\ene

\vspace{0.1in}
\noindent{\bf Proof.} Similar to the proof of Lemma 4.3 in Ref.~\cite{m3}, we here show Eq.~(\ref{f12e}). We multiply Eq.~(\ref{f12eq}) by $\frac{k}{\Sigma_-}e^{4ik^2(t-t')}$ with $0<t'<t$ and integrate along along $\partial D_1^0$ with respect to $dk$ to have
\bee\label{f12ana}
\begin{array}{l}
\d\int_{\partial D_1^0}\frac{k}{\Sigma_-}e^{4ik^2(t-t')}(\tilde{F}_{12}e^{-2ikL})_-dk=
\d\int_{\partial D_1^0}\frac{i k}{2}e^{4ik^2(t-t')}u_0^T\hat{\tilde{G}}_{22}dk
-\d\int_{\partial D_1^0} k^3e^{4ik^2(t-t')}\hat{\tilde{G}}_{12}\bar{\hat{\tilde{\mathcal{G}}}}_{22}^Tdk  \vspace{0.1in}\\
\qquad\qquad\qquad -\d\int_{\partial D_1^0} k e^{4ik^2(t-t')}\left(\hat{\tilde{L}}_{12}-\frac{i}{2}u_0^T\hat{\tilde{G}}_{22}\right)
\left(\bar{\hat{\tilde{\mathcal{L}}}}_{22}^T-\frac{i\alpha}{2}\bar{\hat{\tilde{\mathcal{G}}}}_{12}^Tv_0^T\right)dk \vspace{0.1in}\\
\qquad\qquad\qquad
\d +\d\int_{\partial D_1^0} \frac{k^2\Sigma_+}{\Sigma_-}e^{4ik^2(t-t')}\left[\left(\hat{\tilde{L}}_{12}-\frac{i}{2}u_0^T\hat{\tilde{G}}_{22}\right)\bar{\hat{\tilde{\mathcal{G}}}}_{22}^T
+\left(\bar{\hat{\tilde{\mathcal{L}}}}_{22}^T-\frac{i\alpha}{2}\bar{\hat{\tilde{\mathcal{G}}}}_{12}^Tv_0^T\right)\hat{\tilde{G}}_{12}\right]dk
    \vspace{0.1in}\\
\qquad\qquad\qquad -\d\int_{\partial D_1^0}\frac{2\alpha k^2}{\Sigma_-}e^{4ik^2(t-t')}\left[\left(\hat{\tilde{L}}_{11}-\frac{i}{2}u_0^T\hat{\tilde{G}}_{21}\right)\bar{\hat{\tilde{\mathcal{G}}}}_{21}^T+
\left(\bar{\hat{\tilde{\mathcal{L}}}}_{21}^T-\frac{i\alpha}{2}\bar{\hat{\tilde{\mathcal{G}}}}_{11}^Tv_0^T\right)\hat{\tilde{G}}_{11}\right]dk,
\end{array}
\ene

To further analyse the above equation, the following identities are introduced
\bee\label{con1}
\d\int_{\partial D_1}ke^{4ik^2(t-t')}\hat{F}(t,k)dk
 =\left\{\begin{array}{l}\d \frac{\pi}{2} F(t, 2t'-t),\quad 0<t'<t, \vspace{0.1in}\\
                        \d\frac{\pi}{4}F(t, t),\quad  0<t'=t,
\end{array}\right.
\ene
and
\bee \label{con2}
\d\int_{\partial D_1^0}\frac{k^2}{\Sigma_-}e^{4ik^2(t-t')}\hat{F}(t,k)dk=
2\d\int_{\partial D_1^0}\frac{k^2}{\Sigma_-}\left[\int_0^{t'}e^{4ik^2(t-t')}\hat{F}(t,2\tau-t)d\tau-\frac{F(t, 2t'-t)}{4ik^2}\right]dk,
\ene
which also holds for the case that $\dfrac{k^2}{\Sigma_-}$ is taken place by $\dfrac{k^2\Sigma_+}{\Sigma_-}$ or $k^2$.

It follows from the first integral on the right hand side (RHS) of Eq.~(\ref{f12ana}) and Eq.~(\ref{con1}) that we find
\bes\bee\label{inte1}
&\d\lim_{t'\to t}\d\int_{\partial D_1^0}\frac{i k}{2}e^{4ik^2(t-t')}u_0^T\hat{\tilde{G}}_{22}dk
=\lim_{t'\to t}\frac{i\pi}{2} u_0^T\tilde{G}_{22}(t, 2t'-t)
=\frac{i\pi}{4}u_0^T\tilde{G}_{22}(t, t),\\
& \label{inte2}
\d\lim_{t'\to t}\d\int_{\partial D_1^0}\frac{i k}{2}e^{4ik^2(t-t')}u_0^T\hat{\tilde{G}}_{22}dk
=\d\int_{\partial D_1^0}\frac{i k}{2}u_0^T\hat{\tilde{G}}_{22}dk
=\frac{i\pi}{8}u_0^T\tilde{G}_{22}(t,t),
\ene\ees

Therefore, we know that the first integral on the RHS of Eq.~(\ref{f12ana}) yields the following two terms
\bee
\d \lim_{t'\to t}\int_{\partial D_1^0}\frac{i k}{2}e^{4ik^2(t-t')}u_0^T\hat{\tilde{G}}_{22}dk
=\int_{\partial D_1^0}\frac{i k}{2}u_0^T\hat{\tilde{G}}_{22}dk\Big|_{(\ref{inte1})}
+\int_{\partial D_1^0}\frac{i k}{2}u_0^T\hat{\tilde{G}}_{22}dk\Big|_{(\ref{inte2})},
\ene

Nowadays we study the second integral on the RHS of Eq.~(\ref{f12ana}). It follows from the second integral on the RHS of Eq.~(\ref{f12ana}) and Eq.~(\ref{con2}) that we have
\bee \label{k3} \begin{array}{rl}
-\d\int_{\partial D_1^0} k^3e^{4ik^2(t-t')}\hat{\tilde{G}}_{12}\bar{\hat{\tilde{\mathcal{G}}}}_{22}^Tdk
=&\!\!\!-2\d\int_{\partial D_1^0} k^3\int_0^{t}e^{4ik^2(\tau-t')}\tilde{G}_{12}(t, 2\tau-t)\bar{\hat{\tilde{\mathcal{G}}}}_{22}^Td\tau dk
\vspace{0.1in}\\
=&\!\!\! -2\d\int_{\partial D_1^0} k^3\left[\int_0^{t'}e^{4ik^2(\tau-t')}\tilde{G}_{12}(t, 2\tau-t)d\tau-\frac{\tilde{G}_{12}(t, 2t'-t)}{4ik^2}\right]\bar{\hat{\tilde{\mathcal{G}}}}_{22}^Tdk,
\end{array}
\ene
Therefore we take the limit $t'\to t$ of Eq.~(\ref{k3}) to get
\bee\no
-\lim_{t'\to t}\d\int_{\partial D_1^0} k^3e^{4ik^2(t-t')}\hat{\tilde{G}}_{12}\bar{\hat{\tilde{\mathcal{G}}}}_{22}^Tdk
=-\d\int_{\partial D_1^0} k^3\hat{\tilde{G}}_{12}\bar{\hat{\tilde{\mathcal{G}}}}_{22}^Tdk
+\d\int_{\partial D_1^0}\frac{k}{2i}u_0^T\bar{\hat{\tilde{\mathcal{G}}}}_{22}^Tdk
\ene

Finally, following the proof in Ref.~\cite{m3} we can show the limits $t'\to t$ of the rest three integrals (i.e., the third, fourth and fifth integrals) of Eq.~(\ref{f12ana}) can be deduced by simply making the limit $t'\to t$ inside the every integral, that is, no additional terms
arise in these integrals. For example,
\bee \no
\lim_{t'\to t}\d\int_{\partial D_1^0} k e^{4ik^2(t-t')}\left(\hat{\tilde{L}}_{12}-\frac{i}{2}u_0^T\hat{\tilde{G}}_{22}\right)
\left(\bar{\hat{\tilde{\mathcal{L}}}}_{22}^T-\frac{i\alpha}{2}\bar{\hat{\tilde{\mathcal{G}}}}_{12}^Tv_0^T\right)dk \\
=\d\int_{\partial D_1^0} k \left(\hat{\tilde{L}}_{12}-\frac{i}{2}u_0^T\hat{\tilde{G}}_{22}\right)
\left(\bar{\hat{\tilde{\mathcal{L}}}}_{22}^T-\frac{i\alpha}{2}\bar{\hat{\tilde{\mathcal{G}}}}_{12}^Tv_0^T\right)dk.
\no\ene
Thus we complete the proof of Eq.~(\ref{f12e}). Similarly, we can show that Eqs.~(\ref{f21e}), (\ref{f12e+}) and (\ref{f21e+}) also hold.
$\square$

\vspace{0.1in}
\noindent {\bf Theorem  5.3.} {\it  \it Let $q_{0j}(x)=q_j(x,t=0)=0,\, j=1,0,-1$ be the initial data of Eq.~(\ref{pnls}) on the interval $x\in [0, L]$ and  $T<\infty$. For the Dirichlet problem, the boundary data $u_{0j}(t)$ and $v_{0j}(t)\, (j=1,0,-1)$ on the interval $t\in [0, T)$ are sufficiently smooth and compatible with the initial data $q_{j0}(x)\, (j=1,0,-1)$ at the points $(x_2, t_2)=(0, 0)$ and  $(x_3, t_3)=(L, 0)$, respectively. For the Neumann problem, the boundary data $u_{1j}(t)$ and $v_{1j}(t)\, (j=1,0,-1)$ on the interval $t\in [0, T)$ are sufficiently smooth and compatible with the initial data $q_{0j}(x)\, (j=1,0,-1)$ at the points $(x_2, t_2)=(0, 0)$ and  $(x_3, t_3)=(L, 0)$, respectively. For simplicity, let $n_{33,44}(\mathbb{S})(k)$ have no zeros in the domain $D_1$. Then the spectral functions $S(k)$ and $S_L(k)$ are defined by Eqs.~(\ref{skm}) and (\ref{slm}) with $\Psi(t,k)$ and $\phi(t,k)$ given by Eq.~(\ref{psiglm}) and (\ref{phiglm}).

(i) For the given Dirichlet boundary values $u_0(t)$ and $v_0(t)$, the unknown Neumann boundary values $u_1(t)$ and $v_1(t)$ are given by
\bes
\bee
& \label{u1}
\begin{array}{rl}
u_1^T(t)=&\!\!\left(\!\!\begin{array}{cc}
 u_{11}(t) &  u_{10}(t) \vspace{0.05in}\cr \beta u_{10}(t) & u_{1-1}(t)\end{array} \!\!\right) \vspace{0.1in}\\
 \!\!=&\!\!\! \d\frac{4}{i\pi}\int_{\partial D_1^0}\left\{\frac{\Sigma_+}{\Sigma_-}\left[k^2\hat{\tilde{G}}_{12}(t,t)+
\frac{i}{2}u_0^T(t)\right]
-\d\frac{2\alpha}{\Sigma_-}\left[k^2\bar{\hat{\tilde{\mathcal{G}}}}_{21}^T(t, t)+\frac{i\alpha}{2}v_0^T(t)\right]\right. \vspace{0.1in}\\
&\qquad\qquad \left.+\d\frac{i k}{2}u_0^T\hat{\tilde{G}}_{22}+\frac{k}{2i}u_0^T\bar{\hat{\tilde{\mathcal{G}}}}_{22}^T+\frac{k}{\Sigma_-}[\tilde{F}_{12}(t,k)e^{-2ikL}]_-\right\}dk,
\end{array} \vspace{0.1in}\\
&\label{v1}
\begin{array}{rl}
v_1^T(t)=&\!\!\left(\!\!\begin{array}{cc}
 v_{11}(t) &  v_{10}(t) \vspace{0.05in}\cr \beta v_{10}(t) & v_{1-1}(t)\end{array} \!\!\right)\vspace{0.1in}\\
 =&\!\!\! \d\frac{4}{i\pi}\int_{\partial D_1^0}\left\{-\frac{\Sigma_+}{\Sigma_-}\left[k^2\hat{\tilde{\mathcal{G}}}_{12}(t,t)+\frac{i}{2}v_0^T(t)\right]
+\d\frac{2\alpha}{\Sigma_-}\left[k^2\bar{\hat{\tilde{G}}}_{21}^T(t, t)+\frac{i\alpha}{2}u_0^T(t)\right]\right. \vspace{0.1in}\\
&\qquad\qquad\left.+\d\frac{i k}{2}v_0^T\hat{\tilde{\mathcal{G}}}_{22}+\frac{k}{2i}v_0^T\bar{\hat{\tilde{G}}}_{22}^T
+\frac{\alpha k}{\Sigma_-}\tilde{F}_{21-}(t,k)\right\}dk,
\end{array}
\ene\ees

(ii) For the given Neumann boundary values $u_1(t)$ and $v_1(t)$, the unknown Dirichlet boundary values $u_0(t)$ and $v_0(t)$ are given by
\bes\bee
&\label{u0}
\begin{array}{rl}
u_0^T(t)=\left(\!\!\begin{array}{cc}
 u_{01}(t) &  u_{00}(t) \vspace{0.05in}\cr \beta u_{00}(t) & u_{0-1}(t)\end{array} \!\!\right)
 =&\!\!\!\d\frac{2}{\pi}
\d\int_{\partial D_1^0}\left[\frac{\Sigma_+}{\Sigma_-}\hat{\tilde{L}}_{12}-\frac{2\alpha}{\Sigma_-}\bar{\hat{\tilde{\mathcal{L}}}}_{21}^T
+\frac{1}{\Sigma_-}\left(\tilde{F}_{12}(t, k)e^{-2ikL}\right)_+\right]dk,
\end{array}  \vspace{0.1in}\\
&\label{v0}
\begin{array}{rl}
v_0^T(t)=\left(\!\!\begin{array}{cc}
 v_{01}(t) &  v_{00}(t) \vspace{0.05in}\cr \beta v_{00}(t) & v_{0-1}(t)\end{array} \!\!\right)
 = &\!\!\!\d\frac{2}{\pi}\int_{\partial D_1^0}\left[
\frac{2\alpha}{\Sigma_-}\bar{\hat{\tilde{L}}}^T_{12}-\frac{1}{\Sigma_-}\hat{\tilde{\mathcal{L}}}_{21}
+\frac{\alpha}{\Sigma_-}\tilde{F}_{21+}(t, k)\right]dk,\qquad\quad
\end{array}
\ene\ees
where $\tilde{F}_{12}(t,k)$ and $\tilde{F}_{21}(t,k)$ are defined by Eqs.~(\ref{f12eq}) and (\ref{f21eq}).}

\vspace{0.1in}
\noindent {\bf Proof.} According to the global relation (\ref{gr}) and Proposition 5.1, we can show that the spectral functions $S(k)$ and $S_L(k)$ are defined by Eqs.~(\ref{skm}) and (\ref{slm}) with $\Psi(t,k)$ and $\phi(t,k)$ given by Eq.~(\ref{psiglm}) and (\ref{phiglm}).

 (i) we study the Dirichlet problem. It follows from the global relation (\ref{gr}) with the vanishing initial data
\bee
 c(t,k)=\left(\begin{array}{cc}
  \tilde{c}_{11}(t,k) & \tilde{c}_{12}(t,k)  \vspace{0.1in}\\ \tilde{c}_{21}(t,k) & \tilde{c}_{22}(t,k) \end{array}\right)
  =\mu_2(0, t,k)e^{ikL\hat{\sigma}_4}\mu_3^{-1}(L, t,k),
\ene
that we have
\bes\bee
\label{c12}
&\tilde{c}_{12}(t,k)=\left(\begin{array}{cc}
  c_{13}(t,k) & c_{14}(t,k)  \vspace{0.1in}\\ c_{23}(t,k) & c_{24}(t,k) \end{array}\right)
  =-\alpha\tilde{\Psi}_{11}\bar{\tilde{\phi}}_{21}^T(t,\bar{k})e^{2ikL}+\tilde{\Psi}_{12}\bar{\tilde{\phi}}_{22}^T(t,\bar{k}),
\vspace{0.1in}\\
\label{c21} &
\tilde{c}_{21}(t,k)=\left(\begin{array}{cc}
  c_{31}(t,k) & c_{32}(t,k)  \vspace{0.1in}\\ c_{41}(t,k) & c_{42}(t,k) \end{array}\right)=\tilde{\Psi}_{21}\bar{\tilde{\phi}}_{11}^T(t,\bar{k})-\alpha\tilde{\Psi}_{22}\bar{\tilde{\phi}}_{12}^T(t,\bar{k})e^{-2ikL},
\ene\ees

We substitute Eqs.~(\ref{psiglmg}) and (\ref{phiglmg}) into Eq.~(\ref{c12}) to have
\bee\label{c12g}
-\hat{\tilde{L}}_{12}+\alpha\bar{\hat{\tilde{\mathcal{L}}}}_{21}^Te^{2ikL}=k\hat{\tilde{G}}_{12}
-\alpha k\bar{\hat{\tilde{\mathcal{G}}}}_{21}^Te^{2ikL}+\tilde{F}_{12}(t,k)-\tilde{c}_{12}(t,k),
\ene
where $\tilde{F}_{12}(t,k)$ is given by Eq.~(\ref{f12eq}).

Eq.~(\ref{c12g}) with $k\to -k$ yields
\bee\label{c12gk}
-\hat{\tilde{L}}_{12}+\alpha\bar{\hat{\tilde{\mathcal{L}}}}_{21}^Te^{-2ikL}=-k\hat{\tilde{G}}_{12}
+\alpha k\bar{\hat{\tilde{\mathcal{G}}}}_{21}^Te^{-2ikL}+\tilde{F}_{12}(t, -k)-\tilde{c}_{12}(t, -k),
\ene

It follows from Eqs.~(\ref{c12g}) and (\ref{c12gk}) that we find
\bee
\label{c12g1}
\d\hat{\tilde{L}}_{12}=\frac{k\Sigma_+(k)}{\Sigma_-(k)}\hat{\tilde{G}}_{12}
-\frac{2\alpha k}{\Sigma_-(k)}\bar{\hat{\tilde{\mathcal{G}}}}_{21}^T+\frac{1}{\Sigma_-(k)}\left\{[\tilde{F}_{12}(t, k)-\tilde{c}_{12}(t, k)]e^{-2ikL}\right\}_-.
\ene

We multiply Eq.~(\ref{c12g1})  by $k e^{4ik^2(t-t')}$ with $0<t'<t$ and integrate them along $\partial D_1^0$ with respect to
$dk$, respectively to yield
\bee
\label{c12g22}
\begin{array}{rl}\d\int_{\partial D_1^0}k e^{4ik^2(t-t')}\hat{\tilde{L}}_{12}dk=
&\!\!\!\d\int_{\partial D_1^0} e^{4ik^2(t-t')}\frac{k^2\Sigma_+}{\Sigma_-}\hat{\tilde{G}}_{12}dk
-\d\int_{\partial D_1^0}e^{4ik^2(t-t')}\frac{2\alpha k^2}{\Sigma_-}\bar{\hat{\tilde{\mathcal{G}}}}_{21}^Tdk \vspace{0.1in}\\
&\!\!\!+\d\int_{\partial D_1^0} e^{4ik^2(t-t')}\frac{k}{\Sigma_-}\l[\tilde{F}_{12}(t,k)e^{-2ikL}\r]_-dk,
\end{array}
\ene
where we have used
\bee\no
\d\int_{\partial D_1^0} k e^{4ik^2(t-t')}\tilde{c}_{12-}(t, k)dk
=\d\int_{\partial D_1^0}k e^{4ik^2(t-t')}\l[\tilde{c}_{12}(t, k)e^{-2ikL}\r]_-dk=0
\ene
since these two matrix-valued functions
\bee\no
k e^{4ik^2(t-t')}\tilde{c}_{12-}(t, k),\qquad k e^{4ik^2(t-t')}(\tilde{c}_{12}(t, k)e^{-2ikL})_-
\ene
are bounded and analytic in $D_1^0$.

By using these conditions given by Eqs.~(\ref{con1}) and (\ref{con2}), Eq.~(\ref{c12g22}) can be reduced to
\bee
\label{l22}
\begin{array}{rl}
\d\frac{\pi}{2}\tilde{L}_{12}(t, 2t'-t)=&\!\!\! 2\d\int_{\partial D_1^0}\frac{k^2\Sigma_+}{\Sigma_-}\left[\int_0^{t'}e^{4ik^2(t-t')}
\tilde{G}_{12}(t,2\tau-t)d\tau-\frac{\tilde{G}_{12}(t, 2t'-t)}{4ik^2}\right]dk \vspace{0.1in}\\
&-4\d\int_{\partial D_1^0}\frac{\alpha k^2}{\Sigma_-}\left[\int_0^{t'}e^{4ik^2(t-t')}\bar{\tilde{\mathcal{G}}}_{21}^T(t,2\tau-t)d\tau-\frac{\bar{\tilde{\mathcal{G}}}_{21}^T(t, 2t'-t)}{4ik^2}\right]dk \vspace{0.1in}\\
&+\d\int_{\partial D_1^0}\frac{k}{\Sigma_-}e^{4ik^2(t-t')}\l[\tilde{F}_{12}(t, k)e^{-2ikL}\r]_-dk,
\end{array}
\ene

We consider the limits $t'\to t$ of Eq.~(\ref{l22}) with the initial data (\ref{lgi}) and Proposition 5.2 to find
\bee\label{l22a}
\begin{array}{rl}
\d\frac{\pi}{2}\tilde{L}_{12}(t, t)=&\!\!\! 2\d\lim_{t'\to t}\int_{\partial D_1^0}\frac{k^2\Sigma_+}{\Sigma_-}\left[\int_0^{t'}e^{4ik^2(t-t')}
\tilde{G}_{12}(t,2\tau-t)d\tau-\frac{\tilde{G}_{12}(t, 2t'-t)}{4ik^2}\right]dk \vspace{0.1in}\\
&-4\d\lim_{t'\to t}\int_{\partial D_1^0}\frac{\alpha k^2}{\Sigma_-}\left[\int_0^{t'}e^{4ik^2(t-t')}\bar{\tilde{\mathcal{G}}}_{21}^T(t,2\tau-t)d\tau-\frac{\bar{\tilde{\mathcal{G}}}_{21}^T(t, 2t'-t)}{4ik^2}\right]dk \vspace{0.1in}\\
&+\d\lim_{t'\to t}\int_{\partial D_1^0}\frac{k}{\Sigma_-}e^{4ik^2(t-t')}\l[\tilde{F}_{12}(t, k)e^{-2ikL}\r]_-dk \vspace{0.1in}\\
=&\!\!\! \d\int_{\partial D_1^0}\left\{\frac{\Sigma_+}{\Sigma_-}\left[k^2\hat{\tilde{G}}_{12}(t,t)+\frac{i}{2}\tilde{G}_{12}(t, t)\right]
-\d\frac{2\alpha}{\Sigma_-}\left[k^2\bar{\hat{\tilde{\mathcal{G}}}}_{21}^T(t, t)+\frac{i}{2}\bar{\tilde{\mathcal{G}}}_{21}^T(t, t)\right]\right. \vspace{0.1in}\\
&\qquad\left.+\d\frac{i k}{2}u_0^T\hat{\tilde{G}}_{22}+\frac{ k}{2i}u_0^T\bar{\hat{\tilde{\mathcal{G}}}}_{22}^T+\frac{k}{\Sigma_-}\l[\tilde{F}_{12}(t,k)e^{-2ikL}\r]_-\right\}dk,
\end{array}
\ene
Since the initial data (\ref{lgi}) are of the form
\bee \label{gl22b}
\tilde{L}_{12}(t, t)=\frac{i}{2}u_1^T(t)=\frac{i}{2}\left(\begin{array}{cc}
    u_{11}(t) & u_{10}(t) \vspace{0.1in}\\
    \beta u_{10}(t) & u_{1-1}(t) \end{array}\right),
\ene
then we have Eq.~(\ref{u1}) by using Eqs.~(\ref{l22a}) and (\ref{gl22b}).

To show Eq.~(\ref{v1}) we rewrite Eq.~(\ref{c21}) in the form
\bee \label{c21g}
\bar{\tilde{c}}^T_{21}(t,\bar{k})=\left(\begin{array}{cc}
  \bar{c}_{31}(t,\bar{k}) & \bar{c}_{41}(t,\bar{k})  \vspace{0.1in}\\
  \bar{c}_{32}(t,\bar{k}) & \bar{c}_{42}(t,\bar{k}) \end{array}\right)=\tilde{\phi}_{11}\bar{\tilde{\Psi}}^T_{21}(t,\bar{k})
 -\alpha \tilde{\phi}_{12}\bar{\tilde{\Psi}}_{22}^T(t,\bar{k})e^{2ikL},
\ene

We substitute Eqs.~(\ref{psiglmg}) and (\ref{phiglmg}) into Eq.~(\ref{c21g}) to have
\bee\label{c21gg}
-\bar{\hat{\tilde{L}}}^T_{21}+\alpha\hat{\tilde{\mathcal{L}}}_{12}e^{2ikL}=k\bar{\hat{\tilde{G}}}^T_{21}
-\alpha k\hat{\tilde{\mathcal{G}}}_{12}e^{2ikL}+\tilde{F}_{21}(t,k)-\bar{\tilde{c}}^T_{21}(t,\bar{k}),
\ene
where $\tilde{F}_{21}(t,k)$ is given  by Eq.~(\ref{f21eq}).

Eq.~(\ref{c21gg}) with $k\to -k$ yields
\bee\label{c21gk}
-\bar{\hat{\tilde{L}}}^T_{21}+\alpha\hat{\tilde{\mathcal{L}}}_{12}e^{-2ikL}=-k\bar{\hat{\tilde{G}}}^T_{21}
+\alpha k\hat{\tilde{\mathcal{G}}}_{12}e^{-2ikL}+\tilde{F}_{21}(t,-k)-\bar{\tilde{c}}^T_{21}(t,-\bar{k}),
\ene

It follows from Eqs.~(\ref{c21gg}) and (\ref{c21gk}) that we have
\bee
\label{c21g1}
\d \alpha\hat{\tilde{\mathcal{L}}}_{12}=\frac{2k}{\Sigma_-}\bar{\hat{\tilde{G}}}^T_{21}
-\frac{\alpha k\Sigma_+}{\Sigma_-}\hat{\tilde{\mathcal{G}}}_{12}+\frac{1}{\Sigma_-}[\tilde{F}_{21}(t, k)-\bar{\tilde{c}}_{21}^T(t, \bar{k})]_-
\ene

We multiply Eq.~(\ref{c21g1})  by $k e^{4ik^2(t-t')}$ with $0<t'<t$, integrate them along $\partial D_1^0$ with respect to
$dk$, and use these conditions given by Eqs.~(\ref{con1}) and (\ref{con2}) to yield
\bee
\label{2l2}
\begin{array}{rl}
\d\frac{\alpha\pi}{2}\tilde{\mathcal{L}}_{12}(t, 2t'-t)=&\!\!\! -2\alpha\d\int_{\partial D_1^0}\frac{k^2\Sigma_+}{\Sigma_-}\left[\int_0^{t'}e^{4ik^2(t-t')}
\tilde{\mathcal{G}}_{12}(t,2\tau-t)d\tau-\frac{\tilde{\mathcal{G}}_{12}(t, 2t'-t)}{4ik^2}\right]dk \vspace{0.1in}\\
&+4\d\int_{\partial D_1^0}\frac{ k^2}{\Sigma_-}\left[\int_0^{t'}e^{4ik^2(t-t')}\bar{\tilde{G}}_{21}^T(t,2\tau-t)d\tau-\frac{\bar{\tilde{G}}_{21}^T(t, 2t'-t)}{4ik^2}\right]dk \vspace{0.1in}\\
&+\d\int_{\partial D_1^0}\frac{k}{\Sigma_-}e^{4ik^2(t-t')}\tilde{F}_{21-}(t, k)dk,
\end{array}
\ene
where we have used
\bee\no
\d\int_{\partial D_1^0} \frac{k}{\Sigma_-}e^{4ik^2(t-t')}\bar{\tilde{c}}^T_{21-}(t, \bar{k})dk=0
\ene
since the matrix-valued function
\bee\no
\frac{k}{\Sigma_-}e^{4ik^2(t-t')}\bar{\tilde{c}}^T_{21-}(t, \bar{k})
\ene
is bounded and analytic in $D_1^0$.

We consider the limit $t'\to t$ of Eq.~(\ref{2l2}) with the initial data (\ref{lgi}) and Proposition 5.2 to obtain
\bee\label{2l2a}
\begin{array}{rl}
\d\frac{\alpha\pi}{2}\tilde{\mathcal{L}}_{12}(t, t)=&\!\!\! \d -\alpha\int_{\partial D_1^0}\left\{\frac{\Sigma_+}{\Sigma_-}\left[k^2\hat{\tilde{\mathcal{G}}}_{12}(t,t)+\frac{i}{2}\tilde{\mathcal{G}}_{12}(t, t)\right]
+\d\frac{2}{\Sigma_-}\left[k^2\bar{\hat{\tilde{G}}}_{21}^T(t, t)+\frac{i}{2}\bar{\tilde{G}}_{21}^T(t, t)\right]\right. \vspace{0.1in}\\
&\qquad\quad\left.+\d\frac{i\alpha k}{2}v_0^T\hat{\tilde{\mathcal{G}}}_{22}+\frac{ \alpha k}{2i}v_0^T\bar{\hat{\tilde{G}}}_{22}^T+\frac{k}{\Sigma_-}\tilde{F}_{21-}(t,k)\right\}dk,
\end{array}
\ene

Since the initial data are of the form
\bee \label{2l2b}
\tilde{\mathcal{G}}_{12}(t, t)=\frac{i}{2}v_1^T(t)=\frac{i}{2}\left(\begin{array}{cc}
    v_{11}(t) & v_{10}(t) \vspace{0.1in}\\
    \beta v_{10}(t) & v_{1-1}(t) \end{array}\right),
\ene
then we have Eq.~(\ref{v1}) by using Eqs.~(\ref{2l2a}) and (\ref{2l2b}).

\vspace{0.1in}

 (ii) We now consider the Neumann problem. It follows from Eqs~(\ref{c12g}), (\ref{c12gk}), (\ref{c21gg}) and (\ref{c21gk}) that we have
\bes\bee
&\label{c12g+}
\d\hat{\tilde{G}}_{12}=\frac{1}{k\Sigma_-(k)}\left\{\Sigma_+(k)\hat{\tilde{L}}_{12}-2\alpha \bar{\hat{\tilde{\mathcal{L}}}}_{21}^T+\left[(\tilde{F}_{12}(t, k)-\tilde{c}_{12}(t, k))e^{-2ikL}\right]_+\right\}, \\
&\label{c21g+}
\d\hat{\tilde{\mathcal{G}}}_{12}=\frac{\alpha}{k\Sigma_-(k)}
\left\{2\bar{\hat{\tilde{L}}}^T_{12}-\alpha \Sigma_+(k)\hat{\tilde{\mathcal{L}}}_{12}+\left[\tilde{F}_{21}(t, k)-\bar{\tilde{c}}^T_{21}(t, \bar{k})\right]_+\right\}.\qquad\quad
\ene\ees

We multiply Eqs.~(\ref{c12g+}) and (\ref{c21g+}) by $k e^{4ik^2(t-t')}$ with $0<t'<t$, integrate them along $\partial D_1^0$ with respect to
$dk$, and use these conditions given by Eqs.~(\ref{con1}) and (\ref{con2}) to yield
\bes\bee
&\label{g12+1}
\begin{array}{rl}
\d\frac{\pi}{2}\tilde{G}_{12}(t, 2t'-t)=&\!\!\!
\d\int_{\partial D_1^0}\frac{2\Sigma_+}{\Sigma_-}\left[\int_0^{t'}e^{4ik^2(t-t')}
\tilde{L}_{12}(t,2\tau-t)d\tau-\frac{\tilde{L}_{12}(t, 2t'-t)}{4ik^2}\right]dk \vspace{0.1in}\\
&-\d\int_{\partial D_1^0}\frac{4\alpha}{\Sigma_-}\left[\int_0^{t'}e^{4ik^2(t-t')}\bar{\tilde{\mathcal{L}}}_{21}^T(t,2\tau-t)d\tau-\frac{\bar{\tilde{\mathcal{L}}}_{21}^T(t, 2t'-t)}{4ik^2}\right]dk \vspace{0.1in}\\
&+\d\int_{\partial D_1^0}\frac{1}{\Sigma_-}e^{4ik^2(t-t')}\l[\tilde{F}_{12}(t, k)e^{-2ikL}\r]_+dk,
\end{array}  \\
&\label{g12+2}
\begin{array}{rl}
\d\frac{\pi}{2}\tilde{\mathcal{G}}_{12}(t, 2t'-t)=&\!\!\!
\d\int_{\partial D_1^0}\frac{4\alpha}{\Sigma_-}\left[\int_0^{t'}e^{4ik^2(t-t')}
\bar{\tilde{L}}^T_{12}(t,2\tau-t)d\tau-\frac{\bar{\tilde{L}}^T_{12}(t, 2t'-t)}{4ik^2}\right]dk \vspace{0.1in}\\
&-\d\int_{\partial D_1^0}\frac{2 }{\Sigma_-}\left[\int_0^{t'}e^{4ik^2(t-t')}\tilde{\mathcal{L}}_{21}(t,2\tau-t)d\tau-\frac{\tilde{\mathcal{L}}_{21}(t, 2t'-t)}{4ik^2}\right]dk \vspace{0.1in}\\
&+\d\int_{\partial D_1^0}\frac{\alpha}{\Sigma_-}e^{4ik^2(t-t')}\tilde{F}_{21+}(t, k)dk,
\end{array}
\ene\ees
where we have used
\bee\no
\d\int_{\partial D_1^0} \frac{1}{\Sigma_-}e^{4ik^2(t-t')}(\tilde{c}_{12}(t, k)e^{-2ikL})_+dk=\d\int_{\partial D_1^0} \frac{1}{\Sigma_-}e^{4ik^2(t-t')}\bar{\tilde{c}}^T_{21+}(t, \bar{k})dk=0
\ene
since the matrix-valued functions
\bee\no
\frac{1}{\Sigma_-}e^{4ik^2(t-t')}(\tilde{c}_{12}(t, k)e^{-2ikL})_+, \quad \frac{1}{\Sigma_-}e^{4ik^2(t-t')}\bar{\tilde{c}}^T_{21+}(t, \bar{k})
\ene
are bounded and analytic in $D_1^0$.

We study the limits $t'\to t$ of Eqs.~(\ref{g12+1}) and (\ref{g12+2}) with the initial data (\ref{lgi}) and Proposition 5.2 to find
\bes\bee
&\label{g12g1}
\begin{array}{rl}
\d\frac{\pi}{2}\tilde{G}_{12}(t, t)=&\!\!\!
\d\int_{\partial D_1^0}\left[\frac{\Sigma_+}{\Sigma_-}\hat{\tilde{L}}_{12}-\frac{2\alpha}{\Sigma_-}\bar{\hat{\tilde{\mathcal{L}}}}_{21}^T
+\frac{1}{\Sigma_-}(\tilde{F}_{12}(t, k)e^{-2ikL})_+\right]dk,
\end{array}  \\
&\label{g12g2}
\begin{array}{rl}
\d\frac{\pi}{2}\tilde{\mathcal{G}}_{12}(t, t)=&\!\!\!
\d\int_{\partial D_1^0}\left[\frac{2\alpha}{\Sigma_-}\bar{\hat{\tilde{L}}}^T_{12}
-\frac{1}{\Sigma_-}\hat{\tilde{\mathcal{L}}}_{21}+\frac{\alpha}{\Sigma_-}\tilde{F}_{21+}(t, k)\right]dk,
\end{array}
\ene\ees
Since the initial data are of the form
\bee \label{l22b}
\begin{array}{l}
\tilde{G}_{12}(t, t)=u_0^T(t)=\left(\begin{array}{cc}
     u_{01}(t) & u_{00}(t) \vspace{0.1in}\\
    \beta u_{00}(t) & u_{0-1}(t) \end{array}\right), \vspace{0.1in} \\
\tilde{\mathcal{G}}_{12}(t, t)=v_0^T(t)=\left(\begin{array}{cc}
     v_{01}(t) & v_{00}(t) \vspace{0.1in}\\
    \beta v_{00}(t) & v_{0-1}(t) \end{array}\right),
\end{array}
\ene
then we find Eqs.~(\ref{u0}) and (\ref{v0}) by Eqs.~(\ref{g12g1}) and (\ref{g12g2}). This completes the proof of the Theorem.
$\square$

\vspace{0.1in}
\noindent \subsection*{\it 5.2. The equivalence of the two distinct representations}
\vspace{0.1in}

\quad We here show that the above-mentioned  GLM  representations for the Dirichlet and Neumann boundary data in Theorem 5.3 are equivalent to ones in Theorem 4.2.

\vspace{0.1in}
{\it Case i. From the Dirichlet boundary data to the Neumann boundary data}
\vspace{0.1in}

Eqs.~(\ref{psiglmg}) and (\ref{phiglmg}) imply that
\bee \label{mpsi}
\hat{\tilde{G}}_{12}=\frac{1}{2k}\tilde{\Psi}_{12-},\quad
\hat{\tilde{\mathcal{G}}}_{12}=\frac{1}{2k}\tilde{\mathcal{\phi}}_{12-},\quad
\hat{\tilde{G}}_{22}=\frac{1}{2k}\tilde{\Psi}_{22-},\quad
\hat{\tilde{\mathcal{G}}}_{22}=\frac{1}{2k}\tilde{\mathcal{\phi}}_{22-},
\ene

The substitution of Eqs.~(\ref{f12eq}) and (\ref{mpsi}) into Eq.~(\ref{u1}) yields
\bee
& \label{u1x}
\begin{array}{rl}
u_1^T(t)=&\!\!\! \d\frac{4}{i\pi}\int_{\partial D_1^0}\left\{\frac{\Sigma_+}{\Sigma_-}\left[k^2\hat{\tilde{G}}_{12}(t,t)
 +\frac{i}{2}u_0^T(t)\right]
-\d\frac{2\alpha}{\Sigma_-}\left[k^2\bar{\hat{\tilde{\mathcal{G}}}}_{21}^T(t, t)+\frac{i\alpha}{2}v_0^T(t)\right]\right. \vspace{0.1in}\\
&\left.+\d\frac{i k}{2}u_0^T\hat{\tilde{G}}_{22}+\frac{ k}{2i}u_0^T\bar{\hat{\tilde{\mathcal{G}}}}_{22}^T+\frac{k}{\Sigma_-}\l[\tilde{F}_{12}(t,k)e^{-2ikL}\r]_-\right\}dk \vspace{0.1in}\\
=&\!\!\! \d\frac{4}{i\pi}\int_{\partial D_1^0}\left\{\frac{\Sigma_+}{\Sigma_-}\left[k^2\hat{\tilde{G}}_{12}(t,t)
 +\frac{i}{2}u_0^T(t)\right]
-\d\frac{2\alpha}{\Sigma_-}\left[k^2\bar{\hat{\tilde{\mathcal{G}}}}_{21}^T(t, t)+\frac{i\alpha}{2}v_0^T(t)\right]\right. \vspace{0.1in}\\
&\left.+\d ik u_0^T\hat{\tilde{G}}_{22}+\frac{ k}{2i}u_0^T\bar{\hat{\tilde{\mathcal{G}}}}_{22}^T+\frac{k}{\Sigma_-}\left[\tilde{\Psi}_{12}(\bar{\tilde{\phi}}_{22}^T-\mathbb{I})e^{-2ikL}
-\alpha(\tilde{\Psi}_{11}-\mathbb{I})\bar{\tilde{\phi}}_{21}^T\right]_-\right\}dk \vspace{0.1in}\\
=&\!\!\! \d\int_{\partial D_1^0}\left\{\frac{2\Sigma_+}{i\pi\Sigma_-}
 \left[k\tilde{\Psi}_{12-}+iu_0^T(t)\right]
+\d\frac{4i}{\pi\Sigma_-}\left[\alpha k\bar{\tilde{\mathcal{\phi}}}^T_{21-}+i v_0^T(t)\right]+\frac{2}{\pi}u_0^T\tilde{\Psi}_{22}-\frac{1}{\pi}u_0^T\bar{\tilde{\phi}}_{22}^T\right. \vspace{0.1in}\\
&\left.\d +\frac{4k}{i\pi\Sigma_-}\left[\tilde{\Psi}_{12}(\bar{\tilde{\phi}}_{22}^T-\mathbb{I})e^{-2ikL}
-\alpha(\tilde{\Psi}_{11}-\mathbb{I})\bar{\tilde{\phi}}_{21}^T\right]_- \right\}dk,
\end{array}
\ene
Since
\bee \label{pp}
\begin{array}{c}
\tilde{\Psi}_{11}=\left(\begin{array}{cc}
    \Psi_{11} & \Psi_{12} \vspace{0.1in}\\
    \Psi_{21} & \Psi_{22} \end{array}\right),\quad
    \tilde{\Psi}_{12}=\left(\begin{array}{cc}
    \Psi_{13} & \Psi_{14} \vspace{0.1in}\\
    \Psi_{23} & \Psi_{24} \end{array}\right), \quad
\tilde{\Psi}_{22}=\left(\begin{array}{cc}
    \Psi_{33} & \Psi_{34} \vspace{0.1in}\\
    \Psi_{43} & \Psi_{44} \end{array}\right), \vspace{0.1in}\\
\bar{\tilde{\phi}}^T_{21}=\left(\begin{array}{cc}
    \bar{\phi}_{31} & \bar{\phi}_{41} \vspace{0.1in}\\
    \bar{\phi}_{32} & \bar{\phi}_{42} \end{array}\right),\quad
\bar{\tilde{\phi}}^T_{22}=\left(\begin{array}{cc}
    \bar{\phi}_{33} & \bar{\phi}_{43} \vspace{0.1in}\\
    \bar{\phi}_{34} & \bar{\phi}_{44} \end{array}\right),
\end{array}
\ene
and the integrand in Eq.~(\ref{u1x}) is an odd function about $k$, which makes sure that the contour $\partial D_1^0$ can be replaced by
$\partial D_3^0$, thus we can find the same Neumann boundary data $u_{1j}(t),\, j=1,0,-1$ at $x=0$ given by Eqs.~(\ref{u11})-(\ref{u1-1}) from Eqs.~(\ref{u1x}) and (\ref{pp}). Similarly, we can also find the Neumann boundary data $v_{1j}(t),\,j=1,0,-1$ at $x=L$ given by Eqs.~(\ref{v11})-(\ref{v1-1}) from Eq.~(\ref{v1}).

\vspace{0.1in}
{\it Case ii. From the  Neumann  boundary data to the Dirichlet boundary data}
\vspace{0.1in}

It follows from Eqs.~(\ref{psiglmg}) and (\ref{phiglmg}) that we have
\bee\label{ll}
\hat{\tilde{L}}_{12}=\frac{1}{2}\tilde{\Psi}_{12+}(t,k)+\frac{i}{2}u_0^T\hat{\tilde{G}}_{22},\qquad
\bar{\hat{\tilde{\mathcal{L}}}}^T_{21}=\frac{1}{2}\bar{\tilde{\phi}}^T_{21+}(t,k)+\frac{i}{2}\bar{\hat{\tilde{\mathcal{G}}}}^T_{11}v_0^T,
\ene
The substitution of Eqs.~(\ref{ll}) and (\ref{f12eq}) into Eq.~(\ref{u0}) yields
\bee\label{u0x}
\begin{array}{rl}
u_0^T(t)=&\!\!\!\d\frac{2}{\pi}
\d\int_{\partial D_1^0}\left[\frac{\Sigma_+}{\Sigma_-}\hat{\tilde{L}}_{12}-\frac{2\alpha}{\Sigma_-}\bar{\hat{\tilde{\mathcal{L}}}}_{21}^T
+\frac{1}{\Sigma_-}\left(\tilde{F}_{12}(t, k)e^{-2ikL}\right)_+\right]dk \vspace{0.1in}\\
=&\!\!\!\d
\d\int_{\partial D_1^0}\left\{\frac{\Sigma_+}{\pi\Sigma_-}\tilde{\Psi}_{12+}\!-\!\frac{2\alpha}{\pi\Sigma_-}\bar{\tilde{\phi}}_{21+}^T \right. \vspace{0.1in} \\
& \d\left.\!+\!\frac{2}{\pi\Sigma_-}\left[
\tilde{\Psi}_{12}\l(\bar{\tilde{\phi}}_{22}(t,\bar{k})^T-\mathbb{I}\r)e^{-2ikL}-\alpha(\tilde{\Psi}_{11}-\mathbb{I})\bar{\tilde{\phi}}_{21}^T
\right]_+\right\}dk,
\end{array}
\ene

Since the integrand in Eq.~(\ref{u0x}) is an odd function about $k$, which makes sure that the contour $\partial D_1^0$ can be replaced by
$\partial D_3^0$, thus the substitution of Eq.~(\ref{pp}) into Eq.~(\ref{u0x}) yields the Dirichlet boundary values $u_{0j}(t),\, j=1,2$ again.
Similarly, we can also deduce the Dirichlet boundary values $v_{0j}(t),\, j=1,2$ from Eq.~(\ref{v0}).

 \subsection*{\it 5.3. \, The linearizable boundary conditions}

\quad In the following we investigate the linearizable boundary conditions for the above-mentioned representations.

\vspace{0.1in}
\noindent {\bf Theorem 5.4.} {\it Let $q_j(x, t=0)=q_{0j}(x),\, j=1,0,-1$ be the initial data of the spin-1 GP system (\ref{pnls}) on the interval $x\in [0, L]$, and one of the following boundary data, either

(i) the Dirichlet boundary data $q_j(x=0,t)=u_{0j}(t)=0$ and $q_j(x=L,t)=v_{0j}(t)=0,\, j=1,0,-1,$

or

(ii) the Robin boundary data $q_{jx}(x=0,t)-\chi q_j(x=0,t)=u_{1j}(t)-\chi u_{0j}(t)=0,\, j=1,0,-1$ and $q_{jx}(x=L,t)-\chi q_j(x=L,t)=v_{1j}(t)-\chi v_{0j}(t)=0,\, j=1,0,-1$, where $\chi$ is a real parameter.

Then the eigenfunctions $\Psi(t,k)$ and $\phi(t,k)$ can be given by

(i) \bes\bee\label{psiglm1}
&\Psi(t,k)=\mathbb{I}+\left(\begin{array}{cc} \hat{\tilde{L}}_{11} & \hat{\tilde{L}}_{12} \vspace{0.1in}\\
      \hat{\tilde{L}}_{21} & \hat{\tilde{L}}_{22} \end{array}\right), \\
\label{phiglm1}
&\phi(t,k)=\mathbb{I}+\left(\begin{array}{cc} \hat{\tilde{\mathcal{L}}}_{11} & \hat{\tilde{\mathcal{L}}}_{12} \vspace{0.1in}\\
      \hat{\tilde{\mathcal{L}}}_{21} & \hat{\tilde{\mathcal{L}}}_{22} \end{array}\right),
\ene\ees
where the $4\times 4$ matrix-valued function $L(t, s)=(L_{ij})_{4\times 4}$  satisfies a reduced Goursat system
\bee \label{lge2}
\left\{\begin{array}{l}
\tilde{L}_{11t}+\tilde{L}_{11s}=iu_1^T\tilde{L}_{21}, \vspace{0.1in} \\
\tilde{L}_{12t}-\tilde{L}_{12s}=iu_1^T\tilde{L}_{22}, \vspace{0.1in}\\
\tilde{L}_{21t}-\tilde{L}_{21s}=-i\alpha \bar{u}_1\tilde{L}_{11}, \vspace{0.1in}\\
\tilde{L}_{22t}+\tilde{L}_{22s}=-i\alpha \bar{u}_1\tilde{L}_{12},
\end{array}\right.
\ene
with the initial data (cf. Eq.~(\ref{lgi}))
\bee\label{lgi2}
\tilde{L}_{11}(t, -t)=\tilde{L}_{22}(t, -t)=0, \quad \tilde{L}_{12}(t, t)=\frac{i}{2}u_1^T(t),\quad \tilde{L}_{21}(t, t)=-\frac{i}{2}\alpha \bar{u}_1(t),
\ene
Similarly, the $4\times 4$ matrix-valued function $\mathcal{L}(t, s)=(\mathcal{L}_{ij})_{4\times 4}$  satisfies
the analogous system (\ref{lge2}) with the initial data (\ref{lgi2}) by replacing $u_1(t)$ with $v_1(t)$.}

{\it (ii)
\bes\bee
\label{psiglm2}
&\Psi(t,k)=\d\mathbb{I}+\left(\begin{array}{cc} \hat{\tilde{L}}_{11} & \hat{\tilde{L}}_{12} \vspace{0.1in}\\
      \hat{\tilde{L}}_{21} & \hat{\tilde{L}}_{22} \end{array}\right)+\left(\begin{array}{cc} -\d\frac{i}{2}u_0^T\hat{\tilde{G}}_{21} & k\hat{\tilde{G}}_{12} \vspace{0.1in}\\
 k\hat{\tilde{G}}_{21} & \d\frac{i\alpha}{2}\bar{u}_0\hat{\tilde{G}}_{12}
 \end{array}
 \right),\\
\label{phiglm2}
&\phi(t,k)=\d\mathbb{I}+\left(\begin{array}{cc} \hat{\tilde{\mathcal{L}}}_{11} & \hat{\tilde{\mathcal{L}}}_{12} \vspace{0.1in}\\
      \hat{\tilde{\mathcal{L}}}_{21} & \hat{\tilde{\mathcal{L}}}_{22} \end{array}\right)+\left(\begin{array}{cc} -\d\frac{i}{2}v_0^T\hat{\tilde{\mathcal{G}}}_{21} & k\hat{\tilde{\mathcal{G}}}_{12} \vspace{0.1in}\\
 k\hat{\tilde{\mathcal{G}}}_{21} &\d \frac{i\alpha}{2}\bar{v}_0\hat{\tilde{\mathcal{G}}}_{12}
 \end{array}
 \right),
\ene\ees
where the $4\times 4$ matrix-valued functions $L(t, s)=(L_{ij})_{4\times 4}$ and $G(t, s)=(G_{ij})_{4\times 4}$  satisfy the reduced nonlinear
Goursat system
\bee \label{lge2g}
\label{linearl1}
\left\{\begin{array}{l}
\tilde{L}_{11t}+\tilde{L}_{11s}=i\chi u_0^T\tilde{L}_{21}+\d\frac{1}{2}\left(i\dot{u}_0^T-\alpha u_0^T\bar{u}_0u_0^T\right)\tilde{G}_{21}, \vspace{0.1in} \\
\tilde{L}_{12t}-\tilde{L}_{12s}=i\chi u_0^T\tilde{L}_{22}, \vspace{0.1in}\\
\tilde{L}_{21t}-\tilde{L}_{21s}=-i\chi\alpha \bar{u}_0\tilde{L}_{11}, \vspace{0.1in}\\
\tilde{L}_{22t}+\tilde{L}_{22s}=-i\chi\alpha \bar{u}_0\tilde{L}_{12}-\d\frac{1}{2}\left(i\alpha\dot{\bar{u}}_0+\bar{u}_0u^T_0\bar{u}_0\right)\tilde{G}_{12}, \vspace{0.1in}\\
\tilde{G}_{12t}-\tilde{G}_{12s}=2u_0^T\tilde{L}_{22}, \vspace{0.1in}\\
\tilde{G}_{21t}-\tilde{G}_{21s}=2\alpha \bar{u}_0\tilde{L}_{11},
\end{array}\right.
\ene
with the initial data (cf. Eq.~(\ref{lgi}))
\bee\label{lgi2g}
\left\{\begin{array}{l}
\tilde{L}_{11}(t, -t)=\tilde{L}_{22}(t, -t)=0, \vspace{0.1in}\\
\tilde{L}_{12}(t, t)=\d\frac{i}{2}\chi u_0^T(t),\vspace{0.1in}\\
\tilde{L}_{21}(t, t)=-\d\frac{i}{2}\alpha\chi \bar{u}_0(t),\vspace{0.1in}\\
\tilde{G}_{12}(t, t)=u_0^T(t),\vspace{0.1in}\\
\tilde{G}_{21}(t, t)=\alpha\bar{u}_0(t),
\end{array}\right.
\ene
Similarly, the $4\times 4$ matrix-valued functions $\mathcal{L}(t, s)=(\mathcal{L}_{ij})_{4\times 4}$ and $\mathcal{G}(t, s)=(\mathcal{G}_{ij})_{4\times 4}$  satisfy the similar system (\ref{lge2g}) with the initial data (\ref{lgi2g}) by replacing $u_0(t)$ with $v_0(t)$.}

\vspace{0.1in}
\noindent
{\bf Proof.} Let us proof that the linearizable boundary data can be regarded as the special cases of Proposition 5.1.

{\it Case i.} \, Dirichlet zero boundary data: $q_j(x=0,t)=u_{0j}(t)=0,\, j=1,0,-1$.

It follows from the second one of system (\ref{lge}) that $\tilde{G}_{ij}(t,s),\, i,j=1,2$ satisfy
\bee\label{linearg1}
\left\{\begin{array}{l}
\tilde{G}_{11t}+\tilde{G}_{11s}=iu_1^T\tilde{G}_{21}, \vspace{0.1in}\\
\tilde{G}_{12t}-\tilde{G}_{12s}=iu_1^T\tilde{G}_{22}, \vspace{0.1in}\\
\tilde{G}_{21t}-\tilde{G}_{21s}=-i\alpha \bar{u}_1\tilde{G}_{11}, \vspace{0.1in}\\
\tilde{G}_{22t}+\tilde{G}_{22s}=-i\alpha \bar{u}_1\tilde{G}_{12},
\end{array}\right.
\ene
with the initial data (cf. Eq.~(\ref{lgi}))
\bee
\tilde{G}_{11}(t, -t)=\tilde{G}_{22}(t, -t)=0, \quad \tilde{G}_{12}(t, t)=\tilde{G}_{21}(t, t)=0,
\ene
Therefore, the unique solution of Eq.~(\ref{linearg1}) is trivial, that is, $\tilde{G}_{ij}(t,s)=0,\, i,j=1,2$ such that Eq.~(\ref{psiglm}) reduces to Eq.~(\ref{psiglm1}) and the condition (\ref{lge}) with Eq.~(\ref{lgi}) becomes  Eq.~(\ref{lge2}) with Eq.~(\ref{lgi2}). Similarly, for the Dirichlet zero boundary data $q_j(x=L,t)=v_{0j}(t)=0,\, j=1,0,-1$, we can also have Eq.~(\ref{phiglm1}).

{\it Case ii.} the Robin boundary data
\bee
 q_{jx}(x=0,t)-\chi q_j(x=0,t)=u_{1j}(t)-\chi u_{0j}(t)=0,\quad j=1,0,-1,\ene
imply that the Dirichlet and Neumann boundary data have the linear relation
\bee\label{dnc}
 u_1(t)=\chi u_0(t).
 \ene

Introduce the new $4\times 4$  matrix $Q(t, s)=(Q_{ij})_{4\times 4}$ by
\bee\label{qm}
\left\{\begin{array}{l}
\d\tilde{Q}_{11}(t,s)=\tilde{L}_{11}(t,s)-\frac{i\chi}{2}\tilde{G}_{11}(t,s), \vspace{0.1in}\\
\d\tilde{Q}_{12}(t,s)=\tilde{L}_{12}(t,s)-\frac{i\chi}{2}\tilde{G}_{12}(t,s), \vspace{0.1in}\\
\d\tilde{Q}_{21}(t,s)=\tilde{L}_{21}(t,s)+\frac{i\chi}{2}\tilde{G}_{21}(t,s), \vspace{0.1in}\\
\d\tilde{Q}_{22}(t,s)=\tilde{L}_{22}(t,s)+\frac{i\chi}{2}\tilde{G}_{22}(t,s),
\end{array}\right.
\ene

It follows from Eqs.~(\ref{lge}) and (\ref{qm}) with Eq.~(\ref{dnc}) that $\tilde{Q}_{ij}(t,s),\, \tilde{G}_{ij}(t,s),\, i,j=1,2$ satisfy
\bee\label{linearg2}
\left\{\begin{array}{l}
\tilde{Q}_{11t}-\tilde{Q}_{11s}=\d\left(-\frac{\alpha}{2}u_0^T\bar{u}_0u_0^T+\frac{i}{2}\dot{u}_0^T+\frac{\chi^2}{2}u_0^T\right)\tilde{G}_{21}, \vspace{0.1in}\\
\tilde{Q}_{12t}-\tilde{Q}_{12s}=\d\left(-\frac{\alpha}{2}u_0^T\bar{u}_0u_0^T+\frac{i}{2}\dot{u}_0^T+\frac{\chi^2}{2}u_0^T\right)\tilde{G}_{22}, \vspace{0.1in}\\
\tilde{Q}_{21t}-\tilde{Q}_{21s}=\d\left(-\frac{1}{2}\bar{u}_0u_0^T\bar{u}_0-\frac{i\alpha}{2}\dot{\bar{u}}_0+\frac{\alpha\chi^2}{2}\bar{u}_0\right)\tilde{G}_{11}, \vspace{0.1in}\\
\tilde{Q}_{22t}-\tilde{Q}_{22s}=\d\left(-\frac{1}{2}\bar{u}_0u_0^T\bar{u}_0-\frac{i\alpha}{2}\dot{\bar{u}}_0+\frac{\alpha\chi^2}{2}\bar{u}_0\right)\tilde{G}_{12}, \vspace{0.1in}\\
\tilde{G}_{11t}+\tilde{G}_{11s}=2u_0^T \tilde{Q}_{21}, \vspace{0.1in} \\
\tilde{G}_{12t}-\tilde{G}_{12s}=2u_0^T \tilde{Q}_{22}, \vspace{0.1in}\\
\tilde{G}_{21t}-\tilde{G}_{21s}=2\alpha \bar{u}_0 \tilde{Q}_{11}, \vspace{0.1in}\\
\tilde{G}_{22t}+\tilde{G}_{22s}=2\alpha \bar{u}_0 \tilde{Q}_{12},
\end{array}\right.
\ene
with the initial conditions (cf. Eq.~(\ref{lgi}))
\bee\left\{\begin{array}{l}
\tilde{G}_{11}(t, -t)=\tilde{G}_{22}(t, -t)=0, \vspace{0.1in}\\
\tilde{G}_{12}(t, t)=u_0^T(t),\vspace{0.1in}\\
 \tilde{G}_{21}(t, t)=\alpha\bar{u}_0(t),\vspace{0.1in}\\
\tilde{Q}_{12}(t, t)=\tilde{Q}_{21}(t, t)=0, \vspace{0.1in}\\
\tilde{Q}_{11}(t, -t)=\tilde{Q}_{22}(t, -t)=0,
\end{array}\right.
\ene

Therefore, the unique solution of Eq.~(\ref{linearg1}) is also trivial, that is, $\tilde{Q}_{12}(t,s)=\tilde{Q}_{21}(t,s)
=\tilde{G}_{11}(t,s)=\tilde{G}_{22}(t,s)=0$. As a result, Eq.~(\ref{psiglm}) reduces to Eq.~(\ref{psiglm2}) and the condition
(\ref{lge}) with Eq.~(\ref{lgi}) becomes Eq.~(\ref{lge2g}) with Eq.~(\ref{lgi2g}).

Similarly, for the Robin boundary data  $q_{jx}(x=L,t)-\chi q_j(x=L,t)=v_{1j}(t)-\chi v_{0j}(t)=0,\, j=1,0,-1$, that is, $v_1(t)=\chi v_0(t)$, we can also find  Eq.~(\ref{phiglm2}). $\square$ \\

According to Theorem 5.3 and Theorem 5.4, we have the following Proposition.

\vspace{0.1in}
\noindent{\bf Proposition 5.5.} {\it For the linearizable Dirichlet boundary data $u_0(t)=v_0(t)=0$, we have the Neumann boundary data $u_1(t)$ and $v_1(t)$:
\bes\bee
& u_1^T(t)=\d\frac{4i}{\pi}\d\int_{\partial D_1^0}k\tilde{\Psi}_{12}(\bar{\tilde{\phi}}^T_{22}-\mathbb{I})dk,\vspace{0.1in}\\
& v_1^T(t)=\d\frac{4i}{\pi}\d\int_{\partial D_1^0}k\tilde{\phi}_{12}(\bar{\tilde{\Psi}}^T_{22}-\mathbb{I})dk,
\ene\ees
where
\bee\no
\left\{\begin{array}{l}
\tilde{\Psi}_{12t}+4ik^2\tilde{\Psi}_{12}=iu_1^T(\tilde{\Psi}_{22}+\mathbb{I}), \vspace{0.1in}\\
\tilde{\Psi}_{22t}=-i\alpha\bar{u}_1\tilde{\Psi}_{12}, \vspace{0.1in}\\
\tilde{\phi}_{12t}+4ik^2\tilde{\phi}_{12}=iv_1^T(\tilde{\phi}_{22}+\mathbb{I}), \vspace{0.1in}\\
\tilde{\phi}_{22t}=-i\alpha\bar{v}_1\tilde{\phi}_{12}.
\end{array}\right.
\ene }

\noindent {\bf Remark 5.6}. The analogous analysis of the Fokas unified method will be extended to analyze the IBV problems for other integrable nonlinear evolution PDEs with $4\times 4$ Lax pairs both on the the half-line and the finite interval, such as the three-component NLS equations~\cite{yan3nls}, the three-component coupled derivative NLS equations, etc..

\section{Conclusions and discussions}

\quad In conclusion, we have extended the Fokas method to explore the initial-boundary value problem for the integrable spin-1 GP equations (\ref{pnls})
with a $4\times 4$ Lax pair on the finite interval $x\in [0, L]$. We find that the solution of the system can be generated by means of
the solution of the $4\times 4$ matrix RH problem formulated in the complex $k$-plane. Moreover, the relevant jump matrices of the RH problem can be explicitly found using the spectral functions $\{s(k),\, S(k),\, S_L(k)\}$ related to the initial data and the Dirichlet-Neumann boundary data at $x=0$ and $x=L$, respectively. We present the global relation to deduce two distinct but equivalent types of representations (i.e., one by using the large $k$ of asymptotics of the eigenfunctions and another one in terms of the Gelfand-Levitan-Marchenko (GLM) method) for the Dirichlet and Neumann boundary value problems. In particular, the obtained results for the boundary value problems on the finite interval can reduce to ones on the half-line as the length $L$ approaches to infinity. Finally, we also give the linearizable boundary conditions for the GLM representations. It is still open problem that how to further study the obtained matrix RH problem. The nonlinear steepest descent method~\cite{rh,rh2} and the numerical method~\cite{num} may be chosen to explore it, which will be considered in the future. The idea used in this paper can also be extended to other integrable NLEES with $4\times 4$ Lax pairs on the finite interval.

\vspace{0.2in}
\noindent
{\bf Acknowledgments}
\vspace{0.1in}

 This work was partially supported by the NSFC under Grant No.11571346 and the Youth Innovation Promotion Association, CAS.

\vspace{0.2in}
\noindent {\bf Appendix.} The asymptotic behavior of the eigenfunction $\mu(x, t, k)$ in the Lax pair (\ref{mulax}).

\vspace{0.1in}
We rewrite the matrices in the Lax pair (\ref{mulax}) as
\bee\begin{array}{c}
 \sigma_4=\left(\begin{array}{cc}  \mathbb{I}_{2\times 2} & 0 \vspace{0.1in}\\  0 & -\mathbb{I}_{2\times 2} \end{array} \right),\quad
 U(x,t)=\left(\begin{array}{cc} 0 & \mathbb{Q}^T \vspace{0.1in}\\ \alpha\bar{\mathbb{Q}} & 0 \end{array} \right),\,\ \mathbb{Q}=\left(\begin{array}{cc} q_1 & \beta q_0 \vspace{0.1in}\\  q_0 & q_{-1}  \end{array} \right), \vspace{0.15in}\\
  V(x,t,k)=2kU+V_0, \quad
 V_0=\left(\begin{array}{cc}-i\alpha\mathbb{Q}^T\bar{\mathbb{Q}} & i\mathbb{Q}^T_x \vspace{0.1in}\\
              -i\alpha\bar{\mathbb{Q}}_x & i\alpha\bar{\mathbb{Q}}\mathbb{Q}^T \end{array} \right),
 \end{array}
 \ene
where
 \bee\begin{array}{c}
  \mathbb{Q}^T\bar{\mathbb{Q}}=\left(\begin{array}{cc}
        |q_1|^2+|q_0|^2 & \beta q_1\bar{q}_0+q_0\bar{q}_{-1} \vspace{0.1in}\\
        \beta q_0\bar{q}_1+q_{-1}\bar{q}_0 & |q_{-1}|^2+|q_0|^2 \end{array}
  \right), \vspace{0.15in}\\
  \bar{\mathbb{Q}}\mathbb{Q}^T=\left(\begin{array}{cc}
        |q_1|^2+|q_0|^2 & \beta q_{-1}\bar{q}_0+q_0\bar{q}_{1} \vspace{0.1in}\\
        \beta q_0\bar{q}_{-1}+q_{1}\bar{q}_0 & |q_{-1}|^2+|q_0|^2 \end{array}
  \right),
\end{array}
\ene

Let the eigenfunction $\mu(x,t,k)$ of the Lax pair (\ref{mulax}) be of the form
\bee \label{muex}
\mu(x,t,k)=D^{(0)}(x,t)+\frac{D^{(1)}(x,t)}{k}+\frac{D^{(2)}(x,t)}{k^2}+\frac{D^{(3)}(x,t)}{k^3}+\cdots,
\ene
where the $4\times 4$ matrices $D_j(x,t)$'s are the functions of $(x,t)$ to be determined, then the substitution of Eq.~(\ref{muex}) into the Lax pair (\ref{mulax}) yields the recurrence relations
\bee\label{lax-exp}
\begin{array}{l}
x{\rm -part}: \, \left\{\begin{array}{l}
O(k):\, [\sigma_4, D^{(0)}]=0, \vspace{0.1in} \\
O(k^{-j}):\, D_{x}^{(j)}+i[\sigma_4, D^{(j+1)}]=U^{(j)},\quad j=0,1,2,... \\
\end{array}\right. \vspace{0.1in} \\
t{\rm -part}: \,\left\{\begin{array}{l}
O(k^2): \, [\sigma_4, D^{(0)}]=0, \vspace{0.1in}\\
O(k): \, i[\sigma_4, D^{(1)}]=UD^{(0)}, \vspace{0.1in}\\
O(k^{-j}):\,  D_{t}^{(j)}+2i[\sigma_4, D^{(j+2)}]=2UD^{(j+1)}+V_0D^{(j)},\quad j=0,1,2,... \\
\end{array}\right.
\end{array}
\ene

For  convenience, we write a $4\times 4$ matrix $D^{(j)}=(D^{(j)}_{ls})_{4\times 4}$ as
\bee\label{mde}
\begin{array}{c}
D^{(j)}=\left(\begin{array}{cc} \tilde{D}^{(j)}_{11} & \tilde{D}^{(j)}_{12} \vspace{0.1in}\\ \tilde{D}^{(j)}_{21} & \tilde{D}^{(j)}_{22}\end{array} \right),\quad
\tilde{D}^{(j)}_{11}=\left(\begin{array}{cc} D^{(j)}_{11} & D^{(j)}_{12}  \vspace{0.1in}\\ D^{(j)}_{21} & D^{(j)}_{22} \end{array} \right),\quad
\tilde{D}^{(j)}_{12}=\left(\begin{array}{cc} D^{(j)}_{13} & D^{(j)}_{14}  \vspace{0.1in}\\ D^{(j)}_{23} & D^{(j)}_{24} \end{array} \right),\vspace{0.1in}\\
\tilde{D}^{(j)}_{21}=\left(\begin{array}{cc} D^{(j)}_{31} & D^{(j)}_{32}  \vspace{0.1in}\\ D^{(j)}_{41} & D^{(j)}_{42} \end{array} \right),\quad
\tilde{D}^{(j)}_{22}=\left(\begin{array}{cc} D^{(j)}_{33} & D^{(j)}_{34}  \vspace{0.1in}\\ D^{(j)}_{43} & D^{(j)}_{44} \end{array} \right),
\end{array}\ene

It follows from $O(k^2)$ and  $O(k)$ in the $t$-part of Eq.~(\ref{lax-exp}) that we have
\bee \label{tt}
\left\{\begin{array}{l}
 \tilde{D}_{12}^{(0)}=\tilde{D}_{21}^{(0)}=0, \vspace{0.1in} \\
 \d\tilde{D}_{12}^{(1)}=-\frac{i}{2}\mathbb{Q}^T\tilde{D}_{22}^{(0)}, \vspace{0.1in} \\
\d \tilde{D}_{21}^{(1)}=\frac{i\alpha}{2}\bar{\mathbb{Q}}\tilde{D}_{11}^{(0)},
 \end{array}\right.
 \ene

From $O(k^{-j}),\, j=0,1,2,...$ in the $t$-part of Eq.~(\ref{lax-exp}), we have (cf. Eq.~(\ref{mde}))
\bee \label{t0}
\begin{array}{l}
\left(\begin{array}{cc}
\tilde{D}_{11t}^{(j)}  &  \tilde{D}_{12t}^{(j)}   \vspace{0.1in} \\
 \tilde{D}_{21t}^{(j)} & \tilde{D}_{22t}^{(j)}
\end{array}\right)
+4i\left(\begin{array}{cc} 0 & \tilde{D}_{12}^{(j+2)}    \vspace{0.1in} \\
 -\tilde{D}_{21}^{(j+2)} & 0 \end{array}\right) =2\left(\begin{array}{cc}
  \mathbb{Q}^T\tilde{D}_{21}^{(j+1)}  &  \mathbb{Q}^T\tilde{D}_{22}^{(j+1)}  \vspace{0.1in} \\
   \alpha\bar{\mathbb{Q}}\tilde{D}_{11}^{(j+1)} & \alpha\bar{\mathbb{Q}}\tilde{D}_{12}^{(j+1)}
  \end{array}\right)\vspace{0.1in} \\
\qquad\qquad\qquad +i\left(\begin{array}{cc}
  \mathbb{Q}_x^T\tilde{D}_{21}^{(j)}-\alpha\mathbb{Q}^T\bar{\mathbb{Q}}\tilde{D}_{11}^{(j)}  &
   \mathbb{Q}_x^T\tilde{D}_{22}^{(j)}-\alpha\mathbb{Q}^T\bar{\mathbb{Q}}\tilde{D}_{12}^{(j)}  \vspace{0.1in} \\
   \alpha\bar{\mathbb{Q}}\mathbb{Q}^T\tilde{D}_{21}^{(j)}-\alpha\bar{\mathbb{Q}}_x\tilde{D}_{11}^{(j)}
   & \alpha\bar{\mathbb{Q}}\mathbb{Q}^T\tilde{D}_{22}^{(j)}-\alpha\bar{\mathbb{Q}}_x\tilde{D}_{12}^{(j)}
\end{array}\right),
\end{array}
\ene
which leads to
\bee\label{tpartjg}
\left\{\begin{array}{l}
\tilde{D}_{11t}^{(j)}=2\mathbb{Q}^T\tilde{D}_{21}^{(j+1)}
  +i\left[\mathbb{Q}_x^T\tilde{D}_{21}^{(j)}-\alpha\mathbb{Q}^T\bar{\mathbb{Q}}\tilde{D}_{11}^{(j)}\right], \vspace{0.1in} \\
\tilde{D}_{22t}^{(j)}=2\alpha\bar{\mathbb{Q}}\tilde{D}_{12}^{(j+1)}
+i\left[\alpha\bar{\mathbb{Q}}\mathbb{Q}^T\tilde{D}_{22}^{(j)}-\alpha\bar{\mathbb{Q}}_x\tilde{D}_{12}^{(j)}\right],\vspace{0.1in}\\
\d\tilde{D}_{12}^{(j+2)} =\frac{i}{4} \tilde{D}_{12t}^{(j)}-\frac{i}{2}\mathbb{Q}^T\tilde{D}_{22}^{(j+1)}
  +\frac{1}{4}\left[\mathbb{Q}_x^T\tilde{D}_{22}^{(j)}-\alpha\mathbb{Q}^T\bar{\mathbb{Q}}\tilde{D}_{12}^{(j)}\right],\vspace{0.1in}\\
\d\tilde{D}_{21}^{(j+2)}=-\frac{i}{4}\tilde{D}_{21t}^{(j)}+\frac{i}{2}\alpha\bar{\mathbb{Q}}\tilde{D}_{11}^{(j+1)}
+\frac{1}{4}\left[\alpha\bar{\mathbb{Q}}_x\tilde{D}_{11}^{(j)}-\alpha\bar{\mathbb{Q}}\mathbb{Q}^T\tilde{D}_{21}^{(j)}\right],
\end{array}\right.
\ene

Eq.~(\ref{tpartjg}) with $j=0$  and Eq.~(\ref{tt}) yields
\bee\label{tj0}
\left\{\begin{array}{l}
 \tilde{D}_{11t}^{(0)}=2\mathbb{Q}^T\tilde{D}_{21}^{(1)}-i\alpha \mathbb{Q}^T\bar{\mathbb{Q}}\tilde{D}_{11}^{(0)}=0, \vspace{0.1in}\\
 \tilde{D}_{22t}^{(0)}=2\alpha\bar{\mathbb{Q}}\tilde{D}_{12}^{(1)}+i\alpha \bar{\mathbb{Q}}\mathbb{Q}^T\tilde{D}_{22}^{(0)}=0, \vspace{0.1in}\\
 \d\tilde{D}_{12}^{(2)}=-\frac{i}{2}\mathbb{Q}^T\tilde{D}_{22}^{(1)}+\frac{1}{4}\mathbb{Q}_x^T\tilde{D}_{22}^{(0)}, \vspace{0.1in}\\
\d \tilde{D}_{21}^{(2)}=\frac{i\alpha}{2} \bar{\mathbb{Q}}\tilde{D}_{11}^{(1)}+\frac{\alpha}{4}\bar{\mathbb{Q}}_x\tilde{D}_{11}^{(0)},
 \end{array}\right.    \ene

Eq.~(\ref{tpartjg}) with $j=1$ yields
\bee\label{tj1}
\left\{\begin{array}{l}
 \tilde{D}_{11t}^{(1)}=\d\frac{\alpha}{2}\left(\mathbb{Q}^T\bar{\mathbb{Q}}_x-\mathbb{Q}^T_x\bar{\mathbb{Q}}\right)\tilde{D}_{11}^{(0)}, \vspace{0.1in}\\
 \tilde{D}_{22t}^{(1)}=\d\frac{\alpha}{2}\left(\bar{\mathbb{Q}}\mathbb{Q}^T_x-\bar{\mathbb{Q}}_x\mathbb{Q}^T\right)\tilde{D}_{22}^{(0)}, \vspace{0.1in}\\
 \d\tilde{D}_{12}^{(3)} =\frac{1}{8}\left[\mathbb{Q}^T_t+i\alpha\mathbb{Q}^T\bar{\mathbb{Q}}\mathbb{Q}^T\right]\tilde{D}_{22}^{(0)}
  -\frac{i}{2}\mathbb{Q}^T\tilde{D}_{22}^{(2)}+\frac{1}{4}\mathbb{Q}^T_x\tilde{D}_{22}^{(1)},\vspace{0.1in}\\
\d\tilde{D}_{21}^{(3)}=\frac{1}{8}\left[\alpha\bar{\mathbb{Q}}_t
   -i\bar{\mathbb{Q}}\mathbb{Q}^T\bar{\mathbb{Q}}\right]\tilde{D}_{11}^{(0)}
  +\frac{i\alpha}{2}\bar{\mathbb{Q}}\tilde{D}_{11}^{(2)}+\frac{1}{4}\alpha\bar{\mathbb{Q}}_x\tilde{D}_{11}^{(1)},
   \end{array}\right.    \ene

Eq.~(\ref{tpartjg}) with $j=2$ yields
\bee\label{tj2}
\left\{\begin{array}{rl}
\tilde{D}_{11t}^{(2)}=&\!\!\! 2\mathbb{Q}^T\tilde{D}_{21}^{(3)}
  +i\left[\mathbb{Q}_x^T\tilde{D}_{21}^{(2)}-\alpha\mathbb{Q}^T\bar{\mathbb{Q}}\tilde{D}_{11}^{(2)}\right] \vspace{0.1in}\\
  =&\!\!\! \d\frac{1}{4}\left[\alpha\mathbb{Q}^T\bar{\mathbb{Q}}_t-i(\mathbb{Q}^T\bar{\mathbb{Q}})^2
   +i\alpha\mathbb{Q}^T_x\bar{\mathbb{Q}}_x\right]\tilde{D}_{11}^{(0)}
  +\frac{\alpha}{2}(\mathbb{Q}^T\bar{\mathbb{Q}}_x-\mathbb{Q}^T_x\bar{\mathbb{Q}})\tilde{D}_{11}^{(1)}, \vspace{0.1in} \\
\tilde{D}_{22t}^{(2)}=& \!\!\! 2\alpha\bar{\mathbb{Q}}\tilde{D}_{12}^{(3)}
 +i\left[\alpha\bar{\mathbb{Q}}\mathbb{Q}^T\tilde{D}_{22}^{(2)}-\alpha\bar{\mathbb{Q}}_x\tilde{D}_{12}^{(2)}\right]\vspace{0.1in}\\
 =&\!\!\! \d\frac{1}{4}\left[\alpha\bar{\mathbb{Q}}\mathbb{Q}^T_t+i(\bar{\mathbb{Q}}\mathbb{Q}^T)^2-i\alpha \bar{\mathbb{Q}}_x\mathbb{Q}^T_x\right]\tilde{D}_{22}^{(0)}
   +\frac{\alpha}{2}(\bar{\mathbb{Q}}\mathbb{Q}^T_x-\bar{\mathbb{Q}}_x\mathbb{Q}^T)\tilde{D}_{22}^{(1)},
\end{array}\right.
\ene

Similarly, it follows from $O(k^{-j}),\, j=0,1,2,...$ in the $x$-part of Eq.~(\ref{lax-exp}) that we have
\bee\label{xpartj}
\left(\begin{array}{cc}
\tilde{D}_{11x}^{(j)}  &  \tilde{D}_{12x}^{(j)}  \vspace{0.1in}\\
 \tilde{D}_{21x}^{(j)} & \tilde{D}_{22x}^{(j)}
\end{array}\right)
+2i\left(\begin{array}{cc}
 0 & \tilde{D}_{12}^{(j+1)}   \vspace{0.1in}\\
 -\tilde{D}_{21}^{(j+1)} & 0 \end{array}\right)
 =\left(\begin{array}{cc}
  \mathbb{Q}^T\tilde{D}_{21}^{(j)}  &  \mathbb{Q}^T\tilde{D}_{22}^{(j)} \vspace{0.1in}\\
   \alpha\bar{\mathbb{Q}}\tilde{D}_{11}^{(j)} & \alpha\bar{\mathbb{Q}}\tilde{D}_{12}^{(j)}
\end{array}\right),
\ene
which generates
\bee
\left\{\begin{array}{l}
\tilde{D}_{11x}^{(j)}=\mathbb{Q}^T\tilde{D}_{21}^{(j)}, \vspace{0.1in} \\
\tilde{D}_{22x}^{(j)}=\alpha\bar{\mathbb{Q}}\tilde{D}_{12}^{(j)},\vspace{0.1in}\\
\d\tilde{D}_{12}^{(j+1)} =\frac{i}{2} \tilde{D}_{12x}^{(j)}-\frac{i}{2}\mathbb{Q}^T\tilde{D}_{22}^{(j)},\vspace{0.1in}\\
\d\tilde{D}_{21}^{(j+1)}=-\frac{i}{2}\tilde{D}_{21x}^{(j)}+\frac{i}{2}\alpha\bar{\mathbb{Q}}\tilde{D}_{11}^{(j)},
\end{array}\right.
\ene

Thus, Eq.~(\ref{xpartj}) with $j=0$ yields
\bee \label{xj0}
\left\{\begin{array}{l}
\tilde{D}_{11x}^{(0)}=\tilde{D}_{22x}^{(0)}=0, \vspace{0.1in} \\
\d\tilde{D}_{12}^{(1)}=-\frac{i}{2}\mathbb{Q}^T\tilde{D}_{22}^{(0)}, \vspace{0.1in} \\
\d \tilde{D}_{21}^{(1)}=\frac{i\alpha}{2}\bar{\mathbb{Q}}\tilde{D}_{11}^{(0)},
\end{array}\right.\ene

Eq.~(\ref{xpartj}) with $j=1$ yields
\bee \label{xj1}
\left\{\begin{array}{l}
\d\tilde{D}_{11x}^{(1)}=\mathbb{Q}^T\tilde{D}_{21}^{(1)}=\frac{i\alpha}{2}\mathbb{Q}^T\bar{\mathbb{Q}}\tilde{D}_{11}^{(0)}, \vspace{0.1in} \\
\d\tilde{D}_{22x}^{(1)}=\alpha\bar{\mathbb{Q}}\tilde{D}_{12}^{(1)}=-\frac{i\alpha}{2}\bar{\mathbb{Q}}\mathbb{Q}^T\tilde{D}_{22}^{(0)},\vspace{0.1in}\\
\d\tilde{D}_{12}^{(2)} =\frac{i}{2} \tilde{D}_{12x}^{(1)}-\frac{i}{2}\mathbb{Q}^T\tilde{D}_{22}^{(1)}=
-\frac{i}{2}\mathbb{Q}^T\tilde{D}_{22}^{(1)}+\frac{1}{4}\mathbb{Q}_x^T\tilde{D}_{22}^{(0)},\vspace{0.1in}\\
\d\tilde{D}_{21}^{(2)}=-\frac{i}{2}\tilde{D}_{21x}^{(1)}+\frac{i}{2}\alpha\bar{\mathbb{Q}}\tilde{D}_{11}^{(1)}=
\frac{i\alpha}{2} \bar{\mathbb{Q}}\tilde{D}_{11}^{(1)}+\frac{\alpha}{4}\bar{\mathbb{Q}}_x\tilde{D}_{11}^{(0)},
\end{array}\right.
\ene
and Eq.~(\ref{xpartj}) with $j=2$ yields
\bee \label{xj2}
\left\{\begin{array}{l}
\d\tilde{D}_{11x}^{(2)}=\mathbb{Q}^T\tilde{D}_{21}^{(2)}=\frac{i\alpha}{2} \mathbb{Q}^T\bar{\mathbb{Q}}\tilde{D}_{11}^{(1)}+\frac{\alpha}{4}\mathbb{Q}^T\bar{\mathbb{Q}}_x\tilde{D}_{11}^{(0)}, \vspace{0.1in} \\
\d\tilde{D}_{22x}^{(2)}=\alpha\bar{\mathbb{Q}}\tilde{D}_{12}^{(2)}=
-\frac{i\alpha}{2}\bar{\mathbb{Q}}\mathbb{Q}^T\tilde{D}_{22}^{(1)}+\frac{\alpha}{4}\bar{\mathbb{Q}}\mathbb{Q}_x^T\tilde{D}_{22}^{(0)},
\end{array}\right.
\ene

It follows from Eqs.~(\ref{tj0}) and (\ref{xj0}) and the property of $\mu$ that we have
\bee\label{mu0}
\tilde{D}_{11}^{(0)}=\tilde{D}_{22}^{(0)}=\mathbb{I}_{2\times 2}.
\ene

Thus we can obtain other $D^{(j)},\, j=1,2,3,...$ such that we have the asymptotic behaviour of the eigenfunction $\mu(x, t, k)$ via Eq.~(\ref{muex}).

\end{document}